\documentstyle[12pt,epsfig,francais]{report}
\newcommand{\ds}{\displaystyle}

\newcommand{\be}{\begin{equation}}
\newcommand{\ee}{\end{equation}}
\newcommand{\ba}{\begin{eqnarray}}
\newcommand{\ea}{\end{eqnarray}}

\newcommand{\opb}{${}^{16}O + {}^{208}Pb $ }
\newcommand{\smg}{${}^{32}S + {}^{24}Mg$ }
\newcommand{\clmg}{${}^{35,37}Cl + {}^{24}Mg$ }
\newcommand{\pal}{${}^{31}P + {}^{27}Al$ }
\newcommand{\ssi}{${}^{32}S + {}^{28}Si$ }
\newcommand{\sca}{${}^{32}S + {}^{40}Ca$ }

\title{\bf{Effets de structure dans la diffusion des ions lourds}}
\author{\it{Th\`ese de Doctorat pr\'esent\'ee par}}
\date{Mohamed El Djalil KADI-HANIFI}

\begin{document}
\maketitle
\newpage
$ $
\newpage
\baselineskip=0.7cm
\setcounter{page}{1}
{\Large{\bf{ TABLE DES MATI\`ERES}}}
%\newline
\vskip 30mm
\par
\pagenumbering{roman}
\begin{itemize}
\item[1]. Introduction. \dotfill\ 1

\item[2]. Le Potentiel optique. \dotfill\ 7
\begin{itemize}
\item[2.1]. La th\'eorie de Feshbach. \dotfill\ 7
\item[2.2]. Mod\'elisations de la th\'eorie de Feshbach. \dotfill\ 10
\begin{itemize}
\item[2.2.1]. Mod\`eles ph\'enom\'enologiques et semiph\'enom\'enologiques. 
\dotfill 11
\item[2.2.2]. Mod\`ele de l'approximation de fermeture. \dotfill\ 14
\item[2.2.3]. Troncation des \'etats dans le calcul du potentiel optique. 
\dotfill\ 21
\end{itemize}
\end{itemize}

\item[3]. Absorption au voisinage de la barri\`ere. \dotfill\ 29
\begin{itemize}
\item[3.1]. Donn\'ees exp\'erimentales. \dotfill\ 29
\item[3.2]. Analyse semiph\'enom\'enologique. \dotfill\ 32
\begin{itemize}
\item[3.2.1]. Distributions angulaires et potentiels. \dotfill\ 35
\item[3.2.2]. Distributions radiales de l'absorption. \dotfill\ 44 
\item[3.2.3]. Distributions de spin de l'absorption. \dotfill\ 49
\end{itemize}
\item[3.3]. Analyse avec le potentiel de Feshbach. \dotfill\ 51
\begin{itemize}
\item[3.3.1]. Distributions angulaires et potentiels. \dotfill\ 54
\item[3.3.2]. Distributions radiales de l'absorption. \dotfill\ 64
\item[3.3.3]. Distributions de spin de l'absorption. \dotfill\ 66
\end{itemize}
\end{itemize}

\item[4]. Effets collectifs aux hautes \'energies. \dotfill\ 69
\begin{itemize}
\item[4.1]. Analyse de la diffusion \'elastique \opb. \dotfill\ 70
\item[4.2]. D\'ependance de l'\'energie du potentiel r\'eel. \dotfill\ 74
\item[4.3]. Calculs microscopiques. \dotfill\ 79
\begin{itemize}
\item[4.3.1]. Potentiel imaginaire. \dotfill\ 79
\item[4.3.2]. Diffusion \'elastique \opb. \dotfill\ 81
\end{itemize}
\end{itemize}

\item[5]. Consid\'erations finales. \dotfill\ 87
\begin{itemize}
\item[5.1]. R\'esultats et conclusions. \dotfill\ 87
\item[5.2]. Perspectives futures. \dotfill\ 90
\end{itemize}

\item[] R\'ef\'erences. \dotfill\ 93

\item[] Annexe. \dotfill\ 101
\end{itemize}

\newpage
\setcounter{chapter}{0}
\pagenumbering{arabic}
\chapter{{\large{{\bf{INTRODUCTION}}}}}

La recherche de la solution exacte du probl\`eme pos\'e par la collision de deux noyaux exige la 
r\'esolution d'un syst\`eme d'\'equations coupl\'ees dont le nombre et la complexit\'e 
augmentent avec l'\'energie de la collision. Cela est d\^u \`a 
l'accroissement rapide du nombre de canaux de r\'eaction \'energ\'etiquement 
accessibles, \`a travers lesquels le syst\`eme peut \'evoluer. Une solution 
alternative, formellement \'equivalente, a \'et\'e propos\'ee par 
Feshbach, {\bf(FE58)}, {\bf(FE62)}, 
et peut \^etre obtenue par l'analyse de la voie \'elastique en 
introduisant dans sa description un op\'erateur complexe appell\'e potentiel 
optique, qui simule les effets de toutes les voies explicitement exclues. La 
partie r\'eelle du potentiel optique contient deux termes : le premier, qui est 
local et ind\'ependant de l'\'energie d\'ecrit la diffusion \'elastique directe, tandis 
que le second terme, ou terme de polarisation, qui est non local et d\'ependant 
de l'\'energie, rend compte de la contribution des diff\'erents \'etats du syst\`eme 
comme voies interm\'ediaires d'acc\`es au canal \'elastique. Le 
terme d'absorption du potentiel qui est non local, et d\'ependant de l'\'energie, 
d\'ecrit toutes les voies qui ont \'et\'e explicitement ignor\'ees, car ouvertes, et 
qui, par cons\'equent diminuent le flux de la voie \'elastique. Entre les termes 
non locaux de polarisation et d'absorption existe une relation de dispersion 
indiquant leur origine commune. \'Etant donn\'e qu'habituellement, 
dans les calculs de diffusion \'elastique, on 
utilise des potentiels locaux, il est possible d'\'eliminer la non localit\'e en la 
transformant en une d\'ependance \'energ\'etique additionnelle. Cette transformation 
est responsable du fait que la relation de dispersion ne soit pas satisfaite, 
en g\'en\'eral, entre potentiels locaux. Cependant, sous certaines conditions qui 
affectent le caract\`ere hermitique du propagateur et le type de localisation, 
il a \'et\'e d\'emontr\'e que la relation de dispersion entre potentiels locaux est 
analytique {\bf(PA91)}. La forme la plus simple d'\'evaluer les termes locaux 
du potentiel est bas\'ee sur 
l'utilisation de potentiels ph\'enom\'enologiques qui se caract\'erisent par des 
facteurs de forme tr\`es divers {\bf(HO63)}. Parmis les nombreux facteurs de 
forme existants, celui de Woods-Saxon s'est av\'er\'e \^etre particuli\`erement 
utile. Cependant, bien qu'avec ces potentiels il soit possible d'ajuster les 
donn\'ees exp\'erimentales de diffusion \'elastique de mani\`ere ad\'equate dans la 
majorit\'e des cas, la param\'etrisation ph\'enom\'enologique est loin d'\^etre 
compl\`etement satisfaisante \`a cause de l'existence d'ambig\"uit\'es dans les 
param\`etres du potentiel optique qui rendent toute conclusion physique 
difficile.

L'utilisation de la m\'ethode de la double convolution entre les densit\'es des 
noyaux en interaction dans leurs \'etats fondamentaux et une interaction 
effective obtenue sur des bases physiques a \'et\'e une avanc\'ee importante, en ce 
sens qu'elle a permis de minimiser les ambig\"uit\'es. En particulier, l'usage 
de l'interaction M3Y {\bf(BE77)} a permis d'obtenir un potentiel r\'eel, 
ind\'ependant de l'\'energie, qui, compl\'et\'e par un terme d'absorption du type 
Woods-Saxon, a \'et\'e utilis\'e avec beaucoup de succ\`es dans l'analyse de plusieurs 
syst\`emes. Le plus grand succ\`es obtenu lors de l'analyse de donn\'ees de 
diffusion 
\'elastique avec ce type de potentiels est, sans nul doute, le fait d'avoir 
observ\'e pour la premi\`ere fois l'anomalie de seuil pour le syst\`eme 
$^{32}S+^{40}Ca$ {\bf(BA84)}, post\'erieurement confirm\'ee pour le syst\`eme 
$^{16}O+^{208}Pb$ {\bf(LI85)} et qui a donn\'e lieu \`a d'importants travaux men\'es 
dans le but de comprendre et d'interpr\'eter ce ph\'enom\`ene {\bf(BI91)}, 
{\bf(EL85)}, {\bf(MA86)}et {\bf(NA85)}  qui se produit lorsque 
l'on s'approche de la barri\`ere de Coulomb et qui consiste en une diminution de 
la partie imaginaire du potentiel, dans la r\'egion de sensibilit\'e, accompagn\'ee 
d'une rapide augmentation du potentiel r\'eel obligeant sa renormalisation par un 
facteur d\'ependant de l'\'energie. 

De nombreuses tentatives ont \'et\'e men\'ees  dans le 
but d'obtenir le terme d'absorption du potentiel sur des bases physiques. Il 
existe ainsi la m\'ethode de convolution des densit\'es des noyaux 
avec des interactions effectives complexes {\bf(FA83)} et {\bf(KH81)} 
d\'eduites de calculs de mati\`ere nucl\'eaire qui s'av\`ere \^etre la plus appropri\'ee 
pour les calculs aux \'energies tr\`es loin de la barri\`ere, ou encore les m\'ethodes 
semiclassiques 
{\bf(PO83)} calculant la contribution due aux \'etats in\'elastiques et de 
transfert de nucl\'eons les plus significatifs. Enfin, 
dans le formalisme de Feshbach, il est possible d'\'evaluer la contribution des 
canaux in\'elastiques de plus basses \'energies d'excitation en utilisant des facteurs 
de forme macroscopiques {\bf(KU81)}, {\bf(KU91)}, {\bf(LO77)} ou 
microscopiques {\bf(AN90)}.

Les plus grands succ\`es obtenus dans l'analyse des donn\'ees de diffusion 
\'elastique, nous les devons, sans aucun doute, au mod\`ele dit de fermeture, 
propos\'e par N. Vinh Mau {\bf(VI86)}, {\bf(VI87)} qui permet d'\'evaluer 
globalement le potentiel de Feshbach en assumant certaines hypoth\`eses 
simplificatrices. Ce mod\`ele a \'et\'e utilis\'e avec beaucoup de succ\`es, non 
seulement dans l'analyse des donn\'ees de diffusion \'elastique {\bf(BI91)}, 
{\bf(FE90)} et {\bf(VI91)}, mais aussi pour pr\'edire la fonction d'excitation 
de la fusion aux \'energies proches ou en dessous de la barri\`ere pour les 
syst\`emes sph\'eriques ou faiblement d\'eform\'es {\bf(VI93)}.

Cependant, lorsque l'\'energie de la collision est proche de la barri\`ere de Coulomb 
et que l'absorption, tel qu'il a \'et\'e observ\'e exp\'erimentalement pour les syst\`emes tr\`es d\'eform\'es 
{\bf(SA81)}, est due principalement \`a la contribution d'un nombre r\'eduit de canaux 
in\'elastiques, le mod\`ele de N.Vinh Mau surestime l'absorption. Il 
s'av\`ere qu'il est 
plus ad\'equat d'\'evaluer la contribution due \`a chacun des quelques canaux qui contr\^olent 
l'absorption. Une simple mod\'elisation du potentiel de Feshbach nous a permis d'aborder ces
calculs. \'Etant donn\'e que, pour certains des syst\`emes que nous avons \'etudi\'es, cette contribution
\`a l'absorption est trop faible \`a cause de l'importance des canaux de transfert de nucl\'eons, il
a \'et\'e necessaire de compl\'eter le potentiel imaginaire par un terme de 
surface ph\'enom\'enologique.

En dehors du caract\`ere semi-classique des calculs de la r\'ef\'erence 
{\bf(PO83)}, nos calculs du
potentiel de Feshbach diff\`erent de ces derniers dans la mesure o\`u nous avons inclus les effets de
non-localit\'e. De plus, notre mod\'elisation assume un propagateur complet 
et non adiabatique comme
celui des r\'ef\'erences {\bf(KU81)}, {\bf(KU91)} et {\bf(LO77)}, ce qui 
implique que non seulement nous avons
inclus 
la contribution des processus directs, mais aussi celle 
des processus de
multi \'etapes. Il est finalement important de signaler que nos calculs permettent d'\'evaluer la
contribution de tous les \'etats in\'elastiques d'inter\^et et pas seulement les \'etats RPA comme
c'est le cas dans la r\'ef\'erence {\bf(AN90)}.

Il est donc \'evident que l'int\'er\^et de nos recherches englobe diff\'erents 
aspects. 
En premier lieu, nous pr\'etendons d\'eterminer la relation qui existe entre la structure des noyaux en
interaction et l'absorption exp\'erimentalement observ\'ee \`a des \'energies proches de la barri\`ere de
Coulomb. En d'autres termes, il s'agit de voir quels sont les canaux de reaction qui contr\^olent
l'absorption. 
En second lieu, nous souhaitons localiser la r\'egion du potentiel imaginaire o\`u se produit
l'absorption, ce qui nous permettra de d\'eterminer le r\^ole que jouent les voies de volume dans ce
processus. 
Enfin, en vertu de son caract\`ere fortement collectif, le mod\`ele propos\'e 
par N. Vinh Mau ne peut \^etre valide qu'aux basses \'energies de diffusion. 
Pour cela, l'\'etude de son domaine d'applicabilit\'e en
termes d'\'energie s'av\`ere \^etre d'un grand int\'er\^et et peut nous 
informer sur les conditions sous
lesquelles les processus collectifs cessent de contr\^oler l'absorption. 
Ainsi, 
le travail r\'ealis\'e dans cette th\`ese est divis\'e en cinq chapitres.

Dans le chapitre deux nous d\'ecrivons bri\`evement le probl\`eme que pose l'\'etude de la 
collision entre deux ions et nous proposons sa r\'esolution au moyen de la 
th\'eorie de Feshbach du potentiel optique que nous d\'ecrivons sch\'ematiquement dans la 
limite du couplage faible o\`u la fonction de Green est diagonale. Nous 
obtenons, ainsi, les parties r\'eelle et imaginaire du potentiel qui, 
comme nous l'avons signal\'e 
precedemment, sont non-locales et li\'ees par une relation de 
dispersion. \'Etant 
donn\'e que, dans la pratique, on utilise des potentiels 
locaux, leurs 
composantes r\'eelle et imaginaire peuvent \^etre 
facilement \'evalu\'ees moyennant l'utilisation de diff\'erentes 
mod\'elisations dont nous justifions l'utilit\'e dans cette th\`ese. Celles 
que nous avons utilis\'ees ont \'et\'e :

1)- La mod\'elisation ph\'enom\'enologique dans laquelle les deux parties du 
potentiel sont d\'ecrites par des facteurs de forme du type Woods-Saxon dont les 
param\`etres sont ajust\'es de mani\`ere \`a reproduire les donn\'ees 
exp\'erimentales.

2)- La mod\'elisation semiph\'enom\'enologique dans laquelle la partie r\'eelle du 
potentiel se calcule par double convolution des densit\'es des deux noyaux dans 
leurs \'etats fondamentaux et de l'interaction effective M3Y. Le potentiel ainsi 
obtenu doit \^etre normalis\'e au moyen d'un facteur d\'ependant de 
l'\'energie afin de simuler le terme de polarisation pr\'evu par la th\'eorie 
de Feshbach. La 
partie imaginaire, elle, reste d\'ecrite de fa\c{c}on ph\'enom\'enologique.

3)- Le mod\`ele de l'approximation de fermeture propos\'e par N. Vinh Mau qui 
permet d'\'evaluer globalement la contribution de toutes les voies non-\'elastiques 
du potentiel noyau-noyau moyennant l'introduction d'hypoth\`eses 
simplificatrices que nous d\'ecrivons bri\`evement.

4)- La troncation de canaux. Dans la th\'eorie de Feshbach il est d\'emontr\'e que 
les canaux de r\'eaction dont l'\'energie d'excitation est sup\'erieure \`a 
l'\'energie 
$E_{cm}$ de la collision noyau-noyau ne participent pas \`a l'absorption et, par 
cons\'equent, ne contribuent pas au potentiel imaginaire. Partant de cette id\'ee, 
nous avons construit une mod\'elisation simple du potentiel, adapt\'ee au calcul du 
terme d'absorption aux \'energies proches de la barri\`ere de Coulomb, car ne 
contenant que la contribution des canaux de basse \'energie d'excitation qui sont 
facilement peupl\'es. L'introduction d'hypoth\`eses simplificatrices nous a 
permis d'obtenir des expressions du potentiel imaginaire qui peuvent \^etre 
\'evalu\'ees sans difficult\'e.

Le chapitre trois est consacr\'e \`a l'analyse des collisions noyau-noyau \`a des 
\'energies proches de la barri\`ere de Coulomb. Nous avons d'abord 
r\'ealis\'e des 
calculs semiph\'enom\'eno- logiques qui ont \'et\'e utilis\'es comme 
r\'ef\'erences pour les 
calculs microscopiques post\'erieurs. Une analyse consistante du syst\`eme 
$^{32}S+^{24}Mg$ \`a diff\'erentes \'energies nous a permis de d\'evelopper une 
m\'ethodologie que nous avons appliqu\'ee aux autres syst\`emes \'etudi\'es. Dans 
cette 
analyse, la partie r\'eelle du potentiel a \'et\'e d\'etermin\'ee par 
convolution des 
densit\'es des noyaux en interaction dans leurs \'etats fondamentaux avec 
l'interaction effective M3Y. La d\'ependance \'energ\'etique a \'et\'e introduite au 
moyen d'un facteur de renormalisation qui a \'et\'e d\'etermin\'e par ajustement aux 
donn\'ees exp\'erimentales. La partie imaginaire du potentiel a \'et\'e obtenue au 
moyen d'un terme du type Woods-Saxon dont la profondeur a \'et\'e maintenue fixe et 
dont le rayon aux \'energies les plus basses a \'et\'e d\'eduit de l'\'evolution en 
fonction de l'\'energie observ\'ee aux plus hautes \'energies. Avec ces potentiels,  
nous avons pu reproduire de fa\c{c}on satisfaisante les distributions 
angulaires de diffusion \'elastique et montrer la necessit\'e d'inclure un terme de 
polarisation coulombienne pour bien reproduire les donn\'ees exp\'erimentales aux 
\'energies les plus basses. Nous avons \'egalement calcul\'e les sections efficaces 
de r\'eaction, les distributions de spin et les distributions radiales de 
l'absorption qui nous ont permis de nous assurer que la valeur de la profondeur 
du terme d'absorption n'\'etait pas importante. Utilisant le potentiel r\'eel 
extrait des ajustements pr\'ecedents et le terme 
imaginaire calcul\'e dans le cadre du formalisme de Feshbach incluant les 
\'etats 
collectifs de moindre \'energie d'excitation, nous avons obtenu des pr\'edictions 
totalement compatibles avec les calculs semiph\'enom\'enologiques pr\'ec\'edents pour 
les syst\`emes les plus fortements d\'eform\'es et aux \'energies les plus 
basses. Dans le cas des syst\`emes pour lesquels 
les canaux de transfert contribuent de fa\c{c}on significative \`a 
l'absorption, 
des r\'esultats similaires ont \'et\'e obtenus, moyennant l'introduction d'un terme 
d'absorption de surface ph\'enom\'enologique dont la contribution est d\'etermin\'ee 
par ajustement aux donn\'ees exp\'erimentales.

Dans le chapitre quatre nous abordons l'\'etude des diffusions \'elastiques 
$^{16}O+^{208}Pb$ aux \'energies de 793 MeV et 1503 MeV, tr\`es loin de la 
barri\`ere de Coulomb. Les ajustements avec des potentiels ph\'enom\'enologiques du 
type Woods-Saxon nous permettent de reproduire les donn\'ees et d'obtenir les 
sections efficaces de r\'eaction. Les calculs r\'ealis\'es avec ces potentiels ainsi 
que les potentiels eux-m\^emes sont pris comme r\'ef\'erence. Utilisant 
alors le pouvoir de pr\'ediction de la relation de dispersion, nous 
avons obtenu la d\'ependance \'energ\'etique du potentiel r\'eel 
de convolution 
\`a partir des valeurs des potentiels imaginaires (Woods-Saxon) dont les 
param\`etres ont \'et\'e d\'eduits de l'ajustement aux donn\'ees exp\'erimentales \`a des 
\'energies comprises entre 78 MeV et 312.6 MeV suivant la m\'ethode d\'ecrite dans la 
r\'ef\'erence {\bf(MA86)}. Apr\`es nous \^etre assur\'es de la totale \'equivalence 
entre les potentiels r\'eels ph\'enom\'enologiques et ceux d\'eduits de la relation de 
dispersion, nous avons utilis\'e ces derniers et le mod\`ele propos\'e par N. Vinh 
Mau pour calculer le terme d'absorption. Les pr\'edictions de diffusion 
\'elastique et les sections efficaces de r\'eaction calcul\'ees avec ces potentiels 
microscopiques repr\'esentent un test important qui nous permet de conna\^{\i}tre 
les limites de validit\'e du mod\`ele de l'approximation de fermeture ainsi que 
les processus qui dominent l'absorption aux \'energies de 793 MeV et 1503 MeV.

En conclusion, nous pr\'esentons, au chapitre cinq, les r\'esultats obtenus et 
les cons\'equences qui en d\'ecoulent. Nous proposons en outre certaines 
questions importantes qui restent pos\'ees et qui peuvent faire l'objet de 
recherches futures.

En annexe nous indiquons les distributions angulaires de la diffusion \'elastique 
mesur\'ees par la collaboration Valencia-Strasbourg et analys\'ees dans cette 
th\`ese, dont les valeurs n'ont pas \'et\'e publi\'ees. Les donn\'ees correspondant 
aux diffusions $^{35,37}Cl+^{24}Mg$ \`a diff\'erentes \'energies n'apparaissent 
pas dans l'annexe car elles ont \'et\'e publi\'ees dans la r\'ef\'erence 
{\bf(BA91)}.

\newpage
\chapter{{\Large{{\bf{LE POTENTIEL OPTIQUE}}}}}
\section{\large{{\bf{TH\'EORIE DE FESHBACH}}}}
L'interaction entre deux noyaux A et B est d\'ecrite par 
un Hamiltonien :
\be	H = H_A + H_B + T({\bf r}) + V({\bf r},X_n)	\ee 
o\`u $H_A$ et $H_B$ repr\'esentent les Hamiltoniens "internes" des deux noyaux, 
$T({\bf r})$ est l'\'energie cin\'etique de leur mouvement relatif et 
$V({\bf r},X_n)$ d\'ecrit leur potentiel d'interaction mutuelle. La 
cordonn\'ee ${\bf r}$  est la s\'eparation relative entre les centres de 
masse des deux noyaux.   

La fonction d'onde non antisymetris\'ee, d\'ecrivant un \'etat interne 
quelconque du syst\`eme, peut s'\'ecrire :
\be	\phi_n(X_n) = \Phi_A(X_A) \\ \Phi_B(X_B)	\ee
o\`u $\Phi_A$ et $\Phi_B$ sont les fonctions d'onde  
d\'ecrivant les \'etats internes de chaque noyau et les $X_i$ repr\'esentent 
les variables internes.
 
A partir de $\phi_n$ il est possible de construire la fonction d'onde du 
syst\`eme en interaction, qui est en fait un d\'eveloppement sur un ensemble 
complet d'\'etats internes : 
\be	\psi({\bf r},E_n,X_n) = \sum_{n} {\chi}_n({\bf r},E_n) \\ \phi_n(X_n) \ee
\noindent 
o\`u le symbole $\sum_{n}$ repr\'esente une somme sur les \'etats discrets et 
une int\'egrale sur les \'etats continus. L'indice $n$ se ref\`ere \`a chacun 
des \'etats internes du syst\`eme. Les coefficients ${\chi}_n({\bf r},E_n)$ 
sont les fonctions d'onde du mouvement relatif des deux noyaux qui forment le 
syst\`eme quand ils se trouvent dans un \'etat interne d\'ecrit par 
$\phi_n(X_n)$. Pour des raisons de simplicit\'e, nous n'expliciterons dans ce 
qui suit que la d\'ependance qu'il nous interesse de mettre en relief.
  
Le d\'eveloppement de la fonction d'onde du syst\`eme donn\'e par l'\'equation 
(2.3) pr\'esente deux limitations fondamentales :

a) La dispersion d'inter-\'echange associ\'ee \`a l'indiscernabilit\'e des 
nucl\'eons n'est pas correctement d\'ecrite, vu que la fonction d'onde 
$\phi_n(X_n)$ n'a pas \'et\'e antisym\'etris\'ee. Cependant, ce type de 
r\'eactions, ainsi que celles qui conduisent \`a une r\'eorganisation 
diff\'erente des nucl\'eons qui, initialement, formaient les noyaux A et B, 
peut \^etre d\'ecrit \`a travers des \'etats non li\'es.  

b) La fonction $\phi_n(X_n)$ n'est pas appropri\'ee lorsque le potentiel 
nucl\'eon-nucl\'eon pr\'esente un coeur r\'epulsif car elle ne s'annule pas 
automatiquement quand les noyaux se recouvrent et les nucl\'eons s'approchent 
\`a une distance inf\'erieure au rayon de r\'epulsion du coeur.
 
 La fonction $\psi$  satisfait l'\'equation de Schrodinger : 
$H \psi = E \psi$
et lorsqu'on reporte l'expression de $H$ dans cette \'equation, on aboutit \`a 
un syst\`eme infini d'\'equations coupl\'ees: 
\be (T+V_{mm}+E_m-E) {\chi}_m = -\sum_{n \not= m} {\chi}_n V_{mn}\ \ \ 
 \ \ \ \ \ \ \ \ \forall m  \ee
avec :
\be	V_{mn} = <\phi_m|V|\phi_n>	\ee
$E_m$ est la valeur propre de l'Hamiltonien interne, correspondant \`a la 
fonction propre $\phi_m$, qui satisfait l'\'equation :
\be	(H_A + H_B) \phi_m = E_m \phi_m	\ee
Lorsque l'\'energie d'interaction entre les deux noyaux A et B est tr\`es 
basse, le nombre de canaux ouverts est relativement petit et le syst\`eme 
d'\'equations coupl\'ees (\'equation 2.4) peut \^etre r\'esolu; mais \`a mesure 
que l'\'energie de l'interaction augmente, le nombre de canaux ouverts et, par 
cons\'equent, le nombre d'\'equations coupl\'ees cro\^{\i}t tr\`es rapidement 
et la solution exacte du probl\`eme devient inabordable.
  
La th\'eorie de Feshbach {\bf(FE58)}, {\bf(FE62)} permet de d\'ecoupler le 
syst\`eme d'\'equations diff\'erentielles \`a r\'esoudre, de sorte que si l'on 
n'est int\'eress\'e que par N canaux de r\'eaction, il devient possible de 
ramener le syst\`eme \`a un autre, plus simple, de N \'equations coupl\'ees. 
C'est ainsi que dans le cas de la diffusion \'elastique on obtiendra une seule 
\'equation. On peut arriver \`a ce r\'esultat en utilisant le formalisme des 
op\'erateurs de projection.

 Soient $P$ et $Q$ les op\'erateurs qui projettent la fonction d'onde sur 
les canaux d'int\'er\^et et sur les canaux restants, respectivement, de 
mani\`ere que $Q$=1-$P$. Comme nous ne sommes interess\'es que par la voie 
\'elastique, 
$P$ = $|\phi_0><\phi_0|$  et  $Q$=1-$P$=${\displaystyle \sum_{n\not=0}
|\phi_n><\phi_n|}$. L'\'equation de Schrodinger s'\'ecrit alors: 
$(E-H)(P+Q)|\psi>=0$. 
Moyennant quelques calculs utilisant les 
propri\'et\'es des op\'erateurs de projection on peut d\'ecoupler l'\'equation 
d\'ecrivant la voie \'elastique de toutes les autres, ce qui nous permet 
d'obtenir, pour la voie 
\'elastique, l'\'equation {\bf(FE62)}: 
\be	[E-T_0-<\phi_0|V|\phi_0>-<\phi_0|VQ{1\over E-H_{QQ}+i\eta}QV|\phi_0>]|
{\chi}_0> = 0	\ee
o\`u nous avons adopt\'e la valeur propre $E_0$=0, pour la voie \'elastique et 
$H_{QQ}$ = $QHQ$.   
 
On appelle potentiel optique g\'en\'eralis\'e l'op\'erateur :
\be  V_{opt} = <\phi_0|V|\phi_0>+<\phi_0|VQ{1\over E-H_{QQ}+i\eta}QV|\phi_0> \ee
Le premier terme de cet op\'erateur repr\'esente les transitions directes du 
canal d'entr\'ee \`a la voie de diffusion \'elastique et porte le nom de 
potentiel de convolution. Si l'interaction $V$ est locale et si les effets 
d'inter-\'echange sont n\'egligeables, le terme de convolution sera local 
dans l'espace des configurations. Le second terme englobe la contribution \`a 
la voie \'elastique des canaux qui n'ont pas \'et\'e consid\'er\'es 
explicitement, mais qui peuvent participer comme \'etats interm\'ediaires de 
transition \`a travers desquels le syst\`eme \'evolue jusqu'\`a atteindre la 
voie \'elastique. 
La pr\'esence dans cet 
op\'erateur  du  propagateur  du  mouvement relatif $[E-H_{QQ}+i\eta]^{-1}$ a 
un effet triple sur le potentiel optique :

a) La pr\'esence de $i\eta$ dans le propagateur fait que le potentiel optique 
est complexe.

b) \'Etant donn\'e que la probabilit\'e de transition vers les canaux 
projet\'es par l'op\'erateur $Q$ d\'epend de l'\'energie, le potentiel optique 
pr\'esente \'egalement cette d\'ependance. 

c) Le terme $H_{QQ}$ contient l'op\'erateur \'energie cin\'etique du mouvement 
relatif, impliquant que le potentiel optique n'est pas local. Physiquement, 
cela signifie qu'un syst\`eme qui dans la position ${\bf r'}$ est excit\'e 
par l'interaction vers un des canaux non consid\'er\'es dans $P$, peut se 
propager dans cet \'etat et r\'eappara\^{\i}tre \`a la position ${\bf r}$ dans 
le sous-espace $P$.

Pour n'importe quel Hamiltonien {\bf (JA70)}, l'op\'erateur 
$[E-H_{QQ}+i\eta]^{-1}$ peut s'exprimer sous la forme :
\be {1\over E-H_{QQ}+i\eta}=\int \hspace{-0.5cm} \sum_{n}{|\psi_n><\psi_n|\over E-E_n+i\eta}\ee
\noindent
o\`u $E_n$ est la valeur propre correspondant \`a la fonction propre $\psi_n$ de 
l'Hamiltonien $H_{QQ}$ et le symbole $\ds{\int \hspace{-0.5cm} \sum_n}$ indique une somme sur les 
variables  
discr\`etes et une int\'egration sur les variables continues. Dans la limite du 
couplage faible, nous pouvons exprimer le potentiel optique 
\noindent comme suit :
\be V_{opt}=V_{00}+\sum_{n \not= 0}V_{0n} {|{\chi}_n><
{\chi}_n|\over {E-E_n}}V_{n0}+\int_{E_1}^{\infty}{d \epsilon \over {E-\epsilon
+i\eta}} \sum_{n \not= 0}V_{0n}{|{\chi}_n(\epsilon)><{\chi}_n(\epsilon)}|V_{n0}
\ee 
\noindent
o\`u $V_{0n}=<\phi_0|V|\phi_n>$. Le terme $i\eta$ n'est pas necessaire pour les 
\'etats li\'es vu que pour $E=E_n$ le potentiel optique diverge, donnant lieu 
aux r\'esonances.
  
 Utilisant la relation de Dirac {\bf (SA83)}:
 \be \lim_{\eta \rightarrow {0^+}}{1\over E-\epsilon+i\eta}={\cal{P}}
\left[{1\over E-\epsilon}\right]-i\pi\delta(E-\epsilon)\ee 
\noindent
o\`u $\cal{P}$ repr\'esente la valeur principale de Cauchy, nous pouvons 
obtenir les parties r\'eelle et imaginaire du potentiel optique. \`A partir de 
l'\'equation (2.10):
\be	{\cal{R}}eV_{opt} = V_{00} + \sum_{n \not= 0} V_{0n} {|{\chi}_n><
{\chi}_n|\over {E-E_n}}V_{n0} + {\cal{P}} \int_{E_1}^{\infty}{d \epsilon \over 
{E-\epsilon}}
\sum_{n \not= 0}V_{0n}
{|{\chi}_n(\epsilon)><{\chi}_n(\epsilon)|}V_{n0}	\ee
et
\ba    {\cal{I}}mV_{opt} = \left\{ \begin{array}{ll}
	{\ds -\pi \sum_{n \not= 0}V_{0n}{|{\chi}_n(\epsilon)><{\chi}_n(
	\epsilon)|}V_{n0}}
	& \mbox{si ${\ds E>E_1}$} \\
	{0}
	&  \mbox{si ${\ds E<E_1}$}
	\end{array}
	\right.         \ea

Ainsi, donc, les parties r\'eelle et imaginaire du potentiel optique ont une 
origine commune, cons\'equence du principe de causalit\'e, et qui se traduit 
par une relation de dispersion qui les lie  {\bf(FE62)} :
\be	{\cal{R}}eV_{opt} = V_{00} + \sum_{n \not= 0} V_{0n} {|{\chi}_n><
{\chi}_n|\over {E-E_n}}V_{n0} - {1 \over \pi} {\cal{P}} \int_{E_1}^{\infty}
{{\cal{I}}mV_{opt}(\epsilon) \over {E-\epsilon}} d\epsilon	\ee
Cette expression traduit le fait qu'une onde diffus\'ee ne peut \^etre \'emise 
avant que l'interaction n'ait eu lieu et devient, dans l'approximation de 
mati\`ere nucl\'eaire infinie :
\be	{\cal{R}}eV_{opt} = V_{00} - {1 \over \pi} {\cal{P}} \int_{E_1}^{\infty}
{{\cal{I}}mV_{opt}(\epsilon) \over {E-\epsilon}} d\epsilon	\ee
La relation de dispersion lie des potentiels non locaux dont l'utilisation est 
compliqu\'ee. Dans la pratique, on utilise des potentiels locaux. La 
localisation du potentiel non local transforme sa non localit\'e par une 
d\'ependance suppl\'ementaire en \'energie. Cette d\'ependance n'est pas 
d'origine physique et fait que, en g\'en\'eral, la relation de dispersion n'est 
pas satisfaite entre potentiels locaux.

\vskip 10mm

\section{\large{\bf{ {MOD\'ELISATIONS DE LA TH\'EORIE DE FESHBACH} :}}}
\par
Plusieurs approximations ont \'et\'e propos\'ees pour \'evaluer, de mani\`ere 
approch\'ee, le potentiel optique local calcul\'e \`a partir des 
\'equations (2.12) et (2.13). Dans ce 
chapitre nous exposerons trois mod\'elisations qui seront utilis\'ees dans 
la suite de ce travail. La premi\`ere  consiste \`a remplacer le potentiel 
optique par une fonction d\'ependant de param\`etres \`a ajuster aux donn\'ees 
exp\'erimentales. Les deux autres consistent \`a \'evaluer le potentiel local 
d\'eduit de l'\'equation (2.10) moyennant des hypoth\`eses simplificatrices 
qui permettent 
de prendre en compte de fa\c{c}on globale l'ensemble des canaux ou en ne 
prenant en compte qu'un nombre r\'eduit de voies qui sont celles qui 
contribuent de mani\`ere significative au potentiel.
 
\subsection{Mod\`eles ph\'enom\'enologiques et 
semiph\'enom\'enologiques :}
Dans cette approximation le potentiel optique s'\'ecrit 
$ V(r) = U(r) + iW(r)$. On attribue aux termes $U(r)$ et $W(r)$ une 
forme analytique physiquement raisonnable et d\'ependant d'un ensemble de 
param\`etres \`a ajuster. Il s'agit de potentiels ph\'enom\'enologiques qui 
doivent tenir compte des caract\'eristiques physiques de l'interaction 
nucl\'eon-nucl\'eon et du confinement des nucl\'eons dans les  noyaux. 
Ils doivent donc \^etre attractifs et s'annuler rapidement dans 
la r\'egion superficielle. 

Le facteur de forme le plus souvent utilis\'e pour ces potentiels 
 est celui dit de Woods-Saxon qui a la 
m\^eme forme fonctionnelle que la distribution de Fermi des charges 
nucl\'eaires et qui s'\'ecrit {\bf(WO54)} :
\be	f(r) ={1\over{1+e^{{r-R}\over {a}}}}	\ee
le param\`etre de diffusivit\'e, $ a$, estime l'\'epaisseur superficielle
 du noyau 
et le rayon du potentiel, $R$, est donn\'e par la relation :
\be	R = r_0[ A_P^{1 \over 3} + A_C^{1 \over 3} ]	\ee
le rayon r\'eduit $r_0$ est de l'ordre de 1.2 \`a 1.4 fermi et les nombres de 
masse $A_P$ et $A_C$ se r\'ef\`erent au projectile et \`a la cible, 
respectivement.
  
Dans cette approximation une forme simple du potentiel optique est donn\'ee 
par: 
\be V_{opt}(r) = - V_0 \ f_R(r) - i \ W_0 \ f_I(r) \ee
\noindent
o\`u $V_0$ et $W_0$ repr\'esentent les profondeurs des parties r\'eelle et 
imaginaire du potentiel optique et $f_R(r)$ et $f_I(r)$ leurs facteurs de forme 
respectifs qui peuvent pr\'esenter ou non la m\^eme g\'eom\'etrie.

Comme nous l'avons vu pr\'ecedement, tandis que la partie r\'eelle, donn\'ee 
par l'\'equation (2.12), est domin\'ee par le terme de premier ordre, $V_{00}$, 
la partie imaginaire qui d\'etermine la perte de flux de la voie \'elastique, 
provient uniquement du terme de polarisation qui inclut des processus d'ordre 
sup\'erieur (\'equation 2.13), et que l'on d\'ecrit habituellement moyennant la 
contribution de termes de volume et de surface. Le terme de volume rend compte 
des excitations, et donc, de l'absorption qui se produit pendant la propagation 
du projectile \`a travers la mati\`ere nucl\'eaire. L'existence de modes 
collectifs d'excitation ainsi que de processus p\'eriph\'eriques de transfert, 
conduit \`a une augmentation de l'absorption dans la r\'egion superficielle, ce 
qui justifie le terme de surface.
  
  L'absorption superficielle se param\'etrise de la forme suivante :
\be W_S(r)=-4W_D {df_I(r)\over{dr}}  \ee
o\`u $W_D$ est la profondeur du terme de surface.
 
Cela a l'inconv\'enient d'augmenter le nombre de param\`etres  \`a 
d\'eterminer sans apporter, en g\'en\'eral, d'informations importantes sur le 
potentiel lui-m\^eme. Pour cette raison, il est fr\'equent de n'utiliser que le 
terme de volume et de simuler l'accro\^{\i}ssement de l'absorption en surface 
moyennant un rayon $R_I$ du potentiel imaginaire de volume sup\'erieur au 
rayon du potentiel r\'eel. 
C'est pourquoi nous avons utilis\'e, dans une partie de 
ce travail, le m\^eme facteur de forme mais avec des 
param\`etres diff\'erents pour les partie r\'eelle et imaginaire du potentiel. 
Nous avons \'egalement n\'eglig\'e le terme de spin-orbite, introduit dans les 
analyses de diffusion de nucl\'eons mais qui s'est av\'er\'e tr\`es faible 
dans la diffusion d'ions lourds. 

Il est par contre indispensable d'introduire  dans l'analyse des donn\'ees un 
potentiel coulombien dont la forme correspond \`a celle  de l'interaction 
entre une particule ponctuelle charg\'ee, $Z_P$, et une sph\`ere 
uniform\'ement charg\'ee, $Z_C$, de rayon $R_C$ et de diffusivit\'e nulle 
{\bf(PO76)}. Sa forme analytique est donc :
\ba	 V_C(r) = \left\{ \begin{array}{ll}
{\ds{Z_PZ_Ce^2 \over 2R_C}(3-{r^2 \over R^2_C}}) & \mbox{pour $r \leq R_C$} \\
{\ds{Z_PZ_Ce^2 \over r}}     &  \mbox{pour $r > R_C$}
\end{array}
\right. 	\ea
avec $R_C = r_C [A_P^{1 \over 3} + A_C^{1 \over 3}]$ et $r_C$ le rayon 
r\'eduit.

Ainsi, la forme  du potentiel optique ph\'enom\'enologique que nous avons
utilis\'e est :
\be	V_{tot} = V_C(r) - V_0 \ f_R(r) - i \ W_0 \ f_I(r) 	\ee
 Les param\`etres de ce potentiel doivent \^etre ajust\'es de fa\c{c}on \`a 
assurer une bonne concordance entre les sections efficaces diff\'erentielles 
calcul\'ees et les distributions angulaires mesur\'ees. 
Pour notre part, nous avons utilis\'e la m\'ethode dite du 
$\chi^2$, qui consiste \`a rendre minimale la valeur de la fonction :
\be	{\chi^2 \over n} = {1 \over n} \sum_{i=1}^n \left[ {
{\sigma_{the}(\theta_i)-\sigma_{exp}(\theta_i)} \over {\Delta
\sigma_{exp}(\theta_i)} } \right]^2	\ee
o\`u $n$ est le nombre de donn\'ees exp\'erimentales, $\sigma_{the}$ et 
$\sigma_{exp}$ sont, respectivement, 
les sections efficaces th\'eoriques et exp\'erimentales;   
$\theta_i$ est l'angle de diffusion auquel se mesurent ou se 
calculent les sections efficaces diff\'erentielles et $\Delta \sigma_{exp}$ est 
l'incertitude absolue associ\'ee aux mesures.

Du fait de sa simplicit\'e, la param\'etrisation du potentiel est une m\'ethode 
d'analyse tr\`es utilis\'ee. Elle pr\'esente cependant l'inconv\'enient de 
conduire \`a des param\`etres dont la signification physique n'est pas 
toujours claire. De plus, dans le cas de la diffusion d'ions lourds il est
souvent difficile de choisir entre diff\'erents jeux de param\`etres qui 
permettent de reproduire les donn\'ees exp\'erimentales avec la m\^eme 
qualit\'e.

Afin de limiter ces ambiguit\'es et de r\'eduire le nombre de param\`etres \`a
ajuster, une estimation du potentiel \`a partir de consid\'erations plus
fondamentales a \'et\'e d\'evelopp\'ee : il s'agit du mod\`ele du potentiel de 
convolution {\bf(GR68)}. Dans ce mod\`ele le potentiel optique est calcul\'e  
\`a partir des distributions de mati\`ere nucl\'eaire et de l'interaction 
nucl\'eon-nucl\'eon. Dans sa forme la plus simple, on consid\`ere une 
interaction effective locale, $V_{eff}({\bf r}_{12})$, entre les nucl\'eons 
et on suppose que la collision entre les ions est suffisamment 
rapide pour que leurs densit\'es restent inalt\'er\'ees pendant la r\'eaction. 
De cette fa\c{c}on le potentiel est, en premi\`ere approximation, la somme de 
toutes les interactions nucl\'eon-nucl\'eon qui forment le syst\`eme. Il 
s'\'ecrit :
\be	V_{opt}({\bf r})= \int \rho_P({\bf r}_1) \rho_T({\bf r}_2) 
V_{eff}({\bf r}_{12}) d{\bf r}_1 d{\bf r}_2	\ee
$\rho_i({\bf r}_i)$ repr\'esentant les densit\'es de nucl\'eons pour les deux 
noyaux dans leur \'etat fondamental. La figure 2.1, clarifie les notations des 
coordonn\'ees entrant dans les calculs de convolution. ${\bf r}_{12} = {\bf r} 
+ {\bf r}_2 - {\bf r}_1 $ est un vecteur d\'efinissant la position d'un 
nucl\'eon du noyau projectile relativement \`a un nucl\'eon du noyau cible.

Les diff\'erences entre les nombreux potentiels de convolution qui existent 
sont li\'ees au type d'interaction effective utilis\'ee. Cependant   
les diff\'erentes interactions effectives doivent avoir une base physique 
r\'ealiste afin d'arriver \`a une description unifi\'ee des collisions 
noyau-noyau, qui soit ind\'ependante du syst\`eme consid\'er\'e.

L'interaction effective la plus g\'en\'erale est complexe et  
difficile \`a \'etablir. De ce fait on utilise  des 
approximations dont les formes les plus fr\'equentes sont les potentiels 
centraux ou les formes semi-empiriques.

Dans ce travail, nous avons utilis\'e l'interaction effective appel\'ee 
"interaction M3Y" {\bf(LO75)}, {\bf(GO76)} et {\bf(BE77)}. Sa forme 
fonctionnelle est :
\be	V_{M3Y} = 7999 {e^{-4r_{12}} \over {4r_{12}}} - 2134 {e^{-2.5r_{12}} 
\over {2.5r_{12}}} - 
J(E) \delta({\bf r_{12}})	\ee
%&&&&&&&&&&&&&&&&&&&&&&&&&&&&&&&&&&&&&&&&&&&&&&&&&&&&&&&&&&&&&&&&&&&&&&&&&&&&&
\begin{figure}
\vspace {-2.4cm}
\begin{center}\mbox{\epsfig{file=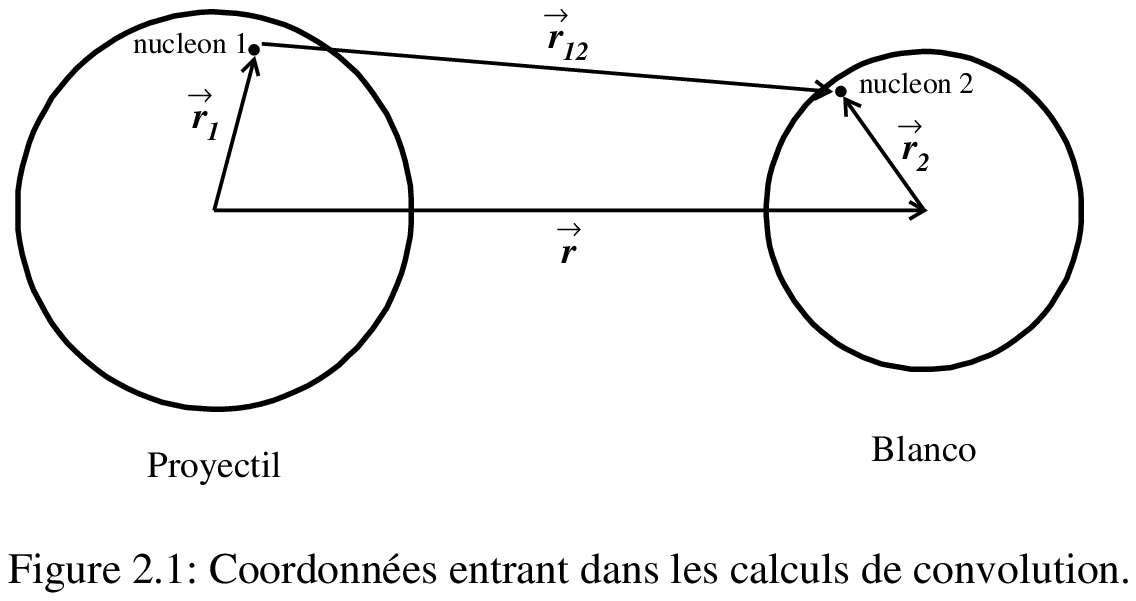}}\end{center}
\vspace {-0.8cm}
\end{figure}
%&&&&&&&&&&&&&&&&&&&&&&&&&&&&&&&&&&&&&&&&&&&&&&&&&&&&&&&&&&&&&&&&&&&&&&&&&&&&&
\noindent
o\`u $r_{12}$ est la distance entre les nucl\'eons consid\'er\'es. 
Cette interaction est ind\'ependante de la densit\'e nucl\'eaire et d\'epend 
tr\`es faiblement de l'\'energie via la profondeur du pseudo-potentiel 
$J(E)\simeq -276 (1-0.005E/A_P)$ MeV.fm$^3$. C'est en fait une moyenne sur un 
intervalle de densit\'es nucl\'eaires et d'\'energies incidentes {\bf(SA79)}. 
L'interaction effective M3Y est r\'eelle. Seule la partie r\'eelle du potentiel 
peut donc \^etre calcul\'ee avec cette force. La partie imaginaire, elle, 
reste ph\'enom\'enologique et donc ses param\`etres doivent \^etre 
d\'etermin\'es comme pr\'ec\'edemment par ajustement aux donn\'ees 
exp\'erimentales.

Les densit\'es nucl\'eaires sont, elles aussi, tr\`es complexes. Nous 
sommes donc conduits, l\`a \'egalement, \`a faire des approximations comme nous 
le verrons plus loin, pour chaque cas. 
 
\subsection{Mod\`ele de l'approximation de fermeture :}
L'expression du potentiel optique donn\'ee par la th\'eorie de Feshbach 
pr\'esente un inconv\'enient majeur lorsqu'il s'agit d'\'evaluer la 
contribution de tous les termes. En effet, il faudrait alors conna\^{\i}tre 
les fonctions d'onde de tous les \'etats excit\'es possibles du syst\`eme 
consid\'er\'e. Une telle \'evaluation est \'evidemment irr\'ealisable et il 
faut utiliser diff\'erentes mod\'elisations de la th\'eorie {\bf (LO77), 
(KU81), (AN90), (KU91)} pour pouvoir calculer la contribution des termes les 
plus importants.

En ce sens, le mod\`ele de l'approximation de fermeture propos\'e par N. Vinh 
Mau {\bf(VI86)} et {\bf(VI87)} a \'et\`e utilis\'e avec succ\`es. Ce mod\`ele 
permet une \'evaluation globale du potentiel optique, moyenant quelques 
hypoth\`eses simplificatrices et certaines approximations.

L'\'equation (2.8) qui d\'ecrit le potentiel d'interaction noyau-noyau, dans 
le formalisme de Feshbach, peut s'\'ecrire:
\be	V({\bf r},{\bf r'})=U_0({\bf r}) \delta({\bf r}-{\bf r'}) + 
\Delta V({\bf r},{\bf r'})	\ee
o\`u $U_0$ peut s'\'ecrire   : 
\be U_0({\bf r})=<\phi_0^P\Phi_0^C|V({\bf r})|\phi_0^P\Phi_0^C > \ee
et, dans la  limite du couplage faible, nous avons : 
\be	\Delta V({\bf r},{\bf r'}) = \sum_{(m,n) \neq (0,0)}V_{mn}^*({\bf r})
	G_{mn}({\bf r},{\bf r'})  V_{mn}({\bf r'})	\ee
avec : 
\be V_{mn}({\bf r}) = <\phi_m^P\Phi_n^C |V({\bf r})| \phi_0^P\Phi_0^C> \ee
et : 
\be	V({\bf r}) = \sum_{i \in P , j \in C}V_{ij}({\bf r})	\ee
o\`u $V_{ij}({\bf r}) = V({\bf r_i}-{\bf r_j}+{\bf r})$ et ${\bf r_i}$, 
${\bf r_j}$ et ${\bf r}$ sont d\'efinis dans la figure 2.2; $\phi_m^P$ et 
$\Phi_n^C$ sont, respectivement, les fonctions d'onde des noyaux projectile et 
cible dans les \'etats $m$ et $n$ avec les \'energies d'excitation $E_m$ et 
$E_n$ , tandis que les fonctions d'onde avec les indices $ m=0$ et $ n=0$ 
repr\'esentent les \'etats fondamentaux; $V({\bf r})$ d\'ecrit l'interaction 
effective entre chaque paire de nucl\'eons appartenant \`a chacun des deux 
noyaux en collision. $G_{mn}({\bf r},{\bf r'})$ est le propagateur du 
mouvement relatif dans le canal ($m,n$). Il est donn\'e par l'expression 
suivante :
\be	G_{mn}({\bf r},{\bf r'}) = \lim_{\eta \rightarrow {0^+}}
	{1\over E-H_{QQ}+i\eta} = \lim_{\eta \rightarrow {0^+}}
        \int
	{{{\chi}_k({\bf r}) {\chi}_k^*({\bf r'})} \over 
	{E-E_m-E_n-{{\hbar^2k^2} \over {2 \mu}}+i \eta} }d^3k	\ee
o\`u ${\chi}_k({\bf r})$ et ${\chi}_k^*({\bf r'})$ sont les fonctions 
d'onde du mouvement relatif des deux noyaux en collision, $\mu$ est la masse 
r\'eduite du syst\`eme et $k$ le moment r\'eduit. Enfin, $U_0({\bf r})$ est 
un potentiel de double convolution, local et ind\'ependant de l' \'energie.   

Lorsque l'\'energie de la collision est assez \'elev\'ee pour que tous les 
canaux in\'elastiques significatifs soient ouverts et puissent contribuer 
\`a la partie imaginaire du potentiel  ou lorsque, au contraire, elle est 
suffisamment basse pour que toutes les voies in\'elastiques soient ferm\'ees 
et que seules des excitations virtuelles puissent avoir lieu, le calcul 
terme \`a terme du potentiel devient prohibitif. C'est alors qu'il s'av\`ere 
utile et qu'il est valide d'\'evaluer globalement le potentiel, moyennant le 
mod\`ele de fermeture qui s'appuie sur les consid\'erations suivantes :
 
1) Le propagateur $G_{mn}({\bf r},{\bf r'})$ d\'efini par l'\'equation (2.30) est 
approxim\'e par le propagateur WKB {\bf(SC69)}:
\be G_{mn}({\bf r},{\bf r'}) \simeq G_{mn}^{WKB}({\bf r},{\bf r'}) = 
         -{\mu\over2\pi \hbar^2}
	{e^{iK_{mn}\left\vert{{\bf r}-{\bf r'}}\right\vert}
	\over {\left\vert{{\bf r}-{\bf r'}}\right\vert}}	\ee 
Ce propagateur WKB peut \^etre exprim\'e en fonction de $R$ et $s$, qui 
repr\'esentent les modules des coordonn\'ees du mouvement relatif et du 
centre de masse, $ {\bf R} = ({\bf r}+{\bf r'})/2$ et 
${\bf s} = {\bf r} - {\bf r'}$ :
\be G_{mn}^{WKB} = -{\mu\over2\pi \hbar^2} {e^{iK_{mn}s}\over {s}}	\ee 
Cette forme du propagateur a \'et\'e utilis\'ee avec succ\`es pour d\'ecrire 
les collisions nucl\'eon-noyau et alpha-noyau \`a basse \'energie {\bf(BO81)}. 
Le moment local $K_{mn}$ \'etant d\'efini par : 
\be	K_{mn}^2(R) = {{2 \mu} \over \hbar^2} \left[ E - E_m -E_n -V_L(R) 
	- V_C(R) \right]	\ee
$V_L(R)$ et $V_C(R)$ \'etant respectivement les potentiels nucl\'eaire local et 
coulombien. Si, comme il est habituel, $V_L(R)$ est un potentiel complexe, le 
propagateur WKB n'est pas hermitique.

2) Le propagateur, m\^eme sous cette forme simple  
 d\'epend explicitement des valeurs propres $E_m$ et $E_n$ \`a 
travers le moment local. Ceci emp\^eche l'\'evaluation globale des termes du 
potentiel optique. Cependant, si nous supposons que tous les \'etats excit\'es 
qui contribuent de mani\`ere importante au potentiel, se concentrent dans un 
\'etroit domaine d'\'energie, nous pouvons remplacer leurs \'energies 
d'excitation par des valeurs moyennes $<E_P>$ (pour le projectile) et $<E_C>$ 
(pour la cible). Sous ces hypoth\`eses, le propagateur WKB peut \^etre 
remplac\'e 
par un propagateur WKB moyenn\'e pour chacun des deux noyaux collisionnants :
\be  G_{mn}^{WKB}(R,s) \simeq G_i^{WKB}(R,s) \ \ \ \ \ \ i=1,2,3 \ee
Dans cette hypoth\`ese, le terme de polarisation donn\'e par l'\'equation 
(2.27) 
ne contient plus que trois termes qui d\'ecrivent les \'etats excit\'es de 
chacun des noyaux projectile ($m\neq 0,n=0$) ou cible ($m=0,n\neq 0$), ou des 
deux ($m\neq 0,n\neq 0$), ce qui correspond \`a i=1,2,3, respectivement.
 
3)  La r\'ef\'erence explicite  aux fonctions d'onde des \'etats excit\'es 
dont les valeurs propres ont \'et\'e moyenn\'ees, peut \^etre \'elimin\'ee. 
En effet, si l'\'energie de la collision est suffisamment grande pour que toutes 
les voies in\'elastiques significatives soient ouvertes ou si elle est 
suffisamment basse pour qu'elles soient toutes ferm\'ees, nous pouvons 
introduire dans l'\'equation (2.27) les relations de fermeture qui, pour chaque 
noyau, peuvent s'\'ecrire :
\be	\sum_{m \neq 0} \left| \phi_m^P><\phi_m^P) \right| = 1 - 
	\left| \phi_0^P><\phi_0^P \right|	\ee
\be	\sum_{n \neq 0} \left| \Phi_n^C><\Phi_n^C \right| = 1 - 
	\left| \Phi_0^C><\Phi_0^C \right|	\ee
De cette fa\c{c}on nous pouvons \'eliminer toute r\'ef\'erence explicite aux 
effets de structure; cependant l'utilisation des relations de fermeture 
implique que tous les canaux soient implicitement pris en consid\'eration.

Ces trois approximations seules ne sont pas suffisantes pour \'evaluer le 
potentiel optique car elles conduisent \`a  des expressions non triviales 
qui pr\'esentent trois types de difficult\'es : 
 
 	 a - Calcul d'int\'egrales triples et quadruples, chose 
qui n'est pas toujours ais\'ee.
 
 	 b -  Les \'el\'ements non diagonaux de la matrice 
densit\'e $\rho_i({\bf r},{\bf r'})$ sont, y compris dans les 
\hskip 10mm 
mod\`eles les 
plus simples, des fonctions compliqu\'ees de ${\bf r}$ et de ${\bf r'}$.

c -  Un potentiel non local n'est traitable que s'il est exprim\'e 
en termes de coordonn\'ees 
\hskip 10mm 
relatives et du centre de masse et non en fonction 
de ${\bf r}$ et de~${\bf r'}$.
  
Il est possible d'\'eviter ces difficult\'es moyennant l'introduction des 
approximations suppl\'ementaires suivantes :

4)-  L'interaction nucl\'eon-nucl\'eon (figure 2.2.a) a souvent \'et\'e d\'ecrite
 au moyen 
d'une force s\'eparable qui permet de reproduire avec beaucoup de succ\`es de 
nombreuses propri\'et\'es li\'ees \`a la structure nucl\'eaire {\bf(BR59)} et 
{\bf(BO75)}. Analytiquement, cette interaction est de  la 
forme :
\be	V_{ij} = - V_0 f({\bf r}_i) f({\bf r}_j)	\ee
o\`u $V_0$ repr\'esente la profondeur de l'interaction et $f(r)$ le facteur de 
forme. Bien que dans notre cas la situation soit diff\'erente, \'etant 
donn\'e que les deux nucl\'eons $i$ et $j$ qui interagissent appartiennent
 \`a chacun des noyaux en collision (figure 2.2.b),  
il semble  raisonnable d'assumer une forme fonctionnelle similaire pour 
d\'ecrire leur interaction, \`a savoir  :
\be	V_{ij}({\bf r}) = - V_0 f({\bf r}_i + {{\bf r} \over 2})
	f({\bf r}_j - {{\bf r} \over 2})	\ee
o\`u ${\bf r}_{i(j)}$  est le rayon vecteur du nucl\'eon $i$($j$) appartenant au 
noyau $1$($2$)  et ${\bf r}$ est le vecteur repr\'esentant la s\'eparation 
entre les centres des deux noyaux (figure 2.2).

Pour des raisons de sym\'etrie et de simplicit\'e, nous adoptons des facteurs 
de forme gaussiens. Ceci conduit \`a r\'e\'ecrire la fonction d'interaction 
de la mani\`ere suivante : 
\be V_{ij}({\bf r}) = - V_0 e^{- \eta \left({\bf r}_i + {\bf r}/2\right)^2} 
                            e^{- \eta \left({\bf r}_j - {\bf r}/2\right)^2} \ee
%&&&&&&&&&&&&&&&&&&&&&&&&&&&&&&&&&&&&&&&&&&&&&&&&&&&&&&&&&&&&&&&&&&&&&&&&&&&&&
\begin{figure}
\vspace {-2.8cm}
\begin{center}\mbox{\epsfig{file=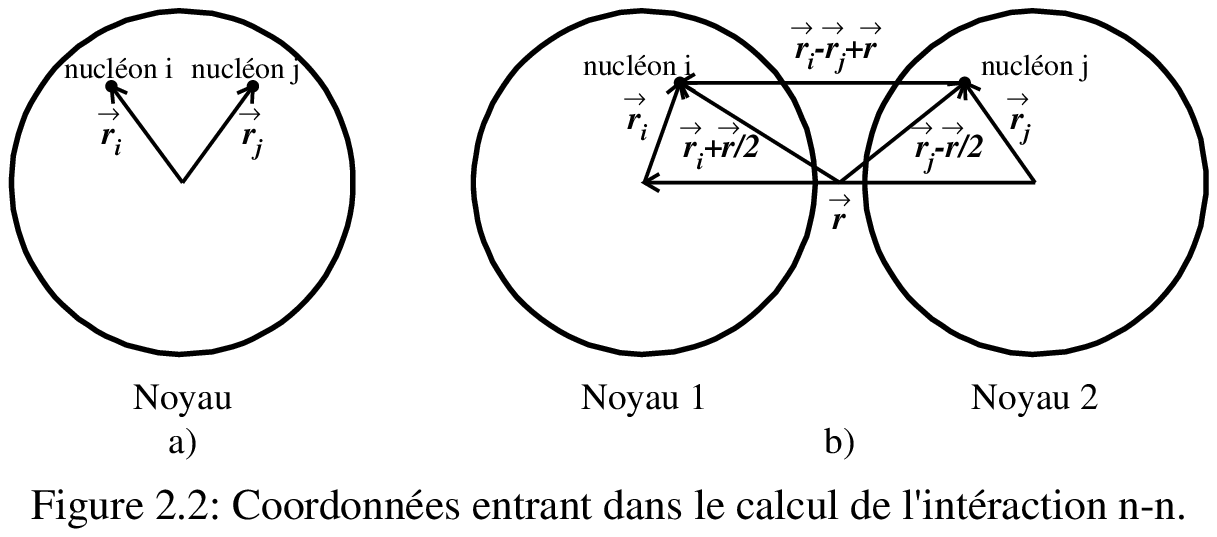}}\end{center}
\vspace {-1.0cm}
\end{figure}
%&&&&&&&&&&&&&&&&&&&&&&&&&&&&&&&&&&&&&&&&&&&&&&&&&&&&&&&&&&&&&&&&&&&&&&&&&&&&&
o\`u $\eta$ est un param\`etre li\'e \`a la port\'ee de l'interaction, 
$r_0=(2/\eta)^{1/2}$ et V$_0$ est l'intensit\'e de l'interaction.
 
5) Pour d\'ecrire les \'el\'ements non diagonaux de la matrice densit\'e, on 
utilise une des formes les plus simples {\bf(SA82)}  qui est l'approximation de 
Slater, bas\'ee sur les expressions obtenues dans les calculs  de mati\`ere 
nucl\'eaire : 
\be	\rho({\bf r}_1,{\bf r}_2) = \rho(R_{12},s_{12}) = 
	\rho(R_{12}) \lambda_1(K_F s_{12})	\ee
o\`u $R_{12}$ et $s_{12}$ repr\'esentent les modules des coordonn\'ees du 
centre de masse et du mouvement relatif de deux nucl\'eons quelconques du noyau 
qui a \'et\'e excit\'e :
\be	 {\bf R}_{12} = {{{\bf r}_1+{\bf r}_2}\over 2} 
	\ \ \ \ \ \ \ \ 
	{\bf s}_{12} = {\bf r}_1 - {\bf r}_2	\ee 
et   
\be	\lambda_1(x) = {3 j_1(x) \over x}	\ee
$j_1(x)$ \'etant la fonction de Bessel de premier ordre. $K_F$ est le moment 
de Fermi donn\'e  par : 
\be	K_F = K_F(R_{12}) = \left[ 3 \pi^2 \rho(R_{12}) \right]^{1 \over 3} \ee
6) Dans le cas o\`u la port\'ee de la non-localit\'e de $\Delta V$ est petite 
devant la port\'ee du potentiel $U_0$, le potentiel local peut \^etre 
approxim\'e par le premier terme de la transformation de Wigner du potentiel 
non local  {\bf(HO80)}, {\bf(PE62)}, {\bf(PE64)} et {\bf(PE80)}. Ainsi on 
\'ecrira :
\be	W_L(R) = 4 \pi {m_r^*(R) \over m} \int_0^{\infty} j_0(ks) 
	{\cal{I}}m \Delta V(R,s) s^2 ds	\ee
$j_0$(ks) repr\'esente la fonction de Bessel d'ordre z\'ero, $ k$ est le 
moment r\'eduit d\'efini par : 
\be	k^2 = {2 \mu \over \hbar^2} \left[E - E_B \right]	\ee
o\`u  $E_B$ = $V_L$(R) + $V_C$, $V_L(R)$ est le potentiel r\'eel total et 
${\ds {m_r^*(R) \over m}}$ est l'op\'erateur de masse effective d\'efini par : 
\be	{m_r^*(R) \over m} = 1 - {d\Delta U_L(E) \over dE}	\ee
\'Etant donn\'e que pour des grandes valeurs de $R$, 
 ${\ds{{d\Delta U_L(E) \over dE} \ll 1}}$, 
({\bf NA85}), ({\bf MA86}) et que les processus de dispersion ne sont 
pas sensibles \`a la valeur du potentiel optique \`a faibles distances, nous 
pouvons consid\'erer que $m_r^*(R)/m\simeq 1$ dans le calcul de $W_L(R)$. 
   
 Une fois que $W_L$(R) a \'et\'e d\'etermin\'e, le terme local \'equivalent 
\`a la partie r\'eelle de $\Delta V(R,s)$, qui repr\'esente les corrections au 
potentiel $U_0$, se calcule selon la relation suivante: 
\be \Delta U_L(R) = 4 \pi \int_0^{\infty} j_0(ks) {\cal{R}}e 
                    \Delta V(R,s) s^2 ds \ee

La forme explicite de ces potentiels locaux d\'epend du caract\`ere hermitique 
ou non hermitique  de l'hamiltonien du  syst\`eme ainsi que de la valeur de 
l'\'energie de la collision par rapport aux barri\`eres d'absorption et de 
Coulomb. En particulier, dans le cas qui nous interesse, lorsque l'\'energie de 
la collision d\'epasse toutes les barri\`eres et que, par cons\'equent, 
l'hamiltonien n'est pas hermitique, le potentiel imaginaire $W_L$ et les 
corrections au potentiel r\'eel $\Delta U_L$ peuvent \^etre \'exprim\'es de la 
forme $W_L = f(W_L,\Delta U_L)$ et $\Delta U_L = g(W_L,\Delta U_L)$, vu que 
$W_L$ et $\Delta U_L$ sont inter-d\'ependants \`a travers des moments locaux 
$K_i$ et $k$. Ceci pose un probl\`eme d'autoconsistance qui recquiert un calcul 
it\'eratif.

Le calcul du potentiel noyau-noyau recquiert la connaissance des valeurs des 
\'energies d'excitation moyennes $<E_P>$ et $<E_C>$ des deux noyaux 
collisionnants, de l'intensit\'e, V$_0$, et de la port\'ee, $r_0$, de 
l'interaction effective et des densit\'es des noyaux cible et projectile dans 
leurs \'etats fondamentaux. Les \'energies moyennes d'excitation, $<E_P>$ et 
$<E_C>$, se d\'eterminent en moyennant les \'energies des \'etats les plus 
fortement excit\'es dans chacun des deux noyaux interagissants. Dans les 
r\'ef\'erences (${\bf VI86}$) et (${\bf BA92}$) sont indiqu\'ees les valeurs 
pour diff\'erents noyaux. Comme il a \'et\'e montr\'e dans la r\'ef\'erence 
(${\bf VI91}$) pour le syst\`eme \opb, l'introduction des \'energies moyennes 
permet de bien reproduire l'absorption due aux voies collectives de plus basses 
\'energies d'excitation, mais surestime l'absorption due aux \'etats collectifs 
de plus hautes \'energies d'excitation, bien que cet effet soit fortement 
att\'enu\'e \`a mesure que l'\'energie de la collision augmente. Dans cette 
m\^eme r\'ef\'erence, il est justifi\'e qualitativement comment le mod\`ele de 
l'approximation de fermeture inclut beaucoup plus que la contribution des 
\'etats de basse et de haute \'energie d'excitation; en particulier, la 
contribution \`a l'absorption due aux processus de transfert d'un ou de 
plusieurs 
nucl\'eons est simul\'ee, principalement, par la contribution des \'etats de plus 
grande \'energie d'excitation. 

En raison de l'introduction des relations de fermeture, toutes les excitations 
des noyaux cible et projectile sont implicitement prises en consid\'eration. 
Cependant, d'autres voies telles que les canaux de diffusion 
nucl\'eon-nucl\'eon ne sont pas inclus dans le mod\`ele. Il s'ensuit qu'\`a 
\'energie \'elev\'ee, 
lorsque ces processus dominent l'absorption, le 
mod\`ele n'est pas en mesure de d\'ecrire de mani\`ere ad\'equate les 
collisions noyau-noyau.
     
La d\'etermination de l'intensit\'e $V_0$ de l'interaction effective et du 
param\`etre $\eta$ li\'e \`a sa port\'ee, se fait une seule fois pour chaque 
syst\`eme. Pour un syst\`eme donn\'e et \`a une \'energie suffisamment 
\'eloign\'ee de la barri\`ere de Coulomb pour que tous les canaux importants 
soient ouverts, on ajuste le potentiel r\'eel total: $U_0(R) + \Delta U_L(R)$ 
au potentiel ph\'enom\'enologique qui reproduit les donn\'ees exp\'erimentales 
de diffusion \'elastique dans le domaine de sensibilit\'e. Pour tous les 
syst\`emes \'etudi\'es, nous avons trouv\'e une valeur de l'intensit\'e de 
l'interaction $V_0 \simeq 58 MeV$ (${\bf FE90}$, ${\bf VI91}$, ${\bf BA91}$, 
${\bf VI93}$). Par contre, la valeur du 
param\`etre $\eta$ est bien distincte selon que le syst\`eme est sym\'etrique 
ou non. Pour les syst\`emes sym\'etriques ou faiblement assym\'etriques, 
nous obtenons toujours une valeur $\eta = 0.7 fm^{-2}$ correspondant \`a une 
port\'ee $r_0 \simeq 1.69 fm$. Pour les syst\`emes fortement assym\'etriques,  
tels que le syst\`eme \opb , nous obtenons $\eta = 0.45 fm^{-2}$ correspondant 
\`a une valeur de la port\'ee $r_0 \simeq 2.1 fm$. Cette diff\'erence est une 
cons\'equence de l'approximation de la force s\'eparable. En effet, 
l'interaction donn\'ee par l'\'equation (2.39) peut s'\'ecrire sous la forme :
\be V_{ij}({\bf r}) = - V_0 e^{- {\eta \over {2}} \left({\bf r}_i + 
                              {\bf r}_j\right)^2} 
                            e^{- {\eta \over {2}}\left({\bf r}_i - {\bf r}_j + 
                              {\bf r}\right)^2} \ee
 Sur la figure 2.2 il est \'evident que le second facteur qui apparait dans 
l'\'equation (2.48) est le facteur de forme gaussien de port\'ee 
$(2/\eta)^{1/2}$ qui d\'ecrit l'interaction entre les nucl\'eons $i$ et $j$, 
tandis que le premier facteur induit une modification artificielle du facteur 
de forme usuel. Si les deux noyaux sont loins l'un de l'autre, ce facteur est 
\'evalu\'e, en moyenne, \`a 
$g_0=e^{-{\eta \over {2}}\left({\bf R_P} - {\bf R_C} 
\right)^2}$ o\`u $R_P$ et $R_C$ sont, respectivement, 
les rayons des noyaux projectile et cible. 
Par ailleurs, si les deux noyaux sont identiques ou s'ils ont 
des rayons proches l'un de l'autre, le facteur $g_0$ tend vers l'unit\'e alors 
que pour des syst\`emes fortement assym\'etriques, il est inf\'erieur \`a 
l'unit\'e ce qui implique necessairement que $V_0$ doit \^etre renormalis\'ee 
et sa valeur augment\'ee afin de compenser la valeur de $g_0$. D'autre part, 
lorsque les noyaux se recouvrent largement, $g_0$ atteint une valeur proche de 
$1$ et la pente du potentiel augmente pour les syst\`emes assym\'etriques. 
Pour corriger cet effet, il faut diminuer le param\`etre $\eta$ dans un 
intervalle qui conduise \`a des valeurs de $r_0$ physiquement raisonnables.

Quant aux densit\'es des noyaux dans leurs \'etats fondamentaux, on utilise 
celles obtenues dans des exp\'eriences de diffusion d'\'electrons ou \`a partir 
de calculs r\'ealis\'es dans le cadre du mod\`ele en couches.

Bien que le mod\`ele de l'approximation de fermeture permette d'\'ecrire le 
potentiel r\'eel comme somme d'un potentiel de double convolution et d'un 
terme de polarisation donn\'e par l'\'equation (2.47), il est mieux adapt\'e au 
calcul du terme d'absorption du potentiel donn\'e par l'\'equation (2.44). Ceci 
est d\^u \`a certaines des approximations que nous avons faites et, en 
particulier, au type de force effective utilis\'ee. En effet, ce type de force 
a \'et\'e amplement utilis\'e dans les \'etudes de structure nucl\'eaire et il 
est donc raisonnable d'y recourir dans la description des \'el\'ements de 
matrice transition de $V_{mn}$, strictement li\'es aux effets de structure. Par 
contre, la dynamique du processus, inclue dans le propagateur, est mieux 
d\'ecrite par d'autres types d'interaction tels que l'interaction effective 
M3Y. Pour cette raison nous avons assum\'e comme potentiel r\'eel 
$U_0(R) + \Delta U_L(R) \simeq N(E) \ \ V_{M3Y}(R)$ que nous avons utilis\'e 
pour l'\'evaluation du propagateur WKB qui intervient dans le calcul du terme 
d'absorption donn\'e par l'\'equation (2.44). 

Notons, pour terminer cette section, que pour des syst\`emes tr\`es d\'eform\'es 
et aux \'energies proches de la barri\`ere de Coulomb, l'absorption est 
contr\^ol\'ee par peu de canaux et l'hypoth\`ese principale du mod\`ele de 
fermeture n'est plus valide. Dans ce cas il est n\'ecessaire de faire appel 
\`a la troisi\`eme mod\'elisation de la th\'eorie de Feshbach expos\'ee dans ce
travail, \`a savoir 
l'\'evaluation, terme \`a terme, de la contribution \`a l'absorption des canaux 
in\'elastiques les plus importants.  

\subsection{Troncation des \'etats dans le calcul du potentiel optique:}
\par
  
Nous rappellerons rapidement le formalisme d\'ej\`a abondamment  d\'ecrit 
dans la litt\'erature {\bf(VI91)}, {\bf(BA91)}, {\bf(BA92)}, {\bf(PA95)}. 
Nous pouvons r\'e\'ecrire le terme de polarisation de l'\'equation (2.27) sous 
la forme:
\be	\Delta V({\bf r},{\bf r'}) =\sum_{i\not=0} 
	V_i^*({\bf r}) G_i({\bf r},{\bf r'}) V_i({\bf r'})	\ee
\noindent o\`u l'indice $i$ repr\'esente tous les canaux in\'elastiques et 
$V_i({\bf r})$ est l'\'el\'ement de matrice de transition qui, pour un 
\'etat i de moment angulaire ($\lambda , \mu$), peut s'\'ecrire :   
\be	V_i({\bf r}) ={1\over\sqrt{2\lambda+1}} 
	f_\lambda ^{(i)}(r) Y_\lambda^\mu ({\bf \hat r}) \ee
\noindent introduisant cette expression dans l'\'equation ant\'erieure, nous obtenons :
\be	\Delta V({\bf r},{\bf r'}) =\sum_{i\not=0} \sum_{\lambda}
	G_i({\bf r},{\bf r'}) f_\lambda ^{(i)*}(r)
	f_\lambda ^{(i)}(r') {P_{\lambda}(cos \ \ {\bf \hat {rr'}}) 
       \over 4 \pi}  \ee
la fonction de Green, $G_i({\bf r},{\bf r'})$, d\'ecrit la propagation du 
syst\`eme  dans le canal $i$ et peut, comme pr\'ec\'edemment, \^etre 
approxim\'ee par le propagateur WKB, 
\be G_i({\bf r},{\bf r'})\simeq G_i^{WKB}({\bf r},{\bf r'}) = 
-{\mu\over2\pi \hbar^2}	{e^{iK_i\left\vert{\bf r}-{\bf r'}\right\vert}
\over {\left\vert{\bf r}-{\bf r'}\right\vert}}=-{\mu\over2\pi \hbar^2}
{e^{iK_is}\over {s}}	\ee  
o\`u $s=\left\vert{\bf r}-{\bf r'}\right\vert$ est la coordonn\'ee relative et  
$K_i$ est le moment local WKB pour le canal in\'elastique $i$ d'\'energie 
d'excitation $E_i$, donn\'e par : 
\be K_i^2={2\mu\over\hbar^2} \left[ E_{cm} - (E_i+U(R)+
i{\cal{I}}m \Delta {V_L}^{in}(R)+V_C(R)) \right] \ee
o\`u $E_{cm}$ est l'\'energie de collision dans le centre de masse, 
$U(R)=U_0(R) + {\cal{R}}e \Delta {V_L}^{in}(R)$ est le potentiel r\'eel local, 
$V_C(R)$ est le potentiel de Coulomb et ${\cal{I}}m \Delta {V_L}^{in}(R)$ est 
le potentiel imaginaire local qui d\'ecrit l'absorption due \`a l'excitation 
des \'etats in\'elastiques et que, par soucis de simplicit\'e, nous noterons 
$W^{in}(R)$. Le moment local $K_i$ peut \^etre 
exprim\'e par $K_i = k_i +  i \kappa_i$, avec :
\be 	k_i^2={\mu\over \hbar^2}
	\left[ E_{cm}-E_B^{i*}(R) +\sqrt{\left[E_{cm}-E_B^{i*}(R)\right]^2
	+\left[W^{in}(R)\right]^2}\right] \ee
\be	\kappa_i ={\mu\over\hbar^2}\cdot {W^{in}(R)\over k_i}
	\hskip 72mm\nonumber\ee
o\`u la hauteur de la barri\`ere du canal $i$ est donn\'ee par : 
$E_B^{i*}(R) = E_i + U(R) + V_C(R)$.
   
Dans la th\'eorie de Feshbach du potentiel optique, il est d\'emontr\'e qu'un 
canal d\'etermin\'e ne participe pas \`a l'absorption si l'\'energie de la 
collision $E_{cm}$ est inf\'erieure \`a l'\'energie seuil: $E_B^{i*}$. Cela 
est d\^u au 
fait que le propagateur est r\'eel. Cette propri\'et\'e doit \^etre satisfaite 
par notre propagateur WKB. En effet, lorsque $E_{cm}<E_B^{i*}$ le canal $i$ 
est \'energ\'etiquement inaccessible et ne peut se peupler; il ne contribue 
donc pas \`a l'absorption et $W_i^{in}(R)=0$. Dans ce cas, $\kappa_i=0$, 
$k_i^2<0$ et $K_i = i\left | k_i \right |$.

Introduisant les composantes $k_i$ et $\kappa_i$ (\'eqs. 2.54 et 2.55) de $K_i$ 
dans l'\'equation (2.52), nous obtenons le propagateur WKB dans tout le 
domaine d'\'energies :
  \ba     G_i^{WKB} = - {\mu\over 2 \pi \hbar^2} {e^{- \kappa_i s }\over s}
	\left[ cos(k_i s) + i sin(k_i s) \right] \ \ \ \ 
	& \mbox{si $ E_{cm} > E_B^{i*}(R)$} \\
	G_i^{WKB} = -{\mu\over 2 \pi \hbar^2} {e^{- \left| k_i \right| s} 
    \over s} \ \ \ \ \ \ \ \ \ \ \ \ \ \ \ \ \ \ \ \ \ \ \ \ \ \ \ \ \ \ \
        & \mbox{si $ E_{cm} < E_B^{i*}(R)$} \ea
Le propagateur du mouvement relatif donn\'e par l'\'equation (2.52) se calcule 
avec l'hamiltonien total $H = H_0 + V_N$, o\`u $H_0$ inclut la somme des 
hamiltoniens des deux noyaux ind\'ependants et de l'interaction de Coulomb entre 
eux, tandis que $V_N$ repr\'esente l'interaction nucl\'eaire cible-projectile 
d\'ecrite par le potentiel optique, $V_N={\cal{R}}eV_N+i{\cal{I}}mV_N$. Dans 
notre 
mod\'elisation, tel qu'il sera indiqu\'e plus loin, ${\cal{R}}eV_N$ est 
consid\'er\'e 
comme le potentiel r\'eel obtenu par ajustement des donn\'ees exp\'erimentales 
et inclut, par cons\'equent, la contribution de tous les processus r\'eels 
(\'energ\'etiquement accessibles) et virtuels (\'energ\'etiquement 
inaccessibles). Par contre, ${\cal{I}}mV_N$ repr\'esente le terme d'absorption du 
potentiel d\^u aux voies ouvertes et qui ont une contribution significative. 
Ainsi donc, notre propagateur $G$ est un propagateur complet et il peut \^etre 
d\'evelopp\'e en termes du propagateur $G_0$ calcul\'e seulement avec 
l'hamiltonien $H_0$ :
 \be G = G_0 + G_0V_NG_0 + G_0V_NG_0V_NG_0 + ..... \ee
et en introduisant cette relation dans l'\'equation (2.49), nous obtenons :
\be   \Delta V({\bf r},{\bf r'}) =\sum_{i\not=0} 
	\left[ V_i^*({\bf r}) G_0({\bf r},{\bf r'}) V_i({\bf r'}) + 
 V_i^*({\bf r}) G_0({\bf r},{\bf r'}) V_N G_0({\bf r},{\bf r'}) 
 V_i({\bf r'}) + ....\right] \ee
o\`u le premier terme d\'ecrit les processus in\'elastiques directs tandis que 
les autres termes correspondent aux processus multi-\'etapes qui d\'ecrivent la 
contribution du noyau compos\'e.
 
 Nous pouvons d\'eduire des \'equations (2.56) et (2.57) que tous les canaux, 
ouverts ou ferm\'es, contribuent \`a la partie r\'eelle du potentiel dans tout 
le domaine radial, tandis que seuls les canaux ouverts contribuent au 
potentiel imaginaire dans le domaine radial o\`u ils sont ouverts. Ceci nous 
pose un probl\`eme difficilement abordable vu que pour calculer la partie 
r\'eelle du potentiel de polarisation nous devons prendre en consid\'eration 
toutes les voies, tandis que pour le calcul de la partie imaginaire nous 
n'avons besoin d'inclure que les voies ouvertes. Notre mod\'elisation 
s'av\`ere donc plus appropri\'ee pour le calcul de la partie imaginaire du 
potentiel et dans tout ce qui suit, nous ne ferons r\'ef\'erence qu'au terme 
d'absorption du potentiel.

Si nous supposons une faible non-localit\'e du potentiel non local, nous 
pouvons obtenir le potentiel local equivalent comme la transform\'ee de 
Fourier du potentiel non local {\bf(PE62)}, {\bf(PE64)}, {\bf(PE80)} et 
{\bf(HO80)}.
\be  W^{in}(R) = \int e^{i{\bf k}\cdot {\bf s}} 
	{\cal{I}}m \Delta V\left({\bf R}+{{\bf s} \over 2},
         {\bf R}-{{\bf s} \over 2} \right) d{\bf s}  \ee
o\`u ${\bf R}$ et ${\bf s}$ sont les coordonn\'ees du centre de masse et 
du mouvement relatif 
\be {\bf R} = {{{\bf r} + {\bf r'}}\over 2} \ \ \ \ et \ \ \ \  {\bf s} = 
{\bf r} - {\bf r'}\ee
et ${\bf k}$ est le moment local d\'efini par :
\be	k^2={2\mu\over \hbar^2} \left[ E_{cm}- E_B^*(R) \right]	\ee
avec :
\be	E_B^*(R) = U(R) + V_C(R)	\ee
$k$ est r\'eel except\'e si $E_{cm}<E_B^*(R)$, auquel cas 
$ k=i\left| k \right| $.

Si nous supposons, pour l'interaction noyau-noyau, une faible d\'ependance 
angulaire entre les vecteurs ${\bf R}$ et ${\bf s}$ (comme cela a \'et\'e 
montr\'e pour le cas de 
l'interaction nucl\'eon-noyau {\bf(BOU81)}), 
il est possible d'\'evaluer 
${\cal{I}}m \Delta V\left({\bf R}+{{\bf s} \over 2},
{\bf R}-{{\bf s} \over 2} \right)$  en choisissant deux situations extr\^emes 
correspondant \`a  ${\bf R}$ et ${\bf s}$ ou, ce qui est \'equivalent, \`a 
$\bf r$ et $\bf r'$, colin\'eaires ou anti-colin\'eaires.
   
Nous pouvons obtenir ais\'ement les relations suivantes :
\ba    {\cal{I}}m \Delta V(R,s) = \left\{ \begin{array}{ll}
	{\ds {1 \over 4 \pi} \sum_{\lambda ,i \not= 0} {\cal{I}}m G_i(R,s)
	f_{\lambda}^{(i)*}(R+{s \over 2}) f_{\lambda}^{(i)}(R-{s \over 2})} 
	& \mbox{si ${\ds R>{s \over 2}}$} \\
	{\ds {1 \over 4 \pi}  \sum_{\lambda ,i \not= 0} (-1)^{\lambda}
	{\cal{I}}m G_i(R,s)
	f_{\lambda}^{(i)*}(R+{s \over 2}) f_{\lambda}^{(i)}(-R+{s \over 2})}   
	&  \mbox{si ${\ds R<{s \over 2}}$}
	\end{array}
	\right.         \ea
 Finalement a partir 
des \'equations (2.60) et (2.64) et vue la faible non localit\'e du potentiel non 
local, $s\simeq 1 fm$, {\bf(BA91)}, nous obtenons 
l'expression :
 \be W^{in}(R) = \sum_{\lambda ,i \not= 0} 
\int_0^{2R} s^2 ds {\cal{I}}m G_i(R,s) j_0(ks) 
f_{\lambda}^{(i)*}(R+{s \over 2}) 
	f_{\lambda}^{(i)}(R-{s \over 2})	\ee
qui est satisfaite pour des valeurs de $R>s/2$.
   
 Le potentiel $W^{in} (R)$ doit \^etre corrig\'e par un 
op\'erateur de masse effective car, dans la d\'efinition du moment local, k, 
nous n'avons inclus que la partie r\'eelle du potentiel optique. Nous 
supposons que cet op\'erateur de masse effective est proche de l'unit\'e 
($ {\ds {m^* \over m}\simeq 1 }$) \`a la surface nucl\'eaire {\bf(VI86)}, 
{\bf(VI87)} et {\bf(VI90)}.

Lorsque nous nous int\'eressons aux \'etats collectifs de plus basses 
\'energies d'excitation, les facteurs de forme radiaux $f_\lambda^{(i)}$ qui 
apparaissent dans l'expression du potentiel de Feshbach sont bien d\'ecrits, 
dans le cadre du mod\`ele collectif, par :
\be	f^{(i)}_\lambda(r)=\beta^{(i)}_{\lambda j}R_j{\partial U^{Cop}(r)\over
	\partial r}=\delta_{\lambda j}^{(i)}{\partial U^{Cop}(r)\over
	\partial r} 	\ee
o\`u $\beta^{(i)}_{\lambda j}$ est l'amplitude de transition de multipolarit\'e 
$\lambda$ dans le canal i, pour le noyau j et  $ \delta_{\lambda j}^{(i)}$ est 
la longueur de d\'eformation correspondante. $R_j$ est le rayon de ce noyau et 
$U^{Cop}(r)$ est le potentiel de Copenhague noyau-noyau {\bf(BR81)} et 
{\bf(BRO81)}.

Il a \'et\'e montr\'e {\bf(BR81)}, que le facteur de forme nucl\'eaire 
calcul\'e \`a partir de l'\'equation pr\'ec\'edente en utilisant le potentiel 
de Copenhague, reproduit les facteurs de forme nucl\'eaires microscopiques RPA 
dans un large domaine radial de la surface nucl\'eaire.

En introduisant la partie imaginaire du propagateur et ces facteurs de forme 
radiaux dans l'expression du potentiel donn\'e par l'\'equation (2.65), nous 
obtenons la contribution des canaux in\'elastiques les plus bas au potentiel 
imaginaire :
\be	W^{in}(R) = - {\mu \over 2 \pi k \hbar^2}
	\sum_{\lambda ,i \not= 0} \delta_{\lambda j}^{(i)2}
	 \int_0^{2R} ds e^{- \kappa_i s}
	sin(ks) sin(k_is) P(R,s)	\ee
quant \`a P(R,s), il  est d\'efini par :
\be	P(R,s) = g(R + {s \over 2}) g(R - {s \over 2})	\ee
avec : 
\be	g(R \pm {s \over 2}) = \left. {\partial U^{Cop}(r)\over
	\partial r} \right|_{r=R\pm {s \over 2}}	\ee
Le facteur de forme nucl\'eaire pour l'excitation mutuelle est donn\'e par {\bf 
(SA83)}:
\be	f^{(ii')}_{\lambda\lambda'}(r)={1\over 4 \pi}\beta^{(i)}_{\lambda j}
       \beta^{(i')}_{\lambda' k}R_jR_k{\partial{}^2 U^{Cop}(r)\over
	\partial{}^2 r}=\delta_{\lambda j}^{(i)}\delta_{\lambda' k}^{(i')}
        {\partial{}^2 U^{Cop}(r)\over \partial{}^2 r} 	\ee
o\`u tous les symboles ont d\'ej\`a \'et\'e d\'efinis et ils se rapportent \`a 
pr\'esent aux noyaux $j,k$ dans leurs \'etats excit\'es respectifs $i,i'$ de 
multipolarit\'es $\lambda, \lambda'$. Dans le cas de l'excitation mutuelle, nous 
pouvons obtenir une expression similaire \`a celle donn\'ee par l'\'equation 
2.67, sans rien faire de plus que remplacer des facteurs de forme par d'autres. 
Dans la pratique, il suffit de remplacer les longueurs de d\'eformation $ 
\delta_{\lambda j}^{(i)2}$ par $ \delta_{\lambda j}^{(i)2}\delta_{\lambda' 
k}^{(i')2}$ et les fonctions $g(R \pm {s \over 2})$ par les fonctions $h(R \pm 
{s \over 2})$, d\'efinies par :
 \be h(R \pm {s \over 2}) =\left. {\partial{}^2  U^{Cop}(r)\over
	\partial{}^2 r} \right|_{r=R\pm {s \over 2}}	\ee
Nous pouvons obtenir une 
indication qualitative sur la forme du potentiel imaginaire due \`a la 
contribution des canaux in\'elastiques ouverts, dans la limite o\`u la 
port\'ee de la non-localit\'e est proche de z\'ero :
\be	W^{in}(R) = \lim_{s \rightarrow 0}
	 {\cal{I}}m \Delta V(R,s) =
	- {\mu \over 8 \pi^2  \hbar^2} \sum_{\lambda, i \not= 0}
	k_i \left| f_\lambda^{(i)}(R) \right|^2	\ee	
Cependant, la port\'ee de la non-localit\'e est faible mais non nulle et cette 
derni\`ere \'equation ne peut \^etre consid\'er\'ee que comme une 
approximation asymptotique. Elle indique que dans la limite o\`u la port\'ee 
de la non-localit\'e est proche de z\'ero, le potentiel imaginaire de 
l'\'equation pr\'ec\'edente peut \^etre obtenu comme la contribution des termes 
proportionnels au  
carr\'e des facteurs de forme radiaux. Avec le type de facteurs de forme 
adopt\'es (\'equation 2.66) nous obtenons :
\be	W^{in}(R) \propto
	\sum_{\lambda ,i \not= 0} k_i \delta_{\lambda j}^{(i)2}
	 \left| {\partial U(R)\over \partial R} \right|^2	\ee
Ceci indique que les canaux in\'elastiques les plus bas ne contribuent qu'\`a 
la surface du potentiel, ce qui est en parfait accord avec les r\'esultats 
obtenus par de Broglia et al. ({\bf{BRO81}}) utilisant des m\'ethodes 
semiclassiques. Les principales diff\'erences entre nos calculs et 
ceux de Broglia et al., en dehors de l'utilisation d'une m\'ethode 
semiclassique, proviennent des points suivants :
  
1. La non localit\'e du potentiel que nous avons correctement prise en 
consid\'eration.
  
2. L'utilisation d'un propagateur complet, entra\^{\i}nant que, dans nos 
calculs, tous les processus, et pas seulement les processus directs, sont pris 
en consid\'eration.
   
L'approximation adopt\'ee, consistant \`a n'inclure dans notre mod\'elisation 
de la th\'eorie de Feshbach que les \'etats collectifs de basse \'energie 
d'excitation et \`a approximer leurs facteurs de forme radiaux par des 
fonctions de surface, signifie que nous n'avons pris en consid\'eration que la 
contribution de surface des \'etats du noyau compos\'e. C'est une approximation 
qui, \`a basse \'energie, ne doit pas affecter les sections efficaces de 
diffusion \'elastique, vu qu'il n'existe aucune sensibilit\'e \`a la valeur du 
potentiel imaginaire \`a l'int\'erieur.
  
Remarquons que, dans le formalisme de Feshbach, la 
partie absorbante du potentiel trouve son origine dans le potentiel de 
polarisation imaginaire dont la structure est compliqu\'ee du fait qu'il 
inclut des processus de second ordre ou plus.  

Dans tout ce qui pr\'ec\`ede nous n'avons pas consid\'er\'e la partie r\'eelle
du potentiel. Il est possible de d\'eduire des expressions similaires pour
 le potentiel de polarisation r\'eel. Cependant nous avons montr\'e que 
 tous les canaux, ouverts ou ferm\'es,  
contribuent au potentiel de polarisation r\'eel tandis que seuls les canaux 
ouverts contribuent au potentiel imaginaire. Par cons\'equent, nous devons 
inclure dans nos calculs tous les canaux, ouverts ou ferm\'es, parce qu'ils 
contribuent tous, non seulement explicitement, au potentiel r\'eel, mais aussi, 
implicitement, aux potentiels r\'eel et imaginaire, \`a travers le moment 
local. Afin d'\'eviter les difficult\'es pos\'ees par ce probl\`eme, nous 
rempla\c{c}ons dans nos calculs le potentiel r\'eel total par le potentiel 
renormalis\'e que nous avions obtenu en ajustant les donn\'ees 
exp\'erimentales. 
Ainsi, la contribution de tous les canaux est prise en compte :
\be	U_0(R) + {\cal{R}}e \Delta V_L(R) \simeq N(E) V_{M3Y} (R)	\ee
Nous arrivons ainsi \`a obtenir le potentiel imaginaire d\^u aux canaux 
in\'elastiques ouverts, explicitement inclus dans les calculs. Cependant il 
faut avoir \`a l'esprit qu'\`a travers les moments locaux $k_i$, 
$\kappa_i$ et k, le potentiel $W^{in}(R)$ d\'epend \`a la fois 
du potentiel r\'eel total et du potentiel imaginaire total de la mani\`ere 
suivante :
\be W^{in}(R) = F \left[ N(E)V_{M3Y}(R) , W^{in}(R)\right] \ee
Cela signifie que, pour effectuer le calcul du potentiel imaginaire, une 
proc\'edure it\'erative est n\'ecessaire.

Pour terminer ce chapitre, il nous reste \`a dire que, tel que cel\`a a \'et\'e 
d\'emontr\'e dans la r\'ef\'erence ({\bf VI91}), la relation de 
dispersion donn\'ee par l'\'equation (2.15) est satisfaite analytiquement entre 
les parties r\'eelle et imaginaire des potentiels non locaux obtenus avec les 
mod\'elisation de Feshbach d\'ecrites dans les parties 2.2.2 et 2.2.3 lorsque 
le propagateur est \'evalu\'e dans l'approximation WKB \`a partir d'un 
hamiltonien hermitique. D'autre part, Pacheco et al. ({\bf PA91}) ont 
d\'emontr\'e que dans le cas de potentiels locaux obtenus en introduisant un 
hamiltonien hermitique dans l'\'evaluation du propagateur, la relation de 
dispersion est satisfaite exactement sous certaines conditions (couplage 
faible, propagateur WKB, localisation par ondes planes et masse effective 
proche de l'unit\'e). Ces r\'esultats apportent un soutien th\'eorique \`a 
toutes les 
analyses ph\'enom\'enologiques bas\'ees sur l'utilisation de potentiels locaux. 
Cependant, si, comme il est habituel, l'hamiltonien qui d\'ecrit l'interaction 
n'est pas hermitique, sa non hermiticit\'e introduit une violation de la 
relation de dispersion dont l'\'evaluation par des m\'ethodes num\'eriques 
({\bf VI91}), indique qu'elle est faible. Son origine r\'eside dans le fait 
que les 
\'equations qui permettent de d\'eterminer les parties r\'eelle et imaginaire 
du potentiel optique forment un syst\`eme d'\'equations coupl\'ees qui ne 
peuvent \^etre \'evalu\'ees s\'epar\'ement et qui necessitent un calcul 
it\'eratif.

\newpage
$ $
\newpage
\setcounter{chapter}{2}
%\setcounter{page}{19}
%\pagenumbering{arabic}
\chapter{{\Large{\bf{ABSORPTION~AU~VOISINAGE~DE~LA~BARRI\`ERE}}}}
\section{\large{\bf {DONN\'EES EXP\'ERIMENTALES :}}}
\par

Une collaboration de plusieurs ann\'ees entre une \'equipe du CRN (actuellement 
IReS) de Strasbourg (France) et une autre de l'IFIC de Valencia (Espagne) a 
permis 
d'accumuler des donn\'ees tr\`es pr\'ecises, sur un large domaine angulaire et 
\`a diff\'erentes \'energies incidentes, de la diffusion \'elastique, 
in\'elastique et des transferts de quelques nucl\'eons dans l'interaction de 
diff\'erents noyaux de la couche sd.

La m\'ethode exp\'erimentale utilis\'ee est bas\'ee sur l'identification par la 
cin\'ematique des produits de r\'eaction. Cette m\'ethode, valable pour les 
r\'eactions \`a deux corps dans la voie de sortie, n\'ecessite la mesure de 
quatre grandeurs. En pratique le dispositif mis au point permettait la mesure 
des \'energies cin\'etiques des noyaux form\'es ainsi que leur angle 
d'\'emission par rapport \`a la direction des noyaux incidents. Il a \'et\'e 
d\'ecrit dans diff\'erents travaux publi\'es (voir en particulier 
{\bf (BI86)}, {\bf (ER72)}, {\bf (HU81)}, {\bf (ST85)}, {\bf (WE72)} et 
{\bf (ZE77)} ).

Nous ne reprendrons ici que les grandes lignes et les points qui ont fait 
l'objet d'une attention particuli\`ere permettant d'optimiser la m\'ethode et 
d'obtenir des donn\'ees de grande qualit\'e.

La d\'etection des produits de r\'eaction \'etait effectu\'ee \`a l'aide de 
deux d\'etecteurs solides (Si) de grandes surfaces (9x48mm), sensibles \`a la 
position et associ\'es \`a un dispositif \'electronique de co\"{\i}ncidence. 
Celui-ci 
permettait de s\'electionner les noyaux provenant d'une m\^eme r\'eaction.  
Les d\'etecteurs \'etaient plac\'es dans une chambre \`a r\'eaction install\'ee 
sur l'une des aires exp\'erimentales de l'acc\'el\'erateur \'electrostatique de 
type tandem qui fut en fonction au CRN jusqu'en 1989. Le dispositif permettait 
de mesurer en m\^eme temps une partie de la distribution angulaire (typiquement 
un intervalle de $40^o$). Pour 
obtenir l'ensemble de la distribution, plusieurs mesures effectu\'ees avec 
des configurations angulaires diff\'erentes des d\'etecteurs \'etaient 
n\'ecessaires.  
La normalisation entre les diff\'erentes portions de distributions angulaires 
\'etait r\'ealis\'ee par recouvrement des zones communes. Le nombre de 
configurations diff\'erentes des d\'etecteurs variait de 4 \`a 6 suivant les 
syst\`emes \'etudi\'es.

Durant les mesures une analyse en ligne d'un \'echantillonage des 
\'ev\'enements enregistr\'es (sur bandes magn\'etiques) permettait le 
contr\^ole de la qualit\'e des donn\'ees. Celle-ci d\'epend de la stabilit\'e 
de l'\'electronique, de la qualit\'e des cibles et des 
d\'etecteurs. La r\'eponse des d\'etecteurs \'etait r\'eguli\`erement 
\'etalonn\'ee \`a l'aide de grilles, form\'ees de fentes tr\`es fines 
r\'eguli\`erement espac\'ees, plac\'ees devant les d\'etecteurs \`a une 
position 
rep\'er\'ee avec pr\'ecision.

La r\'esolution en \'energie du dispositif \'etait assez pauvre. Cela \'etait 
li\'e aux caract\'eristiques des d\'etecteurs mais \'etait aussi inh\'erent \`a 
l'int\'eraction des ions lourds de basse \'energie avec la mati\`ere. Lorsqu'ils 
traversent une \'epaisseur de mati\`ere donn\'ee, les ions perdent une partie 
de leur \'energie et sont d\'evi\'es. Les valeurs de cette perte d'\'energie 
et de cet angle de d\'eviation sont distribu\'ees autour de valeurs moyennes 
qui d\'ependent de l'ion, de la mati\`ere travers\'ee et de l'\'energie 
incidente. La 
largeur des distributions d\'epend des m\^emes causes. Ces effets sont 
d'autant plus importants que les ions sont lourds, les \'energies faibles et 
l'\'epaisseur travers\'ee importante.

  Afin de minimiser ces effets, un soin 
particulier a \'et\'e apport\'e aux cibles. Elles ont \'et\'e choisies tr\`es 
minces (de quelques $\mu$g/cm${}^2$ \`a 20 $\mu$g/cm${}^2$), fabriqu\'ees par 
\'evaporation ou par implantation sur des supports minces de carbone (20 
$\mu$g/cm${}^2$). Leur orientation par rapport au faisceau incident a \'et\'e 
d\'etermin\'ee, pour chaque configuration angulaire des d\'etecteurs, de 
fa\c{c}on \`a minimiser l'\'epaisseur travers\'ee apr\`es la r\'eaction par 
les noyaux d'\'energies les plus basses. Dans tous les cas, elles ont \'et\'e 
plac\'ees de sorte que le faisceau incident rencontre d'abord le support de 
carbone. De cette fa\c{c}on, l'\'energie incidente la plus probable \'etait la 
m\^eme pour tous les \'ev\'enements retenus. Les \'energies incidentes pour 
lesquelles sont donn\'ees les distributions angulaires tiennent compte de la 
perte d'\'energie dans le support de carbone et dans la moiti\'e de 
l'\'epaisseur de la cible utile.

Contrairement \`a leur r\'esolution en \'energie, la r\'esolution angulaire des 
d\'etecteurs \'etait tr\`es bonne \`a condition de les placer \`a une distance 
suffisante. Leur r\'esolution spatiale \'etait meilleure que le millim\`etre. 
Les 
distances choisies pour les d\'etecteurs correspondaient \`a un compromis entre 
:
\begin{itemize}
\item[a)] La r\'esolution angulaire n\'ecessaire. (La s\'eparation des 
diffusions \'elastiques et in\'elastiques est d'autant plus difficile que 
l'\'energie du premier niveau d'excitation de l'un ou l'autre des noyaux 
\'etudi\'es est basse).
\item[b)] L'angle solide couvert par le d\'etecteur. De cette valeur d\'epend 
le nombre de mesures diff\'erentes, donc le temps total pour mesurer une 
distribution angulaire.
\end{itemize}

Rappelons que les effets de dispersion angulaire (straggling angulaire) 
d\'ecrits dans le paragraphe pr\'ec\'edent d\'egradent la r\'esolution 
angulaire du dispositif. Ils ont \'et\'e minimis\'es par les pr\'ecautions 
apport\'ees \`a l'orientation de la cible et \`a son \'epaisseur mais ils sont 
responsables de la limite en r\'esolution angulaire.

Tenant compte des consid\'erations pr\'ec\'edentes sur la r\'esolution en 
\'energie et en angle du dispositif, l'analyse des donn\'ees a \'et\'e faite en 
deux \'etapes. Tout d'abord les grandeurs exp\'erimentales (\'energies et 
angles) ont \'et\'e utilis\'ees pour s\'electionner les syst\`emes de noyaux 
dans la voie de sortie.  
Ensuite, seules les donn\'ees angulaires, beaucoup plus pr\'ecises, \'etaient 
utilis\'ees pour analyser la voie de r\'eaction int\'eressante. De fa\c{c}on 
plus concr\`ete, pour les donn\'ees de diffusion \'elastique utilis\'ees dans 
ce travail, la premi\`ere \'etape consistait \`a s\'electionner les 
\'ev\'enements dont les noyaux dans la voie de sortie correspondaient \`a ceux 
de la voie d'entr\'ee. La seconde \'etape permettait de s\'eparer la voie 
\'elastique des voies in\'elastiques et de d\'eterminer les sections efficaces 
diff\'erentielles, avec un pas angulaire de $0.5^o$ dans le syst\`eme du centre 
de masse.

Le domaine angulaire mesurable est limit\'e aux petits angles par le fort taux 
de comptage qui, d'une part entra\^{\i}ne une destruction des d\'etecteurs, et 
d'autre part perturbe les mesures car le temps de r\'eponse du d\'etecteur est 
trop faible. Il n'\'etait pas possible en g\'en\'eral de mesurer les sections 
efficaces en dessous de $30^o$ dans le syst\`eme du laboratoire.

Une autre difficult\'e, inh\'erente \`a la m\'ethode, est la mesure des 
sections efficaces, g\'en\'eralement tr\`es faibles, aux grands angles. Les 
configurations angulaires des d\'etecteurs pour ces mesures sont tr\`es proches 
de celles utilis\'ees pour les mesures aux petits angles. Lors d'une telle 
mesure, les \'ev\'enements correspondant \`a la diffusion aux petits angles sont 
collect\'es en m\^eme temps que les \'ev\'enements qui correspondent \`a la 
diffusion aux grands angles et dont le rendement peut \^etre 
inf\'erieur de plusieurs ordres de grandeur. Le nombre, rapidement prohibitif, 
d'\'ev\'enements \`a analyser entra\^{\i}ne des erreurs statistiques 
importantes sur les points aux grands angles et m\^eme l'impossibilit\'e de ces 
mesures lorsque le rapport des rendements aux angles avant et arri\`ere 
d\'epasse $10^4$.

Les erreurs sur les points des distributions angulaires tiennent compte des 
erreurs statistiques, des erreurs dues aux recouvrements des portions de 
distributions angulaires mesur\'ees pour des configurations diff\'erentes des 
d\'etecteurs et des erreurs dues \`a la normalisation absolue effectu\'ee aux 
petits angles, en supposant que la diffusion y est purement coulombienne.

En conclusion de cette rapide description de la m\'ethode exp\'erimentale nous 
indiquons les valeurs courantes des grandeurs caract\'eristiques  
exp\'erimentales obtenues : 
\begin{itemize}
\item[-] R\'esolution en \'energie : $\simeq 450 \ keV$
\item[-] R\'esolution angulaire : $\simeq 1^0$ dans le syst\`eme du 
laboratoire.
\item[-] Erreur sur la normalisation absolue : $\simeq 2 \% $
\end{itemize}
 
Toutes les donn\'ees exp\'erimentales utilis\'ees dans ce travail ont \'et\'e
obtenues par la m\'ethode expos\'ee pr\'ec\'edemment \`a l'exception des
donn\'ees de la diffusion de $^{16}O+^{208}Pb$ {\bf (VA75), (LI) et 
(RO)}

\vskip 20mm
\section{\large {\bf{ANALYSE SEMIPH\'ENOM\'ENOLOGIQUE :}}}

La premi\`ere \'etude concerne l'analyse de  diff\'erents syst\`emes \`a 
diff\'erentes \'energies incidentes au voisinage de la barri\`ere de Coulomb.
La hauteur de celle-ci est estim\'ee par la relation {\bf(BR81)} :
\be	E_B = {{Z_P Z_C e^2} \over r_B} (1 - {a \over r_B})	\ee
o\`u $a$ est une constante valant $0.63 fm$  et  $r_B$ est donn\'e 
par la relation {\bf(BR81)} :
 \be	r_B = [ 1.07 ( A_P^{1 \over 3} + A_C^{1 \over 3} ) + 2.72 ]fm	\ee
 
Dans les deux relations pr\'ec\'edentes, l'indice $P$ se r\'ef\`ere au noyau 
projectile tandis que l'indice $C$ concerne le noyau cible. Les lettres $Z$
 et $A$ 
symbolisent le num\'ero atomique et le nombre de masse. Enfin, 
$e$ est la charge \'el\'ementaire.

 La valeur de la hauteur de la barri\`ere de 
Coulomb pour tous les syst\`emes \'etudi\'es est report\'ee dans le tableau 
3.1 
avec les \'energies incidentes pour lesquelles une analyse a \'et\'e 
effectu\'ee dans ce travail. La m\'ethode employ\'ee est d\'ecrite dans la 
section 2.2.1. Le potentiel 
r\'eel de convolution a \'et\'e calcul\'e gr\^ace au code DFPOT 
{\bf(CO82)} tandis que la partie imaginaire a \'et\'e obtenue en assumant un 
facteur de forme radial du type Woods-Saxon dont les param\`etres ont \'et\'e 
d\'etermin\'es en ajustant les donn\'ees de diffusion \'elastique.

\begin{center}
\begin{tabular}{||c|c|c||} \hline \hline
Syst\`eme &   $E_B^{lab} (MeV) $ &$E_{lab}$ (MeV) \\  \hline \hline
$^{16}O+^{208}Pb$ & 81.3 &78.0, 83.0, 87.0 ,90.0, 104.0, 129.5, 192.0 \\ 
$^{31}P+^{27}Al$ & 60.1 & 79.5 \\ 
$^{32}S+^{24}Mg$ &64.9 & 65.0, 75.0, 86.3, 95.0, 110.0 \\ 
$^{32}S+^{28}Si$ & 68.3 & 90.0\\ 
$^{32}S+^{32}S$ & 71.9 & 70.0, 90.0, 97.1, 120.0, 160.0 \\ 
$^{32}S+^{36}S$ & 67.0 &  90.0 \\ 
$^{32}S+^{40}Ca$ & 78.8 &  90.0, 100.0, 110.0, 120.0, 151.5 \\ 
$^{35}Cl+^{24}Mg$ & 71.8 & 79.9, 85.0, 95.0, 110.0, 120.0 \\ 
$^{37}Cl+^{24}Mg$ & 73.8 &  87.9, 98.2, 113.7, 124.0 \\ 
\hline\hline
\end{tabular}
\vskip 3mm
Tableau 3.1 : Hauteurs de la Barri\`ere de Coulomb et 
\'energies incidentes
\end{center}

%\vskip 5mm

Le code ECIS {\bf(RA94)} a \'et\'e utilis\'e pour tous les calculs de mod\`ele 
optique d\'ecrits dans cette th\`ese.
Ainsi, le potentiel total utilis\'e pour les calculs de mod\`ele optique a 
\'et\'e : 
\be	V_T(R) = N(E) V_{M3Y}(R) + i W_{WS}(R) + V_C(R)	\ee
o\`u $V_C(R)$ est le potentiel de Coulomb et $N(E)$ est un facteur de 
renormalisation, d\'ependant de l'\'energie, qui est introduit afin de 
d\'ecrire la partie r\'eelle du terme de polarisation. 
Cette d\'ependance \'energ\'etique, 
connue sous le nom de l'anomalie de seuil, a \'et\'e observ\'ee dans la 
diffusion de diff\'erents syst\`emes {\bf(BA84)}, {\bf(LI85)}, {\bf(DI89)} et 
{\bf(TH85)} et il est, aujourd'hui, parfaitement expliqu\'e {\bf(NA85)}, 
{\bf(MA86)} et {\bf(PA91)}. Il provient des couplages des canaux non 
\'elastiques a la voie \'elastique. La d\'ependance \'energ\'etique du 
potentiel r\'eel est en relation avec l'intensit\'e de ces couplages. Elle 
est directement li\'ee, par une relation de dispersion (eq. 2.15), au taux de 
variation de l'absorption en fonction de l'\'energie, d'une mani\`ere qui 
d\'epend de la structure des deux noyaux en collision. 

Le facteur de renormalisation 
a \'et\'e obtenu en ajustant les donn\'ees de diffusion \'elastique 
et sa pr\'esence change, non seulement la profondeur du potentiel r\'eel, mais 
\'egalement sa pente. Par cons\'equent, la forme de la contribution de 
polarisation n'est pas la m\^eme que celle du potentiel nu. 

Les densit\'es nucl\'eaires utilis\'ees dans les calculs du potentiel par 
convolution M3Y ont \'et\'e obtenues \`a partir des distributions de charge 
des nucl\'eons {\bf(BOR75)} et {\bf(JA74)}, de la mani\`ere standard 
{\bf(SA79)}.
Les distributions de charge nucl\'eaires ont \'et\'e obtenues \`a partir 
d'exp\'eriences de diffusion 
d'\'electrons, except\'e pour le $^{208}Pb$ dont les densit\'es ont \'et\'e 
obtenues par une m\'ethode variationnelle semi-classique et pour le $^{36}S$
 pour lequel il n'existe
pas de donn\'ees de diffusion d'\'electrons. Pour les noyaux 
inclus dans le tableau 3.2 les densit\'es ont \'et\'e param\'etr\'ees par 
une forme parabolique de Fermi \`a quatre param\`etres.
\be \rho_c(r) = \rho_0 [1 + \omega(r^2 / c^2)][1 + exp((r-c)/ z)^{-n}]\ee

\begin{center}
\begin{tabular}{||c|c|c|c|c|c|c||} \hline \hline
Noyau & c(fm)  & z(fm) & $\omega $ & n  & Note &Ref.  \\ \hline \hline
${}^{16}O  $ & 2.608 & 0.513 & -0.051 & 1.00 & & ${\bf VR87}$ \\
${}^{24}Mg  $ &3.192 & 0.604  & -0.249 & 1.00& & $  {\bf LI74}$\\
${}^{27}Al $ & 3.079 & 0.519 &   0.0 & 1.00 & & ${\bf VR87}$  \\
${}^{28}Si$ & 3.239 & 0.574 & -0.149 & 1.00 & & ${\bf BR77}$  \\
${}^{31}P$  & 3.369 & 0.582 & -0.173 & 1.00 & & ${\bf VR87}$  \\
${}^{32}S$  & 3.441 & 0.624 & -0.213 & 1.00 & & ${\bf LI74}$  \\ 
${}^{35}Cl$ & 3.490 & 0.602 & -0.120 & 1.00 & & ${\bf BR80}$  \\ 
${}^{37}Cl $ & 3.554 & 0.588 & -0.130 & 1.00 & & ${\bf VR87}$  \\ 
${}^{40}Ca$ & 3.676 & 0.585 & -0.102 & 1.00 & & ${\bf FR68}$  \\
${}^{208}Pb$ & 7.194 & 0.658 &   0.0 & 1.56 & neutron & ${\bf BR85}$ \\
${}^{208}Pb$ & 6.975 & 0.635 &   0.0 & 1.42 & proton & ${\bf BR85}$ \\
\hline\hline 
\end{tabular}
\vskip 3mm
Tableau 3.2 : Param\`etres des densit\'es de charge exp\'erimentales.
\end{center}
%\begin{center}
%\end{center}
%\par

%\vskip 5mm

Dans le tableau 3.3 sont donn\'es les coefficients d'une param\'etrisation 
du type Bessel-Fourier utilis\'ee dans le cas du $^{36}S$ 
\ba	 \rho_c(r) = \left\{ \begin{array}{ll}
{\ds \sum_{n=1}^{15}a_nj_0\left({{n\pi r}\over R}\right)} & \mbox{si $r 
\leq 8.0 fermi $} \\ {\ds 0} & \mbox{si $r > 8.0 fermi$} \end{array} \right. \ea
$j_0$ repr\'esente la fonction de Bessel d'ordre z\'ero. 

\begin{center}
\begin{tabular}{||c|c|c|c|c|c||}  \hline \hline
$a_1$&0.37032$10^{-1}$&$a_6$&0.61882$10^{-2}$&$a_{11}$&-0.84063$10^{-4}$\\ 
$a_2$&0.57939$10^{-1}$&$a_7$&0.37795$10^{-2}$&$a_{12}$&0.34101$10^{-4}$\\ 
$a_3$&0.10049$10^{-1}$&$a_8$&-0.55272$10^{-3}$&$a_{13}$&-0.11663$10^{-4}$\\ 
$a_4$&-0.19852$10^{-1}$&$a_9$&-0.12904$10^{-3}$&$a_{14}$&0.35204$10^{-5}$\\ 
$a_5$&-0.67176$10^{-2}$&$a_{10}$&0.15845$10^{-3}$&$a_{15}$&-0.95135$10^{-6}$\\ 
\hline\hline
\end{tabular}
\vskip 3mm
Tableau 3.3 : Les coefficients de Bessel-Fourier de la densit\'e de charge 
du $^{36}S$ {\bf(VR87)}. 
\end{center}

\subsection{Distributions angulaires et potentiels}

Aux \'energies proches de la barri\`ere de Coulomb les distributions angulaires
pr\'esentent tr\`es peu de sensibilit\'e aux param\`etres du potentiel
nucl\'eaire. Il est donc utile de  diminuer le nombre de param\`etres. Pour 
cel\`a, 
la profondeur de la partie imaginaire du potentiel  a \'et\'e fix\'ee. Une 
valeur de 60 MeV a \'et\'e choisie car cette valeur est proche de la profondeur 
du potentiel imaginaire calcul\'e avec des mod\`eles microscopiques {\bf(VI86)},
 {\bf(VI87)}, {\bf(VI90)}, {\bf(FE90)} et {\bf(BA92)}. Comme nous le 
v\'erifierons plus tard, la valeur de ce param\`etre n'est importante ni pour 
le calcul de la valeur de l'absorption totale ni pour la d\'etermination de la 
r\'egion du potentiel o\`u a lieu l'absorption.

Dans l'analyse de tous les syst\`emes pour lesquels nous disposons de donn\'ees 
de diffusion \'elastique dans un large domaine d'\'energies, incluyant la 
diffusion sous-coulombienne, nous avons utilis\'e la m\^eme syst\'ematique que 
nous d\'etaillons, dans ce qui suit, pour le syst\`eme ${}^{32}S+{}^{24}Mg$. 
Nous avons 
commenc\'e notre analyse en ajustant les donn\'ees de diffusion \'elastique \`a 
83.6, 95.0 et 110.0 MeV. \'Etant donn\'e que ces \'energies sont suffisamment 
loin de la barri\`ere, il y a une grande sensibilit\'e aux valeurs des 
param\`etres de potentiel optique; par cons\'equent, nous avons recherch\'e les 
valeurs du rayon et de la diffusivit\'e du potentiel imaginaire, ainsi que le 
facteur de renormalisation du potentiel r\'eel qui reproduisent au mieux les 
donn\'ees exp\'erimentales \`a 110.0 MeV. Ces param\`etres ont \'et\'e 
utilis\'es comme valeurs de d\'epart pour l'ajustement correspondant \`a 95.0 
MeV, et ainsi de suite. Les param\`etres qui reproduisent au mieux les 
exp\'eriences sont inclus dans le tableau 3.4.

%%%%%%%%%%%%%%%%%%%%%%%%%%%%%%%%%%%%%%%%%%%%%%%%%%%%%%%%%%%%%%%%%%%%%%
% Tableau S-Mg
\begin{center}
\begin{tabular}{||c|c|c|c|c|c|c|c|c|c|c||} \hline \hline
$E_{lab}$&$N_R$&$W_0$&$R_W$&$a_W$&$<l>$&$<l^2>$&$\chi^2/n$&$\sigma_R$&$\sigma_{2^+}^C$&Note \\
 (MeV) &  & (MeV) & (fm) & (fm) &  & &  &(mb)& (mb)& \\ \hline \hline
110.0&1.27&60&6.75&0.563&22.50&581.0&3.6&1163& & \\
95.0&1.60&60&7.40&0.460&18.70&405.0&2.3&948& & \\
86.3&1.59&60&7.60&0.408&15.80&288.0&0.9&738& & \\
75.0&1.72&60& 7.95&0.370&11.80&165.0&3.4&445& & \\
&1.52&60& 7.95&0.330&- &- &2.2&546&169&a,b\\
65.0&1.91&60& 8.10&0.223&5.94&45.5&5.5&54.7& &a\\
&1.72&60& 8.10&0.093&- &- &1.2&155&121&a,b\\
\hline\hline
\end{tabular}
\end{center}
\vskip 0.1mm
a)$R_W$ fix\'e.
\vskip 0.1mm
\noindent
b) Couplage coulombien de l'\'etat $^{24}Mg(2^+;1.37 MeV)$ \`a la voie \'elastique.
\vskip 3mm
\begin{center}
Tableau 3.4 : Meilleurs param\`etres du mod\`ele optique pour \smg.
\end{center}
%%%%%%%%%%%%%%%%%%%%%%%%%%%%%%%%%%%%%%%%%%%%%%%%%%%%%%%%%%%%%%%%%%%%%%%%%%%%
\newpage

%&&&&&&&&&&&&&&&&&&&&&&&&&&&&&&&&&&&&&&&&&&&&&&&&&&&&&&&&&&&&&&&&&&&&&&&&&&&&&
\begin{figure}[ht]
\begin{center}\mbox{\epsfig{file=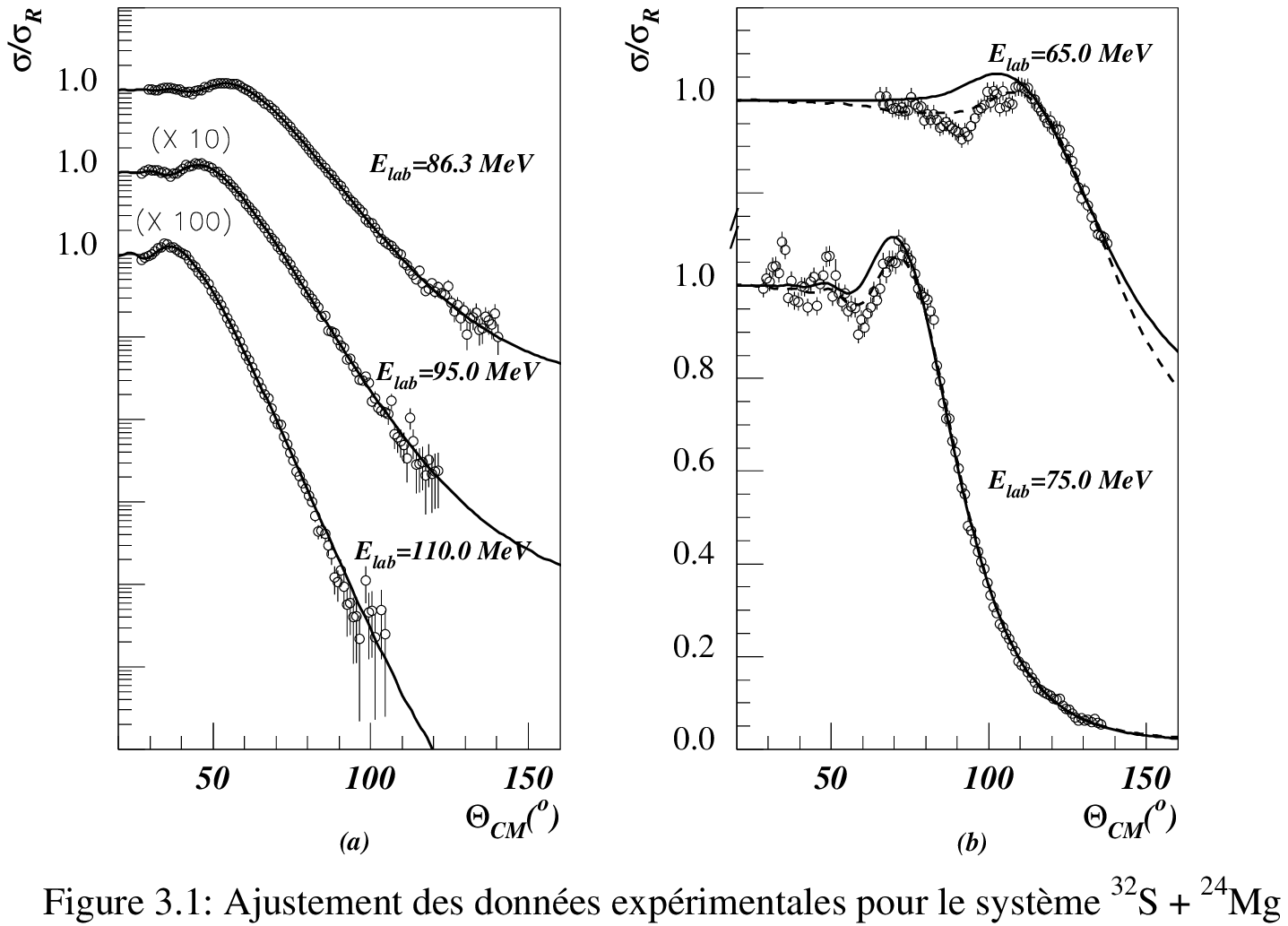}}
%,bbllx=60,bblly=20,bburx=500,bbury=280}}
\end{center}
%\caption{Comparaison des donn\'ees exp\'erimentales aux calculs
%semi-ph\'enom\'enologiques pour le syst\`eme \smg}
%\label{FIG. 3.1}
\end{figure}
%&&&&&&&&&&&&&&&&&&&&&&&&&&&&&&&&&&&&&&&&&&&&&&&&&&&&&&&&&&&&&&&&&&&&&&&&&&&&&
La figure 3.1.a montre les excellents ajustements aux donn\'ees 
exp\'erimentales obtenus aux \'energies les plus hautes, tandis que dans la 
figure 3.1.b sont indiqu\'es les ajustements correspondants aux \'energies les 
plus basses. A 75.0 MeV, nous avons suivi la m\^eme proc\'edure, partant des 
param\`etres optimum obtenus a 86.3 MeV, mais \`a 65.0 MeV la situation est un 
peu plus complexe. En effet, la hauteur de la barri\`ere de Coulomb pour le 
syst\`eme \smg, dans le laboratoire, est approximativement de 64.9 MeV ; 
\'etant donn\'e que l'\'etat excit\'e le plus accessible \'energ\'etiquement 
est le $2^+$ du $^{24}Mg$ situ\'e \`a 1.37 MeV, la hauteur de la barri\`ere 
correspondant \`a cet \'etat est sup\'erieure \`a l'\'energie de la collision, 
tout au moins dans un certain intervalle radial. Ceci justifie la faible 
sensibilit\'e observ\'ee par rapport aux valeurs des param\`etres du potentiel 
optique. Afin de r\'esoudre le probl\`eme pos\'e par ce manque de 
sensibilit\'e, il est necessaire de diminuer le nombre de param\`etres \`a 
ajuster. Pour cela nous avons \'etudi\'e l'\'evolution, avec l'\'energie, du 
rayon du terme d'absorption obtenue lors de l'ajustement des donn\'ees de 
diffusion \'elastique aux quatres \'energies les plus hautes. Ces ajustements 
ont \'et\'e obtenus utilisant diff\'erentes profondeurs du potentiel imaginaire 
dont les valeurs ($W_0$=20, 30, 40, 50, 60 y 100 MeV) \'etaient maintenues 
constantes dans chaque cas.

%\newpage
%&&&&&&&&&&&&&&&&&&&&&&&&&&&&&&&&&&&&&&&&&&&&&&&&&&&&&&&&&&&&&&&&&&&&&&&&&&&&&
\begin{figure}[ht]
\begin{center}\mbox{\epsfig{file=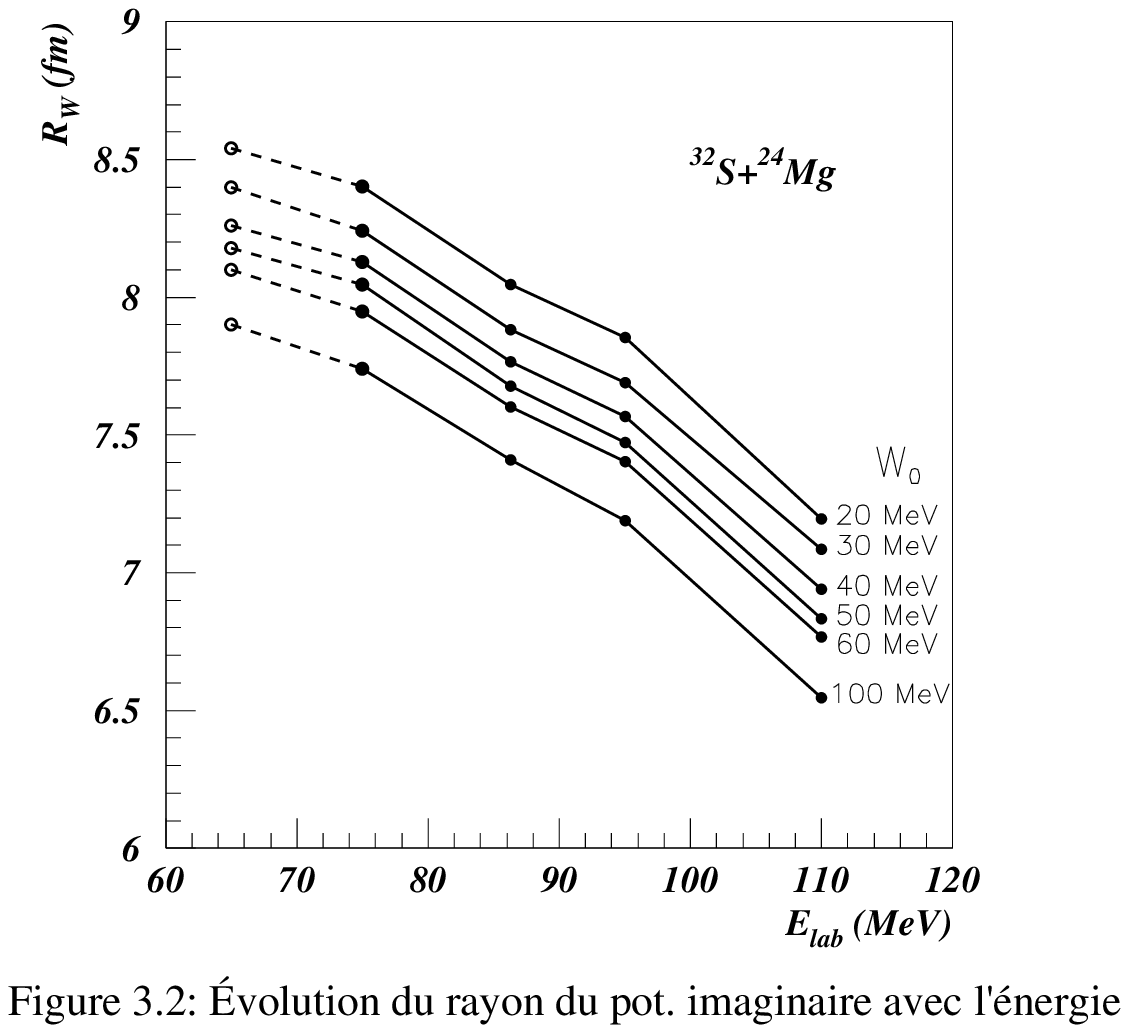}}\end{center}
%,bbllx=60,bblly=20,bburx=500,
%bbury=280}}\end{center}
%\caption{\'Evolution du rayon du potentiel imaginaire avec l'\'energie pour le 
%syst\`eme \smg}
%\label{FIG. 3.2}
\end{figure}
%&&&&&&&&&&&&&&&&&&&&&&&&&&&&&&&&&&&&&&&&&&&&&&&&&&&&&&&&&&&&&&&&&&&&&&&&&&&&&

% \vskip 150mm
Dans la figure 3.2, nous montrons l'\'evolution avec l'\'energie du rayon du 
potentiel imaginaire, $R_W$, correspondant aux meilleurs ajustements aux 
donn\'ees. Suivant cette tendance, nous avons d\'etermin\'e, \`a 65.0 MeV les 
valeurs de $R_W$ qui correspondraient \`a chacune des profondeurs $W_0$ du 
potentiel imaginaire. Ces valeurs du rayon ont \'et\'e maintenues fixes dans 
les diff\'erents calculs du mod\`ele optique.

Il est bien connu que l'absorption due aux effets de polarisation coulombiens 
augmente avec l'\'energie de la collision; cependant, son importance par 
rapport \`a l'absorption due au potentiel nucl\'eaire, diminue tr\`es 
rapidement lorsque l'\'energie de l'interaction cro\^{\i}t. Ceci implique que 
les effets de polarisation coulombienne doivent \^etre pris en consid\'eration 
aux \'energies proches et surtout inf\'erieures \`a celle correspondant \`a la 
barri\`ere de Coulomb; c'est \`a dire: lorsque les collisions p\'eriph\'eriques 
dominent. L'effet de la polarisation de Coulomb se manifeste par l'existence de 
plus d'absorption aux angles de diffusion petits qui correspondent \`a des 
collisions tr\`es p\'eriph\'eriques au cours desquelles les noyaux se 
maintiennent \`a des distances assez importantes pour que les forces 
nucl\'eaires, de courte port\'ee, n'agissent pas.

% \vskip 150mm

Pour cette raison, dans la recherche des param\`etres optiques que nous avons 
laiss\'es libres \`a 65.0 MeV, N(E) et $a_W$, nous n'ajustons que les donn\'ees 
de diffusion \'elastique situ\'ees au del\`a de la r\'egion de l'arc-en-ciel; 
c'est \`a dire celles correspondant \`a l'intervalle angulaire $107.5^0\leq 
\theta_{cm}\leq 137.5^0$, o\`u l'absorption due \`a la polarisation se doit 
d'\^etre minimum. Il n'y aurait aucun sens physique \`a ajuster la partie de la 
distribution angulaire de la diffusion \'elastique qui correspond aux 
collisions p\'eriph\'eriques avec un potentiel nucl\'eaire de courte port\'ee, 
\'etant donn\'e que cette r\'egion est domin\'ee par les effets de polarisation 
coulombienne. Dans la figure 3.1.b, nous repr\'esentons en trait continu, le 
r\'esultat des calculs de diffusion \'elastique obtenus avec les param\`etres 
optiques qui ajustent au mieux les donn\'ees dans l'intervalle angulaire 
pr\'ecedemment mentionn\'e. Les valeurs des param\`etres et de la section 
efficace de r\'eaction se trouvent dans le tableau 3.4.

Dans le but de reproduire de mani\`ere ad\'equate la r\'egion angulaire de la 
distribution \'elastique domin\'ee par les effets de polarisation coulombienne, 
nous avons effectu\'e un calcul simple de canaux coupl\'es dans lequel nous 
n'avons pris en consid\'eration que le couplage du premier \'etat excit\'e du 
$^{24}Mg$($2^+$;1.37 MeV) avec le canal \'elastique et en ne prenant comme 
potentiel de couplage que le potentiel de polarisation de Coulomb.
        
Pour ces ajustements, nous avons recherch\'e la nouvelle valeur du facteur de 
renormalisation du potentiel r\'eel car le fait d'inclure explicitement une 
voie a pour effet d'att\'enuer le potentiel central et, ainsi, le facteur de 
renormalisation doit diminuer. De m\^eme, nous avons ajust\'e la diffusivit\'e 
du terme d'absorption qui, comme pr\'evu, diminue. En effet, la majeure partie 
de l'absorption due \`a la polarisation de Coulomb se produit dans l'intervalle 
angulaire de la diffusion \'elastique qui correspond aux collisions 
p\'eriph\'eriques. Cependant, nous ne pouvons exclure le fait que son effet, 
bien que plus faible, se r\'epercute dans la r\'egion angulaire que nous avons 
ajust\'ee pour d\'eterminer les param\`etres du potentiel optique. Ainsi, en 
incluant explicitement dans les calculs en canaux coupl\'es le potentiel de 
polarisation de Coulomb, de port\'ee sup\'erieure \`a celle du potentiel 
nucl\'eaire, nous \'evitons de forcer ce dernier \`a reproduire ce qui n'est 
pas d'origine nucl\'eaire.

Dans la figure 3.1.b nous repr\'esentons par une ligne pointill\'ee le 
r\'esultat des ajustements obtenus dans les calculs en canaux coupl\'es \`a 65.0 
et \`a 75.0 MeV. Dans le tableau 3.4 nous donnons les valeurs des param\`etres 
ajust\'es et des sections efficaces obtenues dans les calculs en canaux 
coupl\'es. Nous pouvons facilement v\'erifier que bien qu'\`a 75.0 MeV la 
section efficace de r\'eaction correspondant \`a l'\'etat excit\'e du 
$^{24}Mg$($2^+$;1.37 MeV) coupl\'e \`a la voie \'elastique est, en valeur 
absolue, sup\'erieure \`a celle obtenue dans le calcul effectu\'e \`a 65.0 MeV, 
son importance relative par rapport \`a la section efficace totale diminue. En 
effet, \`a 75.0 MeV elle repr\'esente 30\% de la section efficace totale 
de r\'eaction, tandis qu'\`a 65.0 MeV elle repr\'esente 80\%. Une proc\'edure 
\'equivalente a \'et\'e utilis\'ee pour l'ajustement des donn\'ees de 
diffusion \'elastique correspondant aux syst\`emes $^{32}S+^{32}S$ 
et \opb pour lesquels nous disposons de donn\'ees \`a des \'energies 
inf\'erieures \`a la hauteur de la barri\`ere de Coulomb. Les param\`etres 
optiques obtenus en ajustant les distributions angulaires exp\'erimentales et 
les valeurs calcul\'ees de la section efficace totale de r\'eaction sont 
indiqu\'es dans les tableaux 3.5 et 3.6. Dans les figures 3.3.a-b et 
3.4.a-b, nous pr\'esentons les ajustements des donn\'ees pour les deux 
syst\`emes.
%&&&&&&&&&&&&&&&&&&&&&&&&&&&&&&&&&&&&&&&&&&&&&&&&&&&&&&&&&&&&&&&&&&&&&&&&&&&&&
\begin{figure}[ht]
\vspace{-1cm}
\begin{center}\mbox{\epsfig{file=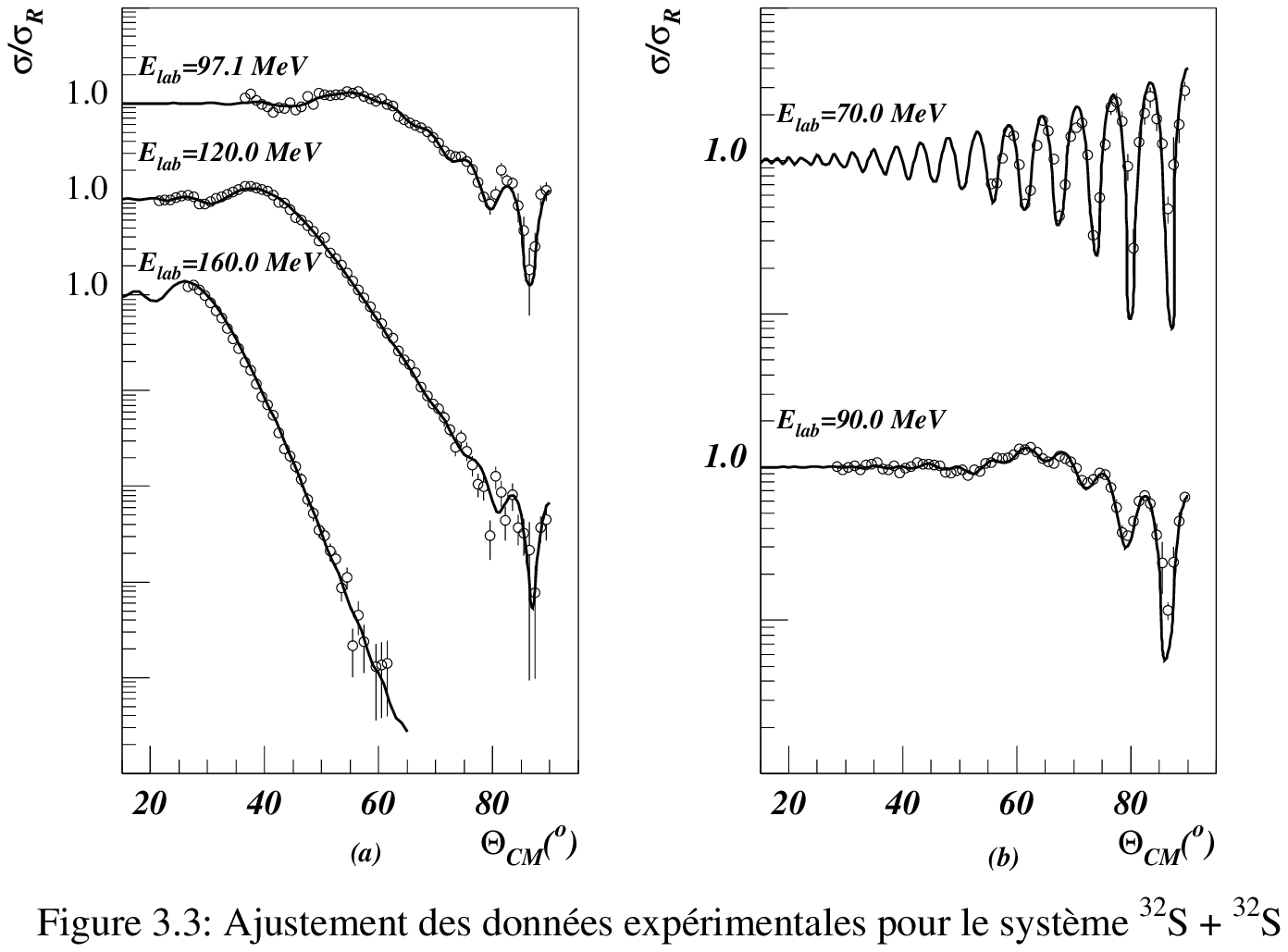}}\end{center}
\end{figure}
%&&&&&&&&&&&&&&&&&&&&&&&&&&&&&&&&&&&&&&&&&&&&&&&&&&&&&&&&&&&&&&&&&&&&&&&&&&&&&
%%%%%%%%%%%%%%%%%%%%%%%%%%%%%%%%%%%%%%%%%%%%%%%%%%%%%%%%%%%%%%%%%%%%%%
% Tableau S32-S32
\begin{center}
\begin{tabular}{||c|c|c|c|c|c|c|c|c||} \hline \hline
$E_{lab}$&$N_R$&$W_0$&$R_W$&$a_W$&$<l>$&$<l^2>$&$\chi^2/n$&$\sigma_R$ \\
 (MeV) &  & (MeV) & (fm) & (fm) &  & & & (mb) \\ \hline \hline
160.0 & 1.160 & 60 & 7.48 & 0.486 &37.1 & 1562.2&2.50 & 1596.0\\
120.0 & 1.453 & 60 & 7.89 & 0.430 &27.8 &882.8 &1.56 & 1240.0\\
97.09 & 1.621 & 60 & 8.31 & 0.310 &20.3 &473.1 &1.54 & 800.0\\
90.0  & 1.672 & 60 & 8.42 & 0.305 &17.4 &349.3 &2.90 & 631.0\\
70.0  & 1.780 & 60 & 8.80 & 0.210 &7.0 &64.2 &4.30 & 24.0\\
\hline\hline
\end{tabular}
\vskip 3mm
Tableau 3.5 : Meilleurs param\`etres du mod\`ele optique pour $^{32}S+^{32}S$.
\end{center}
%&&&&&&&&&&&&&&&&&&&&&&&&&&&&&&&&&&&&&&&&&&&&&&&&&&&&&&&&&&&&&&&&&&&&&&&&&&&&&

%&&&&&&&&&&&&&&&&&&&&&&&&&&&&&&&&&&&&&&&&&&&&&&&&&&&&&&&&&&&&&&&&&&&&&&&&&&&&&
\begin{figure}[ht]
\begin{center}\mbox{\epsfig{file=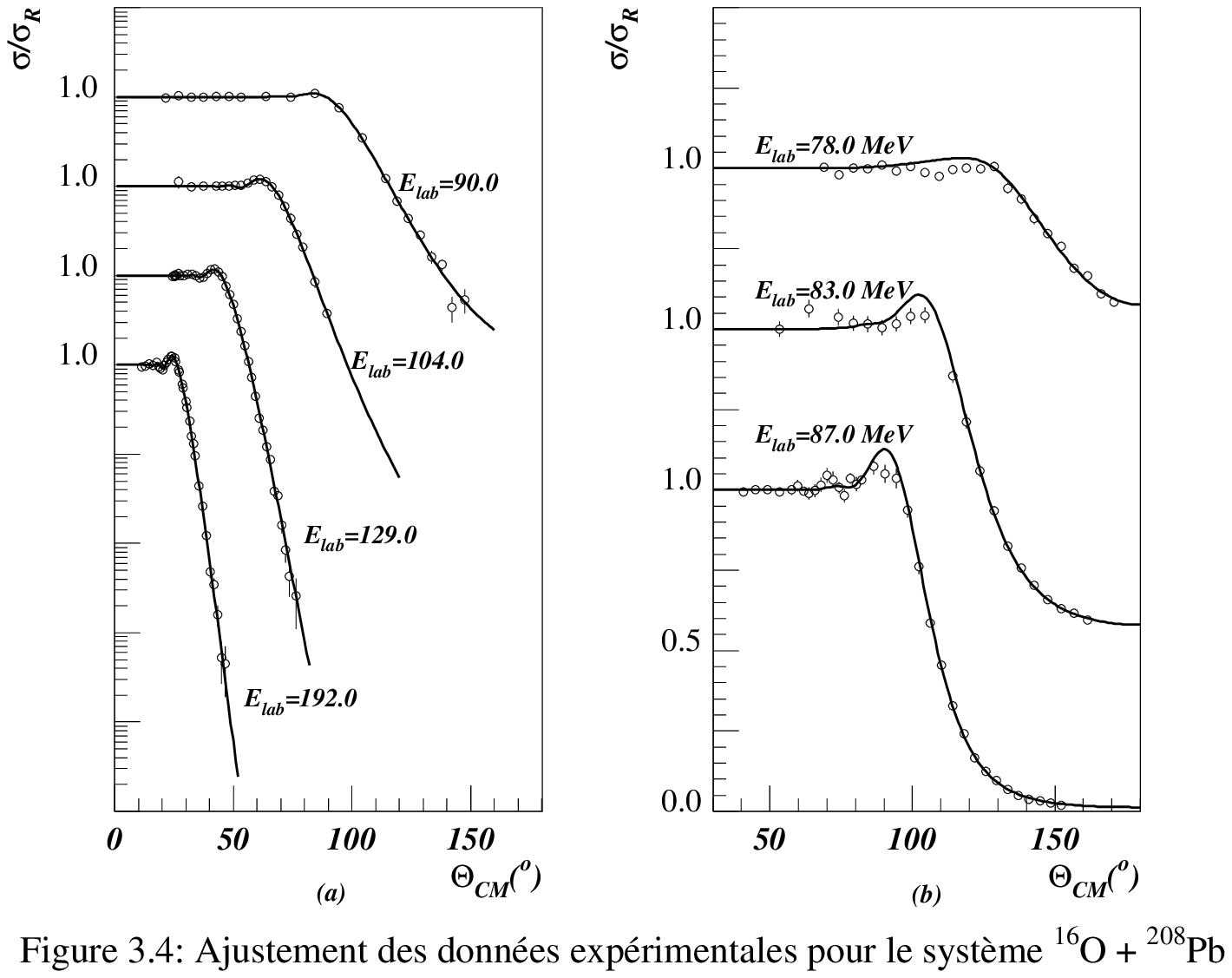}}
%,bbllx=60,bblly=20,bburx=500,
%bbury=280}}
\end{center}
%\caption{Comparaison des donn\'ees exp\'erimentales aux calculs
%semi-ph\'enom\'enologiques pour le syst\`eme \opb}
%\label{FIG. 3.4}
\end{figure}
%&&&&&&&&&&&&&&&&&&&&&&&&&&&&&&&&&&&&&&&&&&&&&&&&&&&&&&&&&&&&&&&&&&&&&&&&&&&&&
%\vspace{2cm}
%\vskip 15mm

%%%%%%%%%%%%%%%%%%%%%%%%%%%%%%%%%%%%%%%%%%%%%%%%%%%%%%%%%%%%%%%%%%%%%%
% tableau O Pb
\begin{center}
\begin{tabular}{||c|c|c|c|c|c|c|c|c||} \hline \hline
$E_{lab}$&$N_R$&$W_0$&$R_W$&$a_W$&$<l>$&$<l^2>$&$\chi^2/n$&$\sigma_R$ \\
 (MeV) &  & (MeV) & (fm) & (fm) & &  & &(mb) \\ \hline \hline
192.0 &0.988 &60 &10.07 &0.607 &72.4 &5945  &1.07 &2905.2\\
129.5 &1.150 &60 &10.19 &0.618 &50.9 &2957.5  &0.49 &2095.2\\
104.0 &1.062 &60 &10.40 &0.530 &35.4 &1442.0  &4.00 &1267.0\\
90.0  &1.272 &60 &10.67 &0.439 &23.8 &663.7  &0.66 &637.2 \\
87.0  &1.607 &60 &11.10 &0.324 &20.2 &475.0  &1.08 &474.3 \\
83.0  &1.801 &60 &11.30 &0.250 &15.2 &176.0  &0.39 &240.3 \\
78.0  &1.901 &60 &11.50 &0.176 &9.9 & 125.0 &0.56 &38.53 \\
\hline\hline
\end{tabular}
\vskip 3mm
Tableau 3.6 : Meilleurs param\`etres du mod\`ele optique pour \opb.
\end{center}
%\begin{center}
%\end{center}
%&&&&&&&&&&&&&&&&&&&&&&&&&&&&&&&&&&&&&&&&&&&&&&&&&&&&&&&&&&&&&&&&&&&&&&&&&&&&&

Les tableaux 3.7, 3.8 et 3.9 et les figures 3.5.a-b, 3.6.a-b et 3.7.a-b 
repr\'esentent, respectivement, les r\'esultats obtenus pour les syst\`emes 
\sca, $^{35}Cl+^{24}Mg$ et $^{37}Cl+^{24}Mg$.

\newpage

%&&&&&&&&&&&&&&&&&&&&&&&&&&&&&&&&&&&&&&&&&&&&&&&&&&&&&&&&&&&&&&&&&&&&&&&&&&&&&
\begin{figure}[ht]
\begin{center}\mbox{\epsfig{file=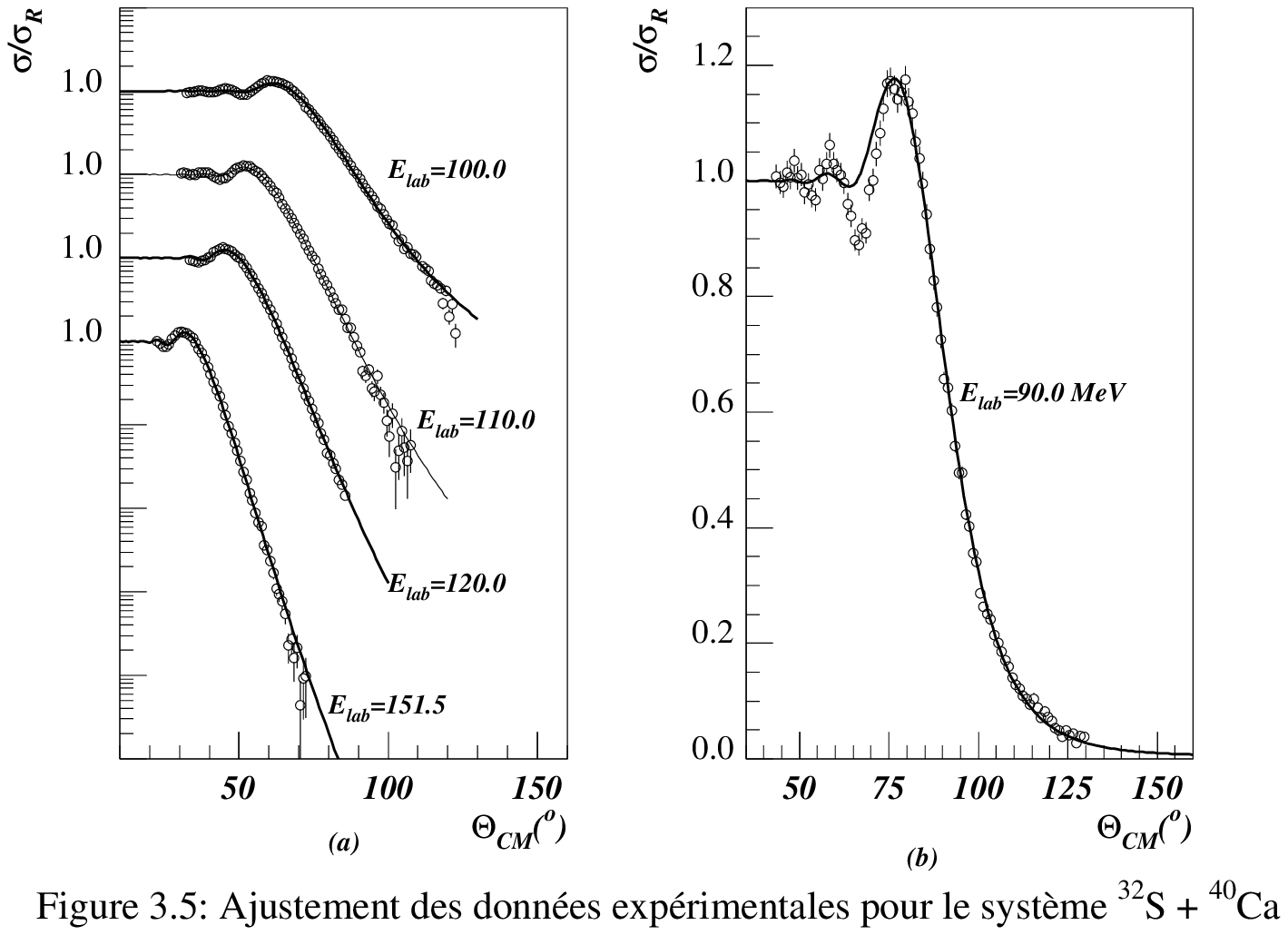}}\end{center}
%,bbllx=60,bblly=20,bburx=500,
%bbury=280}}\end{center}
%\caption{Comparaison des donn\'ees exp\'erimentales aux calculs
%semi-ph\'enom\'enologiques pour le syst\`eme \sca}
%\label{FIG. 3.5}
\end{figure}
%&&&&&&&&&&&&&&&&&&&&&&&&&&&&&&&&&&&&&&&&&&&&&&&&&&&&&&&&&&&&&&&&&&&&&&&&&&&&&
%\vspace{2cm}
%%%%%%%%%%%%%%%%%%%%%%%%%%%%%%%%%%%%%%%%%%%%%%%%%%%%%%%%%%%%%%%%%%%%%%
% Tableau S-Ca
\begin{center}
\begin{tabular}{||c|c|c|c|c|c|c|c|c||} \hline \hline
$E_{lab}$&$N_R$&$W_0$&$R_W$&$a_W$&$<l>$&$<l^2>$&$\chi^2/n$&$\sigma_R$ \\
 (MeV) &  & (MeV) & (fm) & (fm) & & & & (mb) \\ \hline \hline
151.5 & 0.979 & 60 & 7.98 & 0.498 &39.8 &1806.6  &3.7 & 1567.5\\
120.0 & 1.074 & 60 & 8.05 & 0.479 &30.3 & 1052.3 &3.02 & 1153.1\\
110.0 & 1.195 & 60 & 8.24 & 0.455 &26.9 &829.0  &3.09 & 994.7\\
100.0 & 1.435 & 60 & 8.61 & 0.361 &21.8 &548.4  &3.59 & 745.7\\
90.0 & 1.419 & 60 & 8.54 & 0.378 &16.3 & 314.3 &3.59 & 473.0\\
\hline\hline
\end{tabular}
\vskip 3mm
Tableau 3.7 : Meilleurs param\`etres du mod\`ele optique pour \sca.
\end{center}
%&&&&&&&&&&&&&&&&&&&&&&&&&&&&&&&&&&&&&&&&&&&&&&&&&&&&&&&&&&&&&&&&&&&&&&&&&&&&&

Pour ces syst\`emes nous disposons de distributions angulaires mesur\'ees \`a 
diff\'erentes \'energies, toujours sup\'erieures \`a la barri\`ere. 
Pour les analyser, nous avons adopt\'e la m\^eme syst\'ematique que pour les 
diffusions \'elastiques ant\'erieures correspondant aux \'energies les plus 
hautes. Pour chaque syst\`eme nous avons calcul\'e les valeurs de $<l>$, $<l^2>$ et 
$\sigma_R$ qui nous servirons de comparaison avec les valeurs obtenues dans des 
calculs similaires utilisant des potentiels de Feshbach. 
\newpage
%&&&&&&&&&&&&&&&&&&&&&&&&&&&&&&&&&&&&&&&&&&&&&&&&&&&&&&&&&&&&&&&&&&&&&&&&&&&&&
\begin{figure}[ht]
\begin{center}\mbox{\epsfig{file=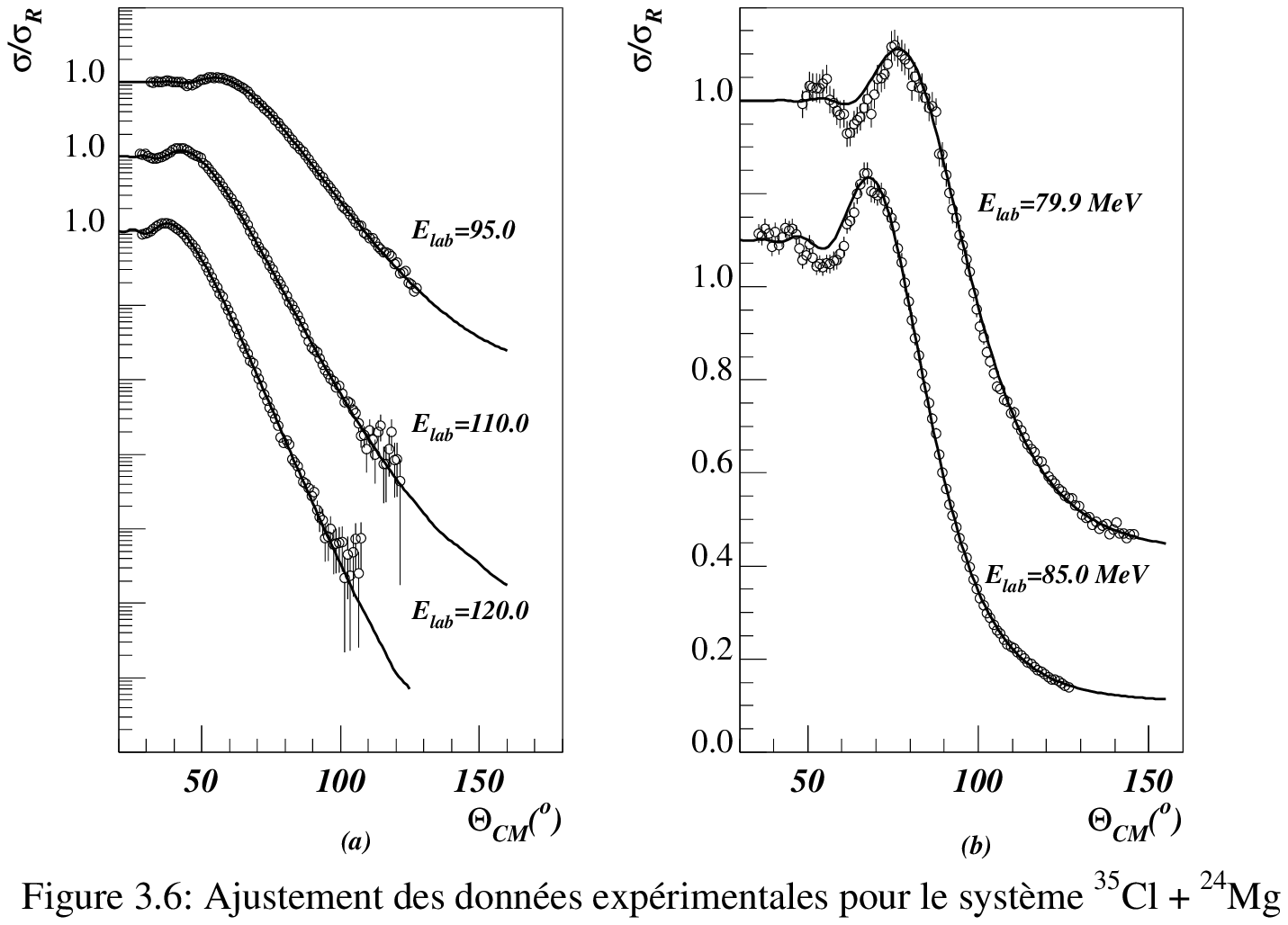}}\end{center}
%,bbllx=60,bblly=20,bburx=500,
%bbury=280}}\end{center}
%\caption{Comparaison des donn\'ees exp\'erimentales aux calculs
%semi-ph\'enom\'enologiques pour le syst\`eme  $^{35}Cl+^{24}Mg$}
%\label{FIG. 3.6}
\end{figure}
%&&&&&&&&&&&&&&&&&&&&&&&&&&&&&&&&&&&&&&&&&&&&&&&&&&&&&&&&&&&&&&&&&&&&&&&&&&&&&
%\vspace{2cm}
%%%%%%%%%%%%%%%%%%%%%%%%%%%%%%%%%%%%%%%%%%%%%%%%%%%%%%%%%%%%%%%%%%%%%%
% Tableau Cl35-Mg
\begin{center}
\begin{tabular}{||c|c|c|c|c|c|c|c|c||} \hline \hline
$E_{lab}$&$N_R$&$W_0$&$R_W$&$a_W$&$<l>$&$<l^2>$&$\chi^2/n$&$\sigma_R$ \\
 (MeV) &  & (MeV) & (fm) & (fm) & &  & & (mb) \\ \hline \hline
120.0 & 0.985 & 60 & 6.88 & 0.574 &23.5 &634.4  &1.26 & 1171.1\\
110.0 & 1.014 & 60 & 7.14 & 0.552 &21.2 & 533.2 &2.02 & 1065.0\\
95.0  & 1.032 & 60 & 7.32 & 0.523 &17.2 & 346.4 &1.59 & 778.7\\
85.0  & 1.252 & 60 & 7.90 & 0.386 &13.1 &202.3  &1.60 & 489.0\\
79.9  & 1.451 & 60 & 8.10 & 0.365 &11.4 & 150.7 &2.53 & 361.9\\
\hline\hline
\end{tabular}
\vskip 5mm
Tableau 3.8 : Meilleurs param\`etres du mod\`ele optique pour 
$^{35}Cl+^{24}Mg$.
\end{center}
%&&&&&&&&&&&&&&&&&&&&&&&&&&&&&&&&&&&&&&&&&&&&&&&&&&&&&&&&&&&&&&&&&&&&&&&&&&&&&

 Pour les syst\`emes \pal, \ssi et $^{32}S+^{36}S$ nous ne disposons de 
distributions angulaires de diffusion \'elastique qu'\`a une seule \'energie 
pour chaque syst\`eme. Les donn\'ees ont \'et\'e ajust\'ees en partant des 
valeurs des param\`etres optiques obtenues dans l'analyse de syst\`emes de 
masses voisines et \`a l'\'energie la plus proche possible. Les param\`etres 
optiques qui ajustent au mieux les donn\'ees sont inclus dans le tableau 3.10, 
tandis que dans la figure 3.8, nous pr\'esentons les ajustements respectifs.

\newpage

%&&&&&&&&&&&&&&&&&&&&&&&&&&&&&&&&&&&&&&&&&&&&&&&&&&&&&&&&&&&&&&&&&&&&&&&&&&&&&
\begin{figure}[ht]
\begin{center}\mbox{\epsfig{file=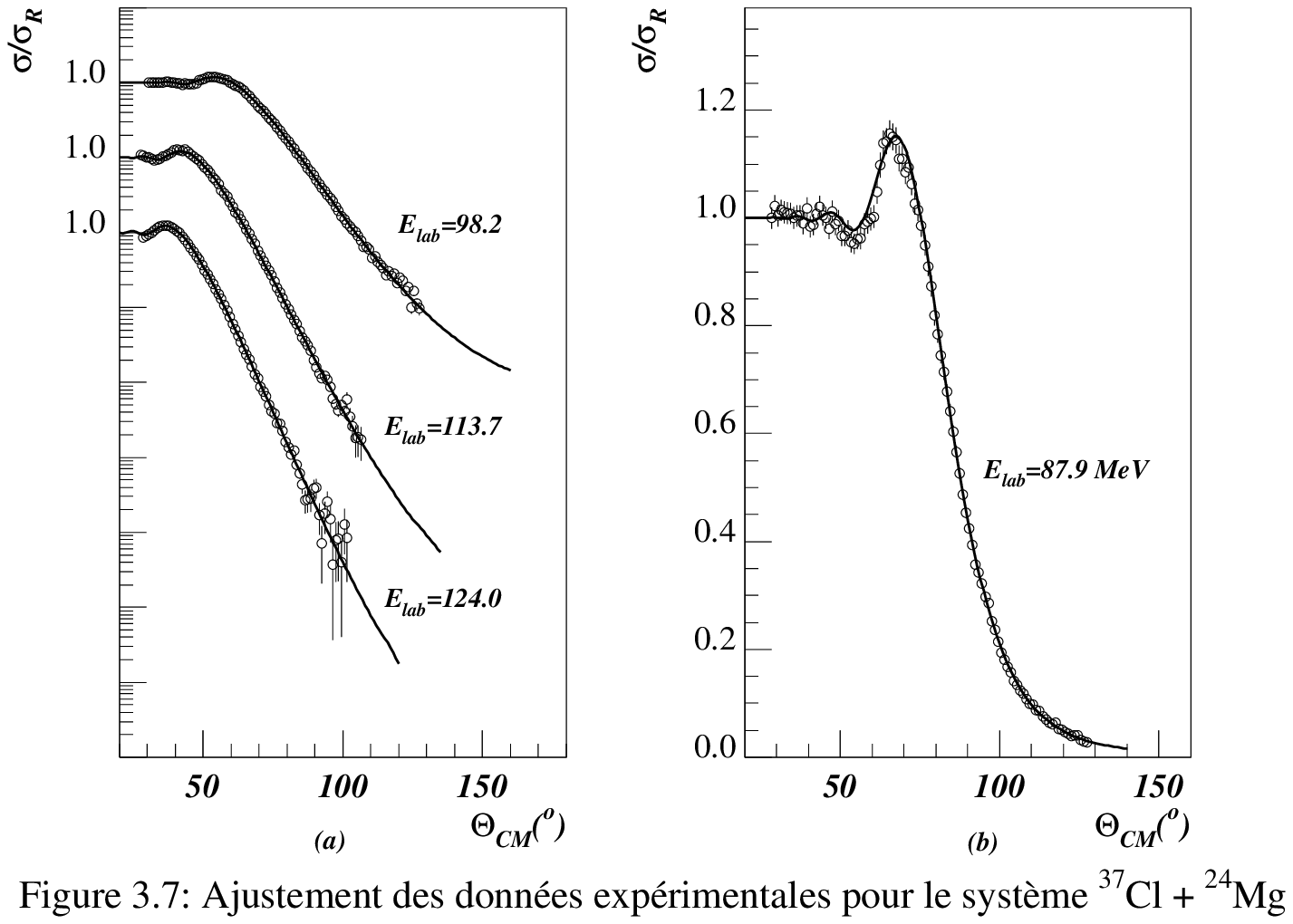}}\end{center}
%,bbllx=60,bblly=20,bburx=500,
%bbury=280}}\end{center}
%\caption{Comparaison des donn\'ees exp\'erimentales aux calculs
%semi-ph\'enom\'enologiques pour le syst\`eme $^{37}Cl+^{24}Mg$}
%\label{FIG. 3.7}
\end{figure}
%&&&&&&&&&&&&&&&&&&&&&&&&&&&&&&&&&&&&&&&&&&&&&&&&&&&&&&&&&&&&&&&&&&&&&&&&&&&&&
%\vspace{1.5cm}
%%%%%%%%%%%%%%%%%%%%%%%%%%%%%%%%%%%%%%%%%%%%%%%%%%%%%%%%%%%%%%%%%%%%%%
%Tableau Cl37-Mg
\begin{center}
\begin{tabular}{||c|c|c|c|c|c|c|c|c||} \hline \hline
$E_{lab}$&$N_R$&$W_0$&$R_W$&$a_W$&$<l>$&$<l^2>$&$\chi^2/n$&$\sigma_R$ \\
 (MeV) &  & (MeV) & (fm) & (fm) & &  & & (mb) \\ \hline \hline
124.0 & 1.150 & 60 & 7.14 & 0.567 &24.6 & 694.7 &1.46 & 1256.1\\
113.7 & 1.192 & 60 & 7.25 & 0.548 &22.2 & 565.5 &1.47 & 1110.5\\
98.2  & 1.293 & 60 & 7.51 & 0.491 &17.6 & 360.6 &0.95 & 808.3\\
87.9  & 1.366 & 60 & 7.93 & 0.382 &13.4 & 211.4 &0.90 & 508.0\\
\hline\hline
\end{tabular}
\vskip 3mm
Tableau 3.9 : Meilleurs param\`etres du mod\`ele optique pour 
$^{37}Cl+^{24}Mg$.
\end{center}
%%%%%%%%%%%%%%%%%%%%%%%%%%%%%%%%%%%%%%%%%%%%%%%%%%%%%%%%%%%%%%%%%%%%%%
%%%%%%%%%%%%%%%%%%%%%%%%%%%%%%%%%%%%%%%%%%%%%%%%%%%%%%%%%%%%%%%%%%%%%%
% Tableau P-Al, S-Si, S-S36
\begin{center}
\begin{tabular}{||c|c|c|c|c|c|c|c|c|c||} \hline \hline
Syst\`eme&$E_{lab}$&$N_R$&$W_0$&$R_W$&$a_W$&$<l>$&$<l^2>$&$\chi^2/n$&$\sigma_R$   \\
 &(MeV) &  & (MeV) & (fm) & (fm) &  & &  &(mb) \\ \hline \hline
\pal&79.5 & 1.217 & 60 & 7.38 & 0.471 &16.6 & 318.1 & 1.80 & 737.0\\
\ssi&90.0 & 1.410 & 60 & 7.85 & 0.406 &17.3 & 345.1 &1.15 & 749.0\\
$^{32}S+^{36}S$&90.0 & 1.710 & 60 & 7.97 & 0.501 &22.8 &597.0  &1.75 & 943.0\\
\hline\hline
\end{tabular}
\vskip 3mm
Tableau 3.10 : Meilleurs param\`etres du mod\`ele optique.
\end{center}
%&&&&&&&&&&&&&&&&&&&&&&&&&&&&&&&&&&&&&&&&&&&&&&&&&&&&&&&&&&&&&&&&&&&&&&&&&&&&&

%&&&&&&&&&&&&&&&&&&&&&&&&&&&&&&&&&&&&&&&&&&&&&&&&&&&&&&&&&&&&&&&&&&&&&&&&&&&&&
\begin{figure}[ht]
\begin{center}\mbox{\epsfig{file=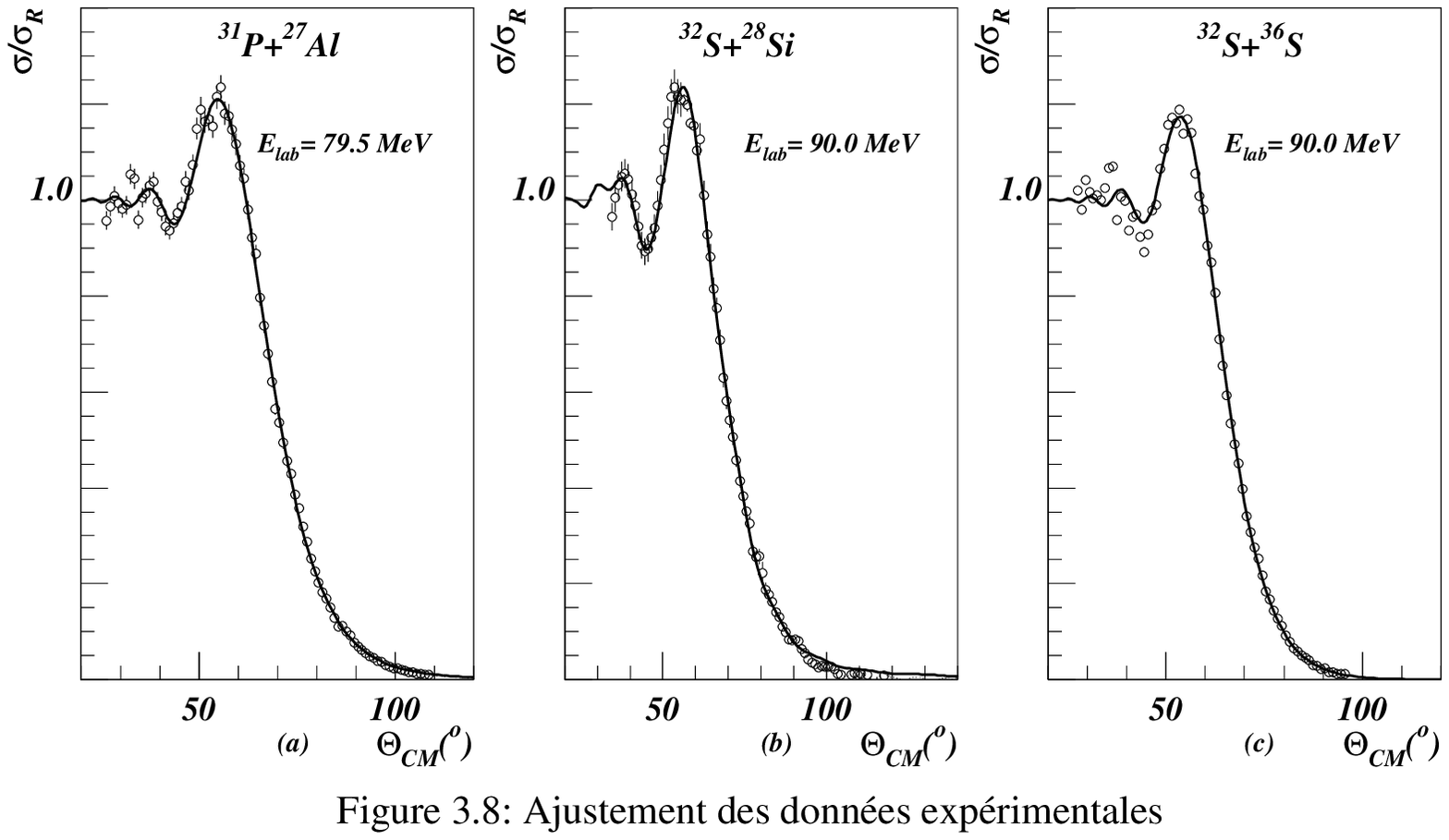,bbllx=60,bblly=20,bburx=500,
bbury=280}}\end{center}
\end{figure}
%&&&&&&&&&&&&&&&&&&&&&&&&&&&&&&&&&&&&&&&&&&&&&&&&&&&&&&&&&&&&&&&&&&&&&&&&&&&&&

\subsection{Distributions radiales de l'absorption}

L'analyse pr\'ec\'edente a l'avantage de bien reproduire les donn\'ees 
exp\'erimentales. Cependant il est difficile d'en tirer des conclusions 
g\'en\'erales sur le comportement de l'absorption. Nous introduisons 
ci-apr\`es une repr\'esentation beaucoup mieux adapt\'ee \`a cette \'etude. La 
section efficace totale de r\'eaction, $\sigma_R$, s'\'ecrit {\bf(SA83)} : 
\be 	\sigma_R=-{1\over{(2I_P+1)(2I_T+1)}}{2\over{\hbar v_0}} 
	<\chi^+_0\vert {\cal{I}}m \Delta V_L\vert \chi^+_0> \ee
o\`u $I_P$ et $I_T$ sont les spins intrins\`eques du projectile et de la cible 
dans leurs \'etats fondamentaux,  $v_0$ et $\chi^+_0$ sont la 
vitesse relative et la fonction  d'onde du mouvement relatif dans le canal 
\'elastique \`a une \'energie $E_{cm}$ du centre de masse, telle que :
 \be	\left( E_{cm} - T - V_C - V_L \right) \chi^+_0 = 0	\ee
o\`u $T$ est l'\'energie cin\'etique et $V_L=U_0+{\cal{R}}e\Delta V_L+
i{\cal{I}}m\Delta V_L$.
 
Introduisant le d\'eveloppement en onde partielle de $\chi^+_0$ :
\be	\chi^+_0={1\over{kr}}\sum_li^l(2l+1)\chi_l(r) P_l(cos \theta)	\ee
nous pouvons r\'e\'ecrire la section efficace totale sous la forme suivante : 
\be	\sigma_R= \int_0^{\infty} \sigma(r) dr \ee
o\`u   $\sigma(r)$   vaut : 
\be	\sigma(r)=-{1\over{(2I_P+1)(2I_T+1)}}{8\pi\over{k^2 \hbar v_0}}
        \sum_l (2l+1)\vert\chi_l(r)\vert^2 {\cal{I}}m \Delta V_L(r)	\ee
$k$ etant le nombre d'onde, $\chi_l(r)$ la partie radiale de la fonction d'onde 
partielle du mouvement relatif et ${\cal{I}}m \Delta V_L(r)$, le potentiel 
d'absorption ph\'enom\'enologique ou celui calcul\'e dans le cadre du 
formalisme de Feshbach selon les mod\'elisations d\'ecrites dans le chapitre 
pr\'ec\'edent.
  
Nous avons utilis\'e le code ECIS {\bf(RA94)} pour calculer les fonctions 
d'onde $\chi_l(r)$  pour les diff\'erents potentiels imaginaires d\'ecrits 
pr\'ec\'edemment. Ceci  nous a permis de tracer les graphes des fonctions 
$\sigma (r)$ en fonction de $r$, repr\'esentant la distribution radiale de 
l'absorption. Pour notre syst\`eme de r\'ef\'erence, $^{32}S+^{24}Mg$, la 
figure 3.9 repr\'esente les distributions radiales d'absorption 
correspondantes obtenues avec les param\`etres optiques inclus dans le tableau 
3.4.
 
%&&&&&&&&&&&&&&&&&&&&&&&&&&&&&&&&&&&&&&&&&&&&&&&&&&&&&&&&&&&&&&&&&&&&&&&&&&&&&
\begin{figure}[ht]
\begin{center}\mbox{\epsfig{file=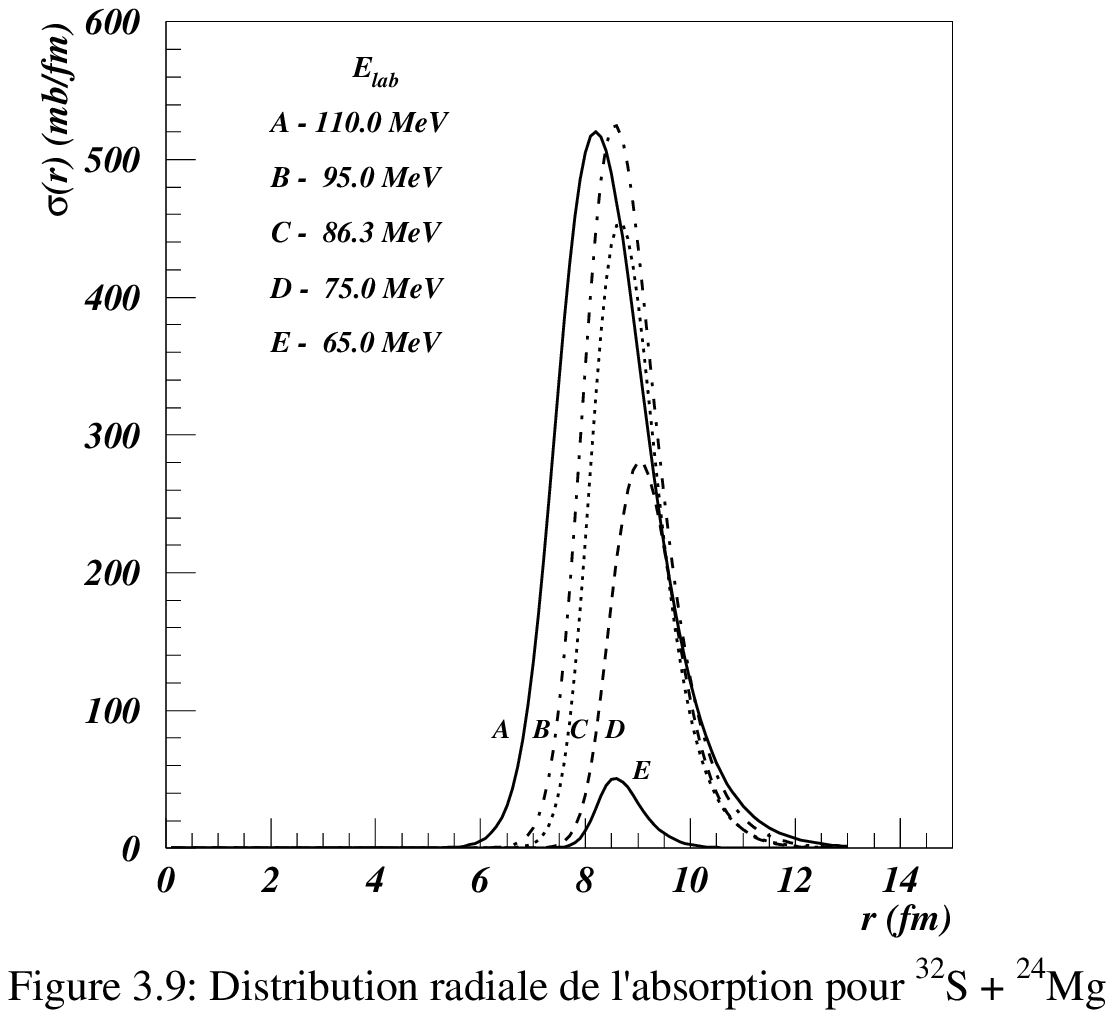}}\end{center}
\end{figure}
%&&&&&&&&&&&&&&&&&&&&&&&&&&&&&&&&&&&&&&&&&&&&&&&&&&&&&&&&&&&&&&&&&&&&&&&&&&&&&
   
L'\'etude de cette distribution radiale de l'absorption nous permet de tirer 
les conclusions g\'en\'erales suivantes :
  
a) L'absorption se trouve localis\'ee dans un \'etroit domaine de la surface du 
potentiel nucl\'eaire. L'origine de ce confinement peut facilement se 
comprendre \`a partir de l'\'equation 3.10 en tenant compte du fait qu'aux 
grandes valeurs du rayon, le potentiel imaginaire de courte port\'ee 
annihile la fonction d'onde tandis qu'aux faibles valeurs du rayon, la fonction 
d'onde, fortement absorb\'ee \`a la surface nucl\'eaire, \`a cause du 
ph\'enom\`ene de forte absorption qui domine les interactions entre noyaux 
lourds, annihile le potentiel imaginaire. Il en r\'esulte que l'absorption est 
concentr\'ee dans un domaine r\'eduit de la surface nucl\'eaire.

b)  Lorsque l'\'energie de la collision est \'elev\'ee, le domaine o\`u a lieu 
l'absorption est large. Mais lorsque l'\'energie diminue, ce domaine devient 
de plus en plus \'etroit.

c) Pour les \'energies sup\'erieures \`a la barri\`ere de Coulomb, l'absorption 
p\'eriph\'erique commence approximativement \`a la m\^eme valeur du rayon, pour 
un syst\`eme donn\'e. Mais lorsque l'\'energie diminue, l'absorption dans la 
r\'egion interne du noyau est fortement r\'eduite.

d) Pour une \'energie en dessous de la barri\`ere de Coulomb, l'absorption 
est concentr\'ee dans un domaine tr\`es \'etroit de la surface nucl\'eaire et 
nous observons un \'ecart dans l'absorption radiale entre cette \'energie et 
une 
autre situ\'ee au-dessus de la barri\`ere de Coulomb. Cet \'ecart est li\'e 
\`a un 
changement brusque de la diffusivit\'e du potentiel imaginaire lorsque 
l'\'energie de la collision passe en dessous de la barri\`ere. Le m\^eme 
comportement, quoique plus prononc\'e, est observ\'e avec des potentiels 
microscopiques. Il est d\^u \`a l'absence d'absorption en dessous de la 
barri\`ere de Coulomb. 

Dans cette repr\'esentation, l'aire sous la courbe correspond \`a la section 
efficace totale de r\'eaction que nous trouvons \`a l'avant derni\`ere colonne 
du tableau 3.4 
%%%%%%%%%%%%%%%%%%%%%%%%%%%%%%%%%%%%%%%%%%%%%%%%%%%%%%%%%%%%%%%%%%%%%%
% Tableau S-Mg
\begin{center}
\begin{tabular}{||c|c|c|c|c|c|c||} \hline \hline
$E_{lab}$&$N_R$&$W_0^a$&$R_W$&$a_W$&$\chi^2/n$&$\sigma_R$ \\
 (MeV) &  & (MeV) & (fm) & (fm) &  & (mb) \\ \hline \hline
95.0&1.61&100&7.19&0.454&2.3&941  \\
&1.60&60&7.40&0.460&2.3&948  \\
&1.57&30&7.69&0.466&2.3&949  \\
65.0&1.88&100& 7.90$^a$&0.240&5.5&55.5 \\
&1.91&60& 8.10$^a$&0.223&5.5&54.7 \\
&1.99&30& 8.40$^a$&0.185&5.5&54.0 \\
\hline\hline
\end{tabular}
\vskip 0.1mm
a)$W_0$ et/ou $R_W$ a \'et\'e maintenu fixe.
\vskip 3mm
Tableau 3.11 : Param\`etres equivalents du mod\`ele optique pour \smg.
\end{center}
%%%%%%%%%%%%%%%%%%%%%%%%%%%%%%%%%%%%%%%%%%%%%%%%%%%%%%%%%%%%%%%%%%%%%%%%%%%%
\newpage
%&&&&&&&&&&&&&&&&&&&&&&&&&&&&&&&&&&&&&&&&&&&&&&&&&&&&&&&&&&&&&&&&&&&&&&&&&&&&&
\begin{figure}[ht]
\begin{center}\mbox{\epsfig{file=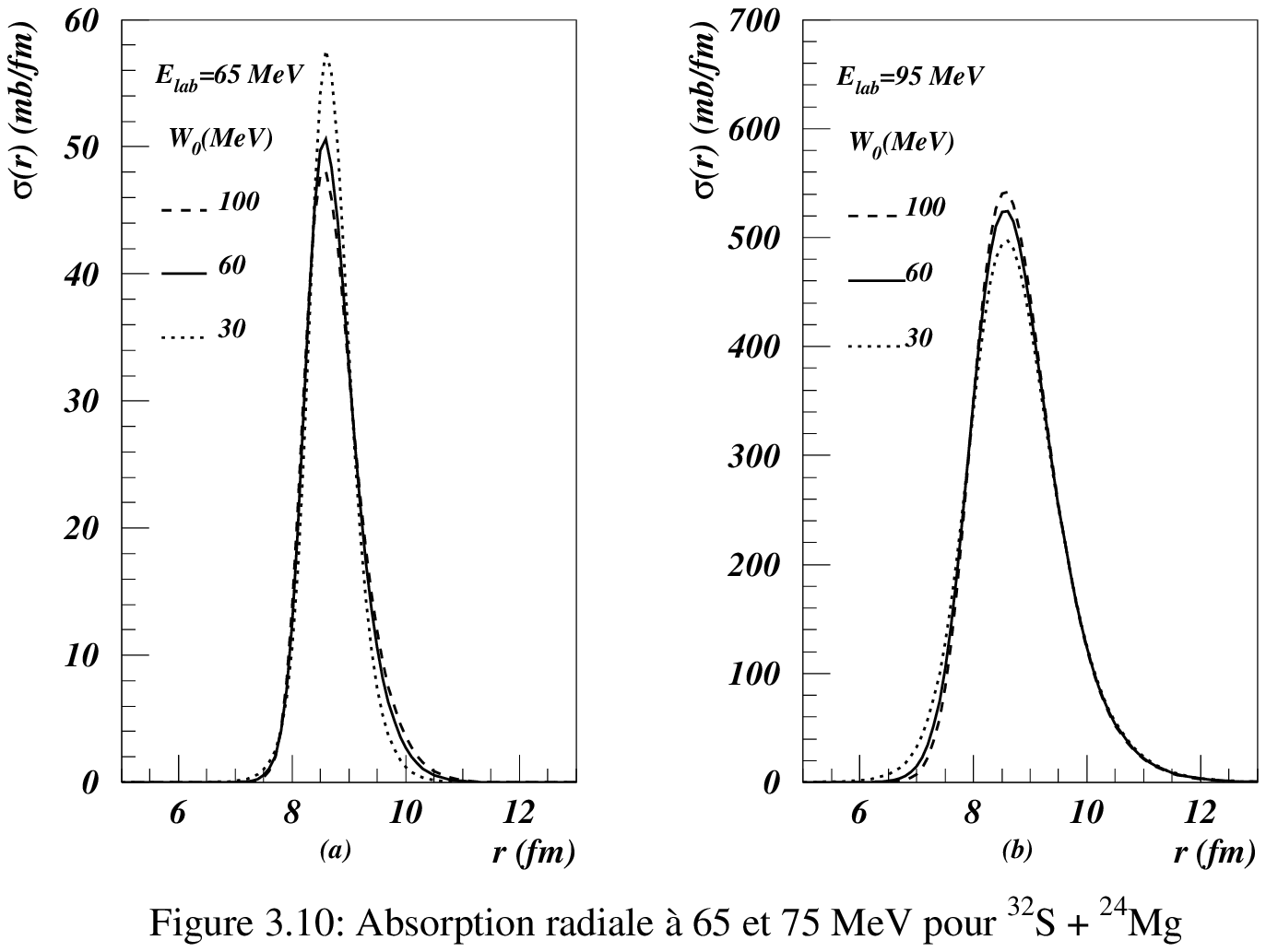}}
\end{center}
\end{figure}
%&&&&&&&&&&&&&&&&&&&&&&&&&&&&&&&&&&&&&&&&&&&&&&&&&&&&&&&&&&&&&&&&&&&&&&&&&&&&&

La repr\'esentation de la distribution radiale de l'absorption permet de 
comprendre le manque de sensibilit\'e aux valeurs des param\`etres du 
potentiel optique observ\'e aux \'energies proches ou en dessous de la 
barri\`ere de Coulomb. En effet, \`a ces \'energies, le domaine radial o\`u 
l'absorption a lieu est tr\`es \'etroit et des potentiels imaginaires avec des 
g\'eom\'etries diff\'erentes peuvent avoir les valeurs ad\'equates dans ce 
domaine et bien reproduire l'absorption totale exp\'erimentale.

 Nous pouvons \'egalement utiliser cette repr\'esentation pour montrer que la 
profondeur du potentiel imaginaire, $W_0$, utilis\'ee dans l'ajustement des 
donn\'ees exp\'erimentales n'est pas importante pour la localisation du 
domaine radial o\`u l'absorption a lieu. Nous l'avons fait pour le syst\`eme
$^{32}$S + $^{24}$Mg. La figure 3.10 repr\'esente les 
fonctions $\sigma (r)$ \`a deux \'energies diff\'erentes  et pour diff\'erentes 
valeurs du potentiel optique inclues dans le tableau 3.11. Nous constatons que 
l'absorption a lieu dans la m\^eme r\'egion de la surface nucl\'eaire bien que 
la profondeur du potentiel imaginaire varie de 30 \`a 100 MeV.

Dans la figure 3.11 nous avons inclus les distributions radiales de 
l'absorption pour tous les syst\`emes \'etudi\'es. Les r\'esultats obtenus, 
dans tous les cas confirment les conclusions auxquelles nous sommes arriv\'es 
pour le syst\`eme $^{32}S+^{24}Mg$.
%&&&&&&&&&&&&&&&&&&&&&&&&&&&&&&&&&&&&&&&&&&&&&&&&&&&&&&&&&&&&&&&&&&&&&&&&&&&&&
\begin{figure}
\begin{center}\mbox{\epsfig{file=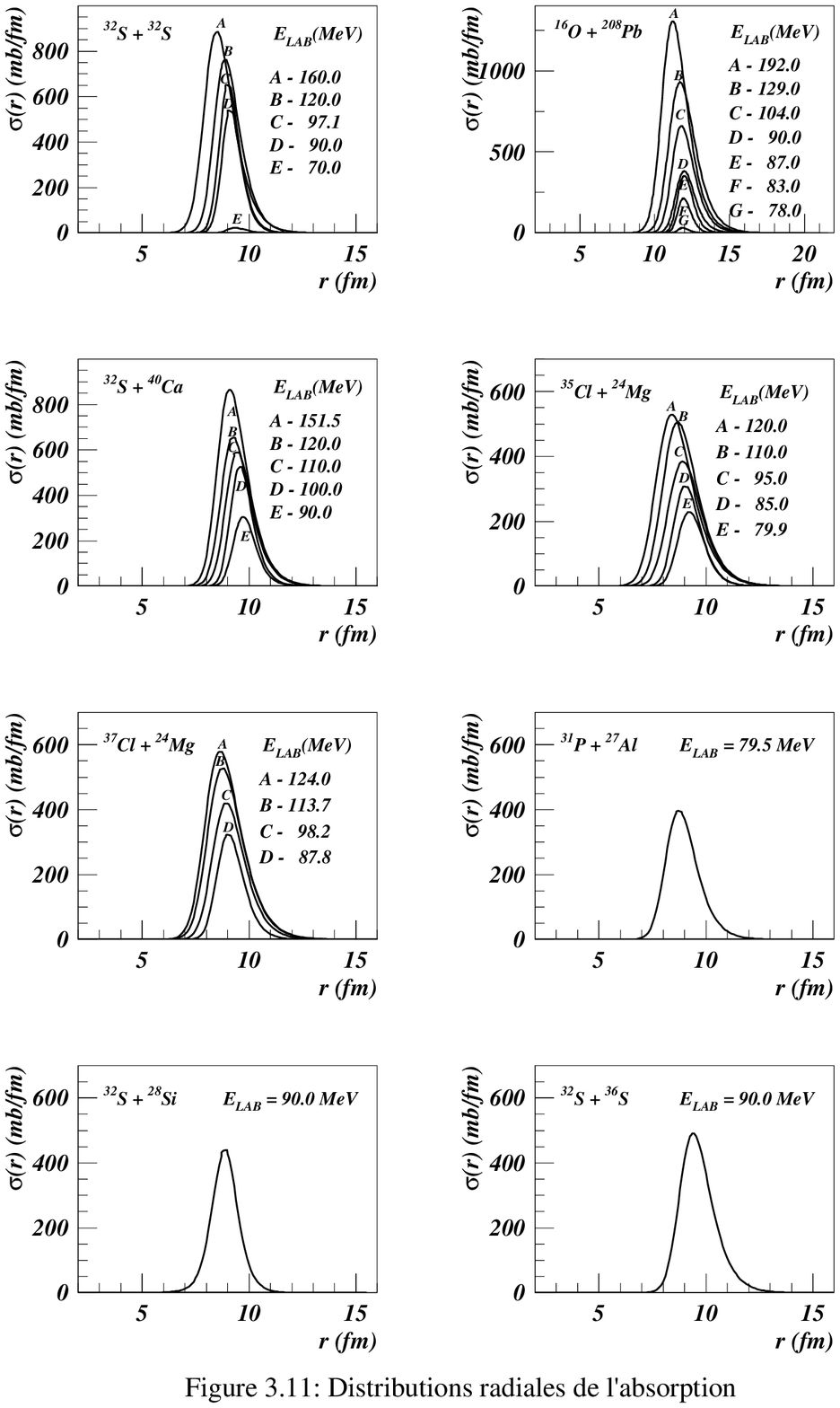}}\end{center}
\end{figure}
%&&&&&&&&&&&&&&&&&&&&&&&&&&&&&&&&&&&&&&&&&&&&&&&&&&&&&&&&&&&&&&&&&&&&&&&&&&&&&
 
\newpage

\subsection{Distributions de spin de l'absorption}
 
Outre la repr\'esentation de la distribution radiale de l'absorption, il est 
int\'eressant d'\'etudier la distribution de spin qui s'est av\'er\'ee tr\`es 
sensible au mod\`ele utilis\'e, en particulier dans l'\'etude de la fusion au 
voisinage de la barri\`ere de Coulomb.

Nous pouvons r\'e\'ecrire la section efficace totale de r\'eaction de la 
mani\`ere suivante :
\be 	\sigma_R= \sum_l \sigma_R (l)   \ee
avec 
\be 	\sigma_R (l)=-{1\over{(2I_P+1)(2I_T+1)}}{8\pi\over{k^2 \hbar 
	v_0}}(2l+1) \int_0^{\infty} \vert\chi_l(r)\vert^2 {\cal{I}}m \Delta 
	V_L(r)dr 	\ee

La fonction $\sigma_R (l)$ repr\'esente la contribution de chaque onde 
partielle \`a l'absorption totale. La figure 3.12, repr\'esente les 
distributions de spin calcul\'ees pour notre syst\`eme de r\'eference, 
$^{32}S+^{24}Mg$, avec les potentiels semiph\'enom\'enologiques du 
tableau 3.4. La d\'ependance \'energ\'etique des valeurs moyennes de $< l >$ 
et de $< l^2 >$ est \'egalement indiqu\'ee. 

%&&&&&&&&&&&&&&&&&&&&&&&&&&&&&&&&&&&&&&&&&&&&&&&&&&&&&&&&&&&&&&&&&&&&&&&&&&&&&
\begin{figure}[ht]
\begin{center}\mbox{\epsfig{file=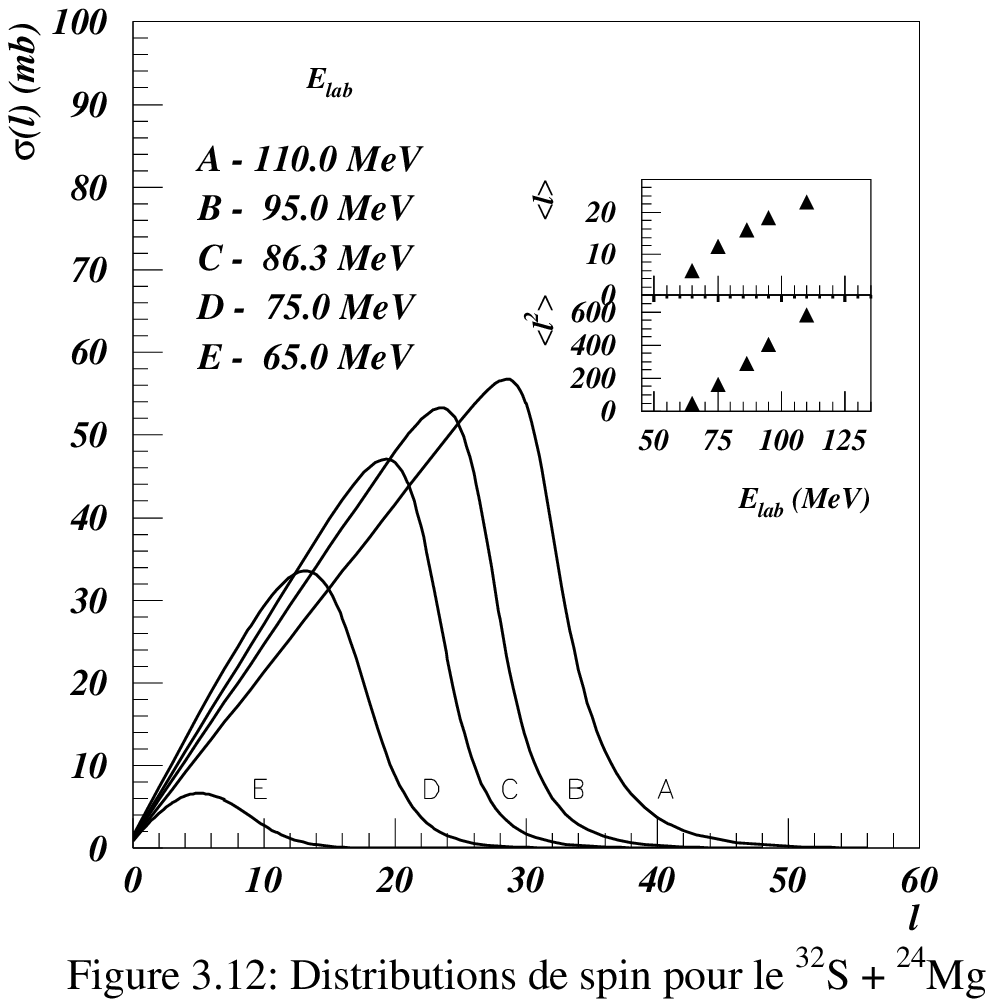}}\end{center}
\end{figure}
%&&&&&&&&&&&&&&&&&&&&&&&&&&&&&&&&&&&&&&&&&&&&&&&&&&&&&&&&&&&&&&&&&&&&&&&&&&&&&

%&&&&&&&&&&&&&&&&&&&&&&&&&&&&&&&&&&&&&&&&&&&&&&&&&&&&&&&&&&&&&&&&&&&&&&&&&&&&&
\begin{figure}
\begin{center}\mbox{\epsfig{file=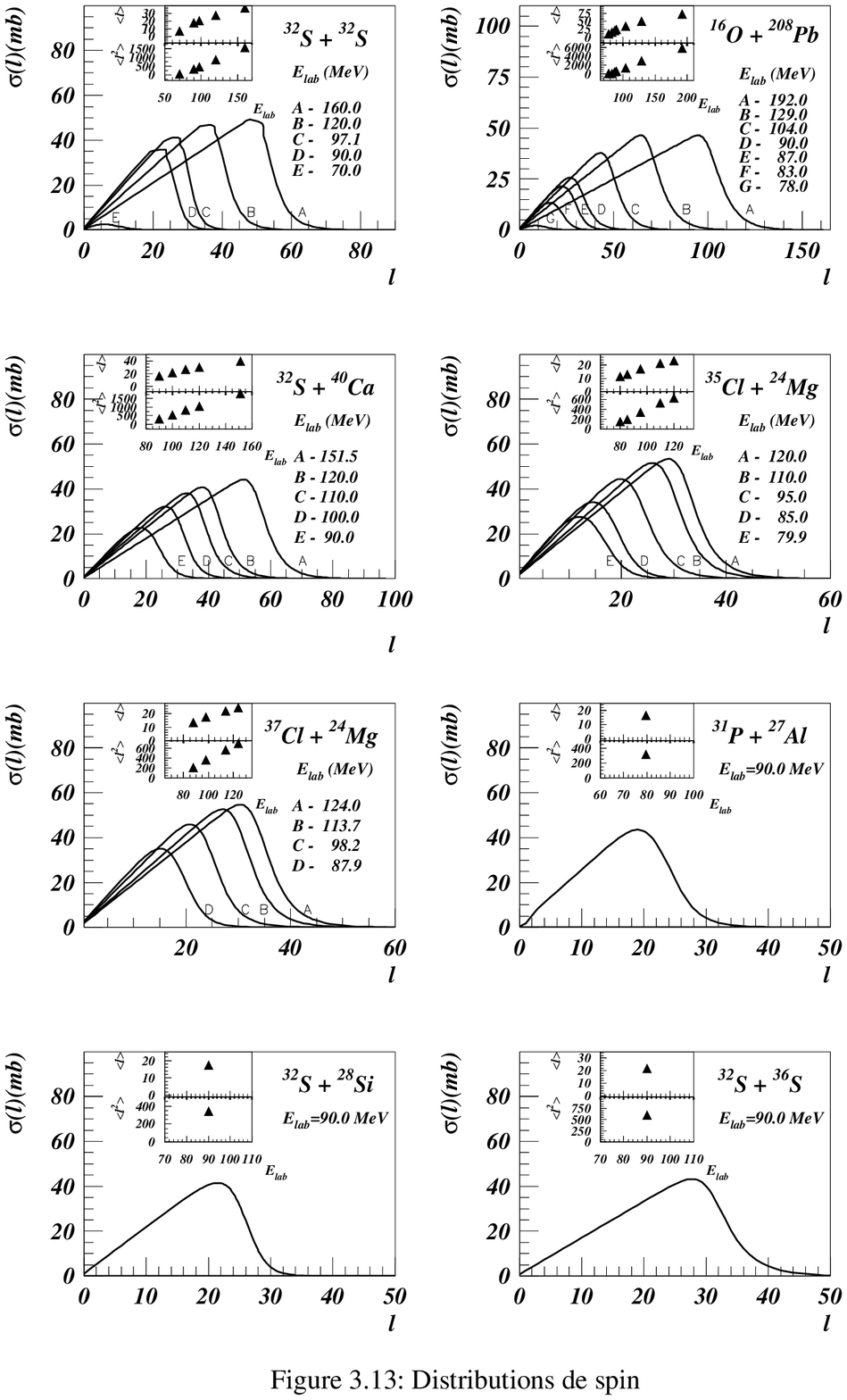}}\end{center}
\end{figure}
%&&&&&&&&&&&&&&&&&&&&&&&&&&&&&&&&&&&&&&&&&&&&&&&&&&&&&&&&&&&&&&&&&&&&&&&&&&&&&

Dans cette repr\'esentation, l'aire sous chaque courbe correspond aussi \`a la 
section efficace totale de r\'eaction que nous trouvons \`a l'avant derni\`ere 
colonne du tableau 3.4 

Dans la figure 3.13 nous montrons les distributions de spin et la variation 
avec l'\'energie des valeurs moyennes $< l >$ et $< l^2 >$ pour les autres 
syst\`emes \'etudi\'es. \'Etant donn\'e que les potentiels 
semiph\'enom\'enologiques ont \'et\'e ajust\'es de mani\`ere \`a reproduire les 
donn\'ees exp\'erimentales, ils incluent tous les effets (d\'ependance de 
l'\'energie, de la densit\'e, inter-\'echange de nucl\'eons, etc.). Par 
cons\'equent les calculs r\'ealis\'es avec ces potentiels nous serviront de 
r\'ef\'erence pour tester la validit\'e de l'\'etude microscopique que nous 
abordons \`a la section suivante.   

\vskip 20mm
\section{\large {\bf{ANALYSE AVEC LE POTENTIEL DE FESHBACH :}}}

Nous avons jusqu'\`a pr\'esent effectu\'e une \'etude du comportement de 
l'absorption pour diff\'erents syst\`emes de noyaux. Nous avons mis en
\'evidence l'importance des processus p\'eriph\'eriques, \'etant donn\'e que 
l'absorption se produit dans une r\'egion \'etroite de la surface du potentiel 
nucl\'eaire. Nous nous int\'eressons maintenant \`a d\'eterminer quelles sont 
les voies de r\'eaction pr\'epond\'erantes et ceci en relation avec la 
structure des noyaux en pr\'esence. C'est dans ce but que nous utilisons les 
mod\`eles microscopiques d\'ecrits dans le chapitre 2 de cette th\`ese.

Aux \'energies voisines de la barri\`ere de Coulomb il a \'et\'e montr\'e 
({\bf VIN93}) que le mod\`ele de l'approximation de fermeture peut d\'ecrire de 
mani\`ere ad\'equate la diffusion \'elastique entre noyaux sph\'eriques ou peu 
d\'eform\'es. Par contre, le mod\`ele ne d\'ecrit pas correctement les 
interactions entre noyaux tr\`es d\'eform\'es lorsque l'absorption est 
control\'ee par un nombre r\'eduit de canaux ({\bf BA91}),({\bf BA92}), vu que 
son hypoth\`ese fondamentale, qui remplace pour chaque noyau l'\'energie 
d'excitation de tous les \'etats par une valeur moyenne, surestime 
l'absorption. Pour 
ces syst\`emes tr\`es d\'eform\'es, il s'av\`ere plus ad\'equat d'\'evaluer la 
contribution d'un nombre r\'eduit de canaux qui contr\^olent l'absorption.

En ce sens, l'\'etude des 
syst\`emes constitu\'es de noyaux en int\'eraction qui poss\`edent des 
\'etats de 
basse \'energie d'excitation tr\`es d\'eform\'es, est tr\`es attrayante. En ce 
sens, les syst\`emes $^{32}S+^{24}Mg$ et \clmg sont d'excellents candidats, 
mais tandis que pour le premier nous disposons de donn\'ees exp\'erimentales 
\`a des \'energies sup\'erieures et inf\'erieures \`a la barri\`ere de Coulomb, 
nous ne disposons, pour les seconds, que de diffusions \'elastiques mesur\'ees 
\`a des \'energies proches, bien que sup\'erieures, \`a 
celle correspondant \`a la barri\`ere. De plus, le syst\`eme $^{32}S+^{24}Mg$ 
pr\'esente des caract\'eristiques interessantes vu que ses deux
%&&&&&&&&&&&&&&&&&&&&&&&&&&&&&&&&&&&&&&&&&&&&&&&&&&&&&&&&&&&&&&&&&&&&&&&&&&&&&
\begin{figure}
\vspace{-.5cm}
\begin{center}\mbox{\epsfig{file=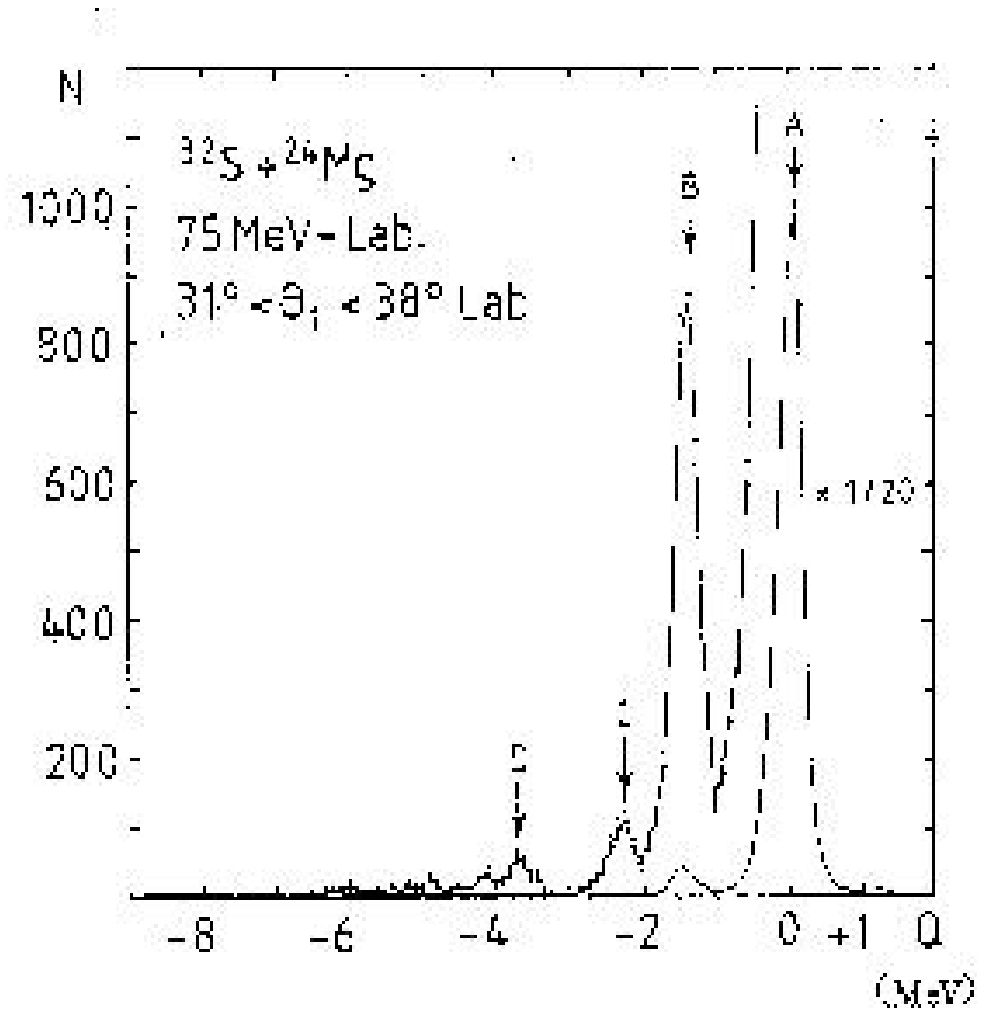,bbllx=140,bblly=300,bburx=455,
bbury=570,height=10cm,width=12cm,}}\end{center}
\end{figure}
\begin{center}
\vspace {-0.2cm}
Figure 3.14 : Spectre exp\'erimental de la diffusion de \smg
\end{center}
%&&&&&&&&&&&&&&&&&&&&&&&&&&&&&&&&&&&&&&&&&&&&&&&&&&&&&&&&&&&&&&&&&&&&&&&&&&&&&
\noindent  premiers 
\'etats excit\'es $2_1^+$, c'est \`a dire les \'etats (1.37 MeV-$^{24}Mg$) et 
(2.23 MeV-$^{32}S$) sont tr\`es d\'eform\'es et fortement peupl\'es y compris 
\`a des \'energies inf\'erieures \`a la barri\`ere, eu \'egard aux effets de 
polarisation coulombienne.

Les analyses ph\'enom\'enologiques nous indiquent que m\^eme pour les 
syst\`emes sph\'eriques, l'absorption observ\'ee aux \'energies proches de la 
barri\`ere est un processus fondamentalement superficiel et bien que le 
mod\`ele de l'approximation de fermeture soit capable de reproduire de 
mani\`ere 
satisfaisante les donn\'ees exp\'erimentales mesur\'ees pour ces syst\`emes, 
nous ne pouvons obtenir d'informations sur l'importance, pour l'absorption, de 
certaines voies d\'etermin\'ees car il \'evalue, en moyenne, la contribution de 
toutes les voies.
 
Dans cette optique, nous avons inclus l'\'etude de syst\`emes de 
caract\'eristiques tr\`es diff\'erentes qui vont de ceux form\'es par des 
noyaux sph\'eriques \`a couches ferm\'ees \`a ceux constitu\'es de noyaux 
d\'eform\'es \`a couches ouvertes.
   
 De ces consid\'erations, on peut s'attendre \`a ce qu'un calcul dans le 
mod\`ele de Feshbach, qui prenne en compte explicitement les niveaux les plus 
bas des noyaux en int\'eraction, permette de reproduire de fa\c{c}on 
satisfaisante les donn\'ees exp\'erimentales. C'est l'objet de la pr\'esente 
partie de notre travail.

Dans la figure 3.14, nous pr\'esentons un spectre exp\'erimental 
caract\'eristique de la diffusion \smg \`a 75.0 MeV. Sur ce spectre, nous 
pouvons ais\'ement identifier la diffusion \'elastique (A); la diffusion 
in\'elastique (B) dans laquelle le $^{24}Mg$ reste dans son premier \'etat 
excit\'e ($2^+, 1.37 MeV$); la diffusion in\'elastique (C) dans laquelle le 
$^{32}S$ est observ\'e dans son premier \'etat d'excitation ($2^+, 2.23 MeV$) 
et l'excitation mutuelle des deux \'etats (D). Le reste du spectre inclut 
d'autres processus qui contribuent tr\`es faiblement \`a l'absorption. Pour le 
syst\`eme \smg nous avons \'etudi\'e les diffusions \'elastiques \`a 65.0 
et \`a 75.0 MeV qui correspondent, respectivement, \`a des \'energies 
inf\'erieure et sup\'erieure \`a la barri\`ere de Coulomb. Pour les autres 
syst\`emes nous avons analys\'e les diffusions \'elastiques \`a une seule 
\'energie. Cette \'energie a \'et\'e choisie de fa\c{c}on \`a \^etre la plus 
proche possible de la barri\`ere et \`a ce que les donn\'ees de diffusion 
\'elastique correspondantes aient la meilleur qualit\'e possible. Comme cela a 
\'et\'e indiqu\'e ant\'erieurement (section 2.2.3), 
nous avons utilis\'e les parties r\'eelles des potentiels 
semiph\'enom\'enologiques obtenus dans la section pr\'ec\'edente pour effectuer 
les calculs microscopiques des potentiels imaginaires de chaque syst\`eme. Ces 
potentiels imaginaires seront compar\'es aux potentiels ph\'enom\'enologiques 
qui nous servent de r\'ef\'erences.
\begin{center}
\begin{tabular}{||c|c|c|c|c|c|c|c||} \hline \hline
Syst\`eme & $E_{lab}$  & $N_R$ & $W_0$ & $R_W$  & $a_W$ &$\chi^2/n$ &$\sigma_R$ 
  \\
  & (MeV) &  & (MeV) & (fm) & (fm) &  & (mb) \\ \hline \hline
${}^{16}O + {}^{208}Pb$&87.0&1.607&60&11.10&0.324&1.08&474\\
${}^{31}P + {}^{27}Al$&79.5&1.217&60&7.378&0.471&1.80&737\\
${}^{32}S + {}^{28}Si$&90.0&1.410&60&7.853&0.406&1.15&749\\
${}^{32}S + {}^{32}S$&90.0&1.672&60&8.416&0.305&2.90&631\\
${}^{32}S + {}^{36}S$&90.0&1.710&60&7.967&0.501&1.75&943\\
${}^{32}S + {}^{40}Ca$&90.0&1.419&60&8.539&0.378&3.59&473\\
${}^{35}Cl + {}^{24}Mg$&85.0&1.252&60&7.897&0.386&1.60&489\\
${}^{37}Cl + {}^{24}Mg$&87.9&1.366&60&7.929&0.382&0.90&508\\
\hline\hline
\end{tabular}
\vskip 5mm
Tableau 3.12 : Meilleurs param\`etres du mod\`ele optique.
\end{center}

Le tableau 3.12 indique, pour chaque syst\`eme, l'\'energie incidente 
consid\'er\'ee et rappelle  les param\`etres d'ajustement du mod\`ele optique 
obtenus dans l'\'etude semiph\'enom\'enologique.
\newpage
\subsection{Distributions angulaires et potentiels :}
  
 Sur la base du formalisme pr\'esent\'e \`a la Section 2.2.3 du 
pr\'ec\'edent chapitre, nous avons utilis\'e le code FESHIOP {\bf(PAC95)} pour \'evaluer le 
potentiel imaginaire d\^u \`a la contribution des \'etats excit\'es list\'es, 
pour chaque syst\`eme, dans le tableau 3.13 et pour lesquels les \'energies 
ainsi que les informations spectroscopiques sont indiqu\'ees. Les param\`etres 
de d\'eformation utilis\'es dans le calcul des facteurs de forme radiaux, ont 
\'et\'e d\'eduits de donn\'ees exp\'erimentales sur les temps de 
vie des \'etats excit\'es {\bf(EN78)}. Pour les noyaux rotationnels avec moment 
quadripolaire intrins\`eque, les valeurs du param\`etre de d\'eformation ont 
\'et\'e corrig\'es de fa\c{c}on ad\'equate ({\bf SA89}).
%&&&&&&&&&&&&&&&&&&&&&&&&&&&&&&&&&&&&&&&&&&&&&&&&&&&&&&&&&&&&&&&&&&&&&&&&&&&&&
\begin{figure}
\vspace{-1cm}
\begin{center}\mbox{\epsfig{file=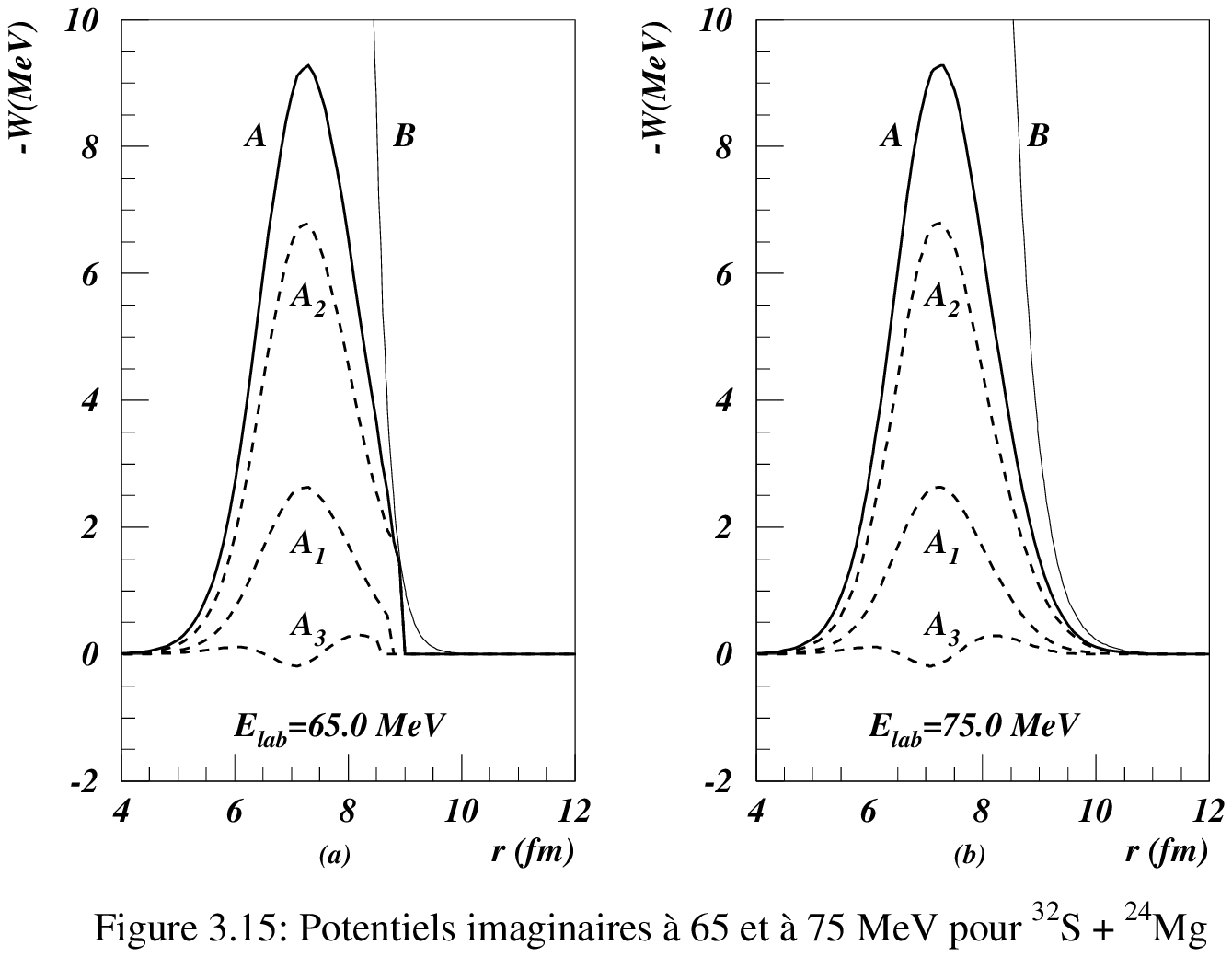}}\end{center}
\end{figure}
%&&&&&&&&&&&&&&&&&&&&&&&&&&&&&&&&&&&&&&&&&&&&&&&&&&&&&&&&&&&&&&&&&&&&&&&&&&&&&
Les \'etats in\'elastiques inclus dans les calculs ont \'et\'e choisis de 
mani\`ere \`a ce qu'ils donnent une contribution d'au moins 1\% \`a la section 
efficace de r\'eaction. Dans la figure 3.15.a-b nous comparons les potentiels 
imaginaires ph\'enom\'enologiques (ligne continue B) obtenus par ajustement des 
donn\'ees de diffusion \'elastique \smg \`a 65.0 et \`a 75.0 MeV aux calculs 
effectu\'es dans le cadre du formalisme de Feshbach (ligne continue A). Les 
diff\'erentes composantes du potentiel imaginaire sont repr\'esent\'ees par des 
lignes discontinues;  $A_1, A_2$ et $A_3$ repr\'esentent, respectivement, les 
contributions \`a l'absorption dues \`a l'excitation des \'etats $2_1^+,2.23 
MeV$ du $^{32}S$, $2_1^+,1.37 MeV$ du $^{24}Mg$ et l'excitation mutuelle de ces 
deux \'etats.
\begin{center}
\begin{tabular}{||c|c|c|c|c|c|c||} \hline \hline
$Noyau_{g.s.}$ & $J^\pi$ & $E_J^\pi (MeV)$ & $E_\lambda$ & $\beta_N$ 
& Branching Ratio & Mixing Ratio   \\ \hline \hline
$^{16}O_{0^+}$ & $3^-$ & 6.130 & E3 & 0.71 & 100.0 &  \\
  & $2^+$ & 6.920 & E2 & 0.38 & 100.0 &  \\ \hline
$^{24}Mg_{0^+}$ & $2^+$ & 1.370 & E2 & 0.6 & 100.0 &  \\
  & $2^+$ & 4.250 & E2 & 0.14 & 78.9 &  \\ \hline
$^{27}Al_{{5 \over 2}^+}$ & ${3 \over 2}^+$ & 1.014 & E2+M1 & 0.12 & 97.0 & 
+0.351 \\
  & ${7 \over 2}^+$ & 2.211 & E2+M1 & 0.24 & 100.0 & -0.468 \\
  & ${9 \over 2}^+$ & 3.004 & E2 & 0.19 & 88.6 &  \\ \hline
$^{28}Si_{0^+}$ & $2^+$ & 1.780 & E2 & 0.41 & 100.0 &  \\
  & $3^-$ & 6.880 & E3 & 0.40 & 100.0 &  \\ \hline
$^{31}P_{{1 \over 2}^+}$ & ${3 \over 2}^+$ & 1.266 & E2+M1 & 0.15 & 100.0 & 
+0.300 \\
  & ${5 \over 2}^+$ & 2.234 & E2 & 0.22 & 100.0 &   \\
  & ${7 \over 2}^-$ & 4.431 & E3 & 0.31 & 10.0 &  \\ \hline
$^{32}S_{0^+}$ & $2^+$ & 2.230 & E2 & 0.31 & 100.0 &  \\
  & $2^+$ & 4.280 & E2 & 0.12 & 86.0 &  \\
  & $3^-$ & 5.010 & E3 & 0.44 & 3.1 &  \\ \hline
$^{35}Cl_{{3 \over 2}^+}$ & ${5 \over 2}^+$ & 1.760 & E2+M1 & 0.17 & 100.0 & 
+2.850 \\
  & ${7 \over 2}^+$ & 2.650 & E2 & 0.10 & 90.6 &  \\
  & ${7 \over 2}^-$ & 3.160 & E3 & 0.10 & 90.0 &  \\ \hline
$^{36}S_{0^+}$ & $2^+$ & 3.290 & E2 & 0.16 & 100.0 &  \\ \hline
$^{37}Cl_{{3 \over 2}^+}$ & ${5 \over 2}^+$ & 3.090 & E2+M1 & 0.13 & 100.0 & 
+1.500 \\
  & ${7 \over 2}^-$ & 3.100 & E3+M2 & 0.16 & 100.0 &  +0.180 \\
  & ${9 \over 2}^-$ & 4.010 & E3 & 0.17 & 31.0 &  \\ \hline
$^{40}Ca_{0^+}$ & $3^-$ & 3.740 & E3 & 0.40 & 100.0 &  \\
  & $2^+$ & 3.900 & E2 & 0.12 & 100.0 &  \\ \hline
$^{208}Pb_{0^+}$ & $3^-$ & 2.600 & E3 & 0.11 & 100.0 &  \\
\hline\hline
\end{tabular}
\vskip 5mm
Tableau 3.13 : \'Etats collectifs inclus dans les calculs de Feshbach. 
\end{center}

%&&&&&&&&&&&&&&&&&&&&&&&&&&&&&&&&&&&&&&&&&&&&&&&&&&&&&&&&&&&&&&&&&&&&&&&&&&&&&
\begin{figure}
\vspace{-1cm}
\begin{center}\mbox{\epsfig{file=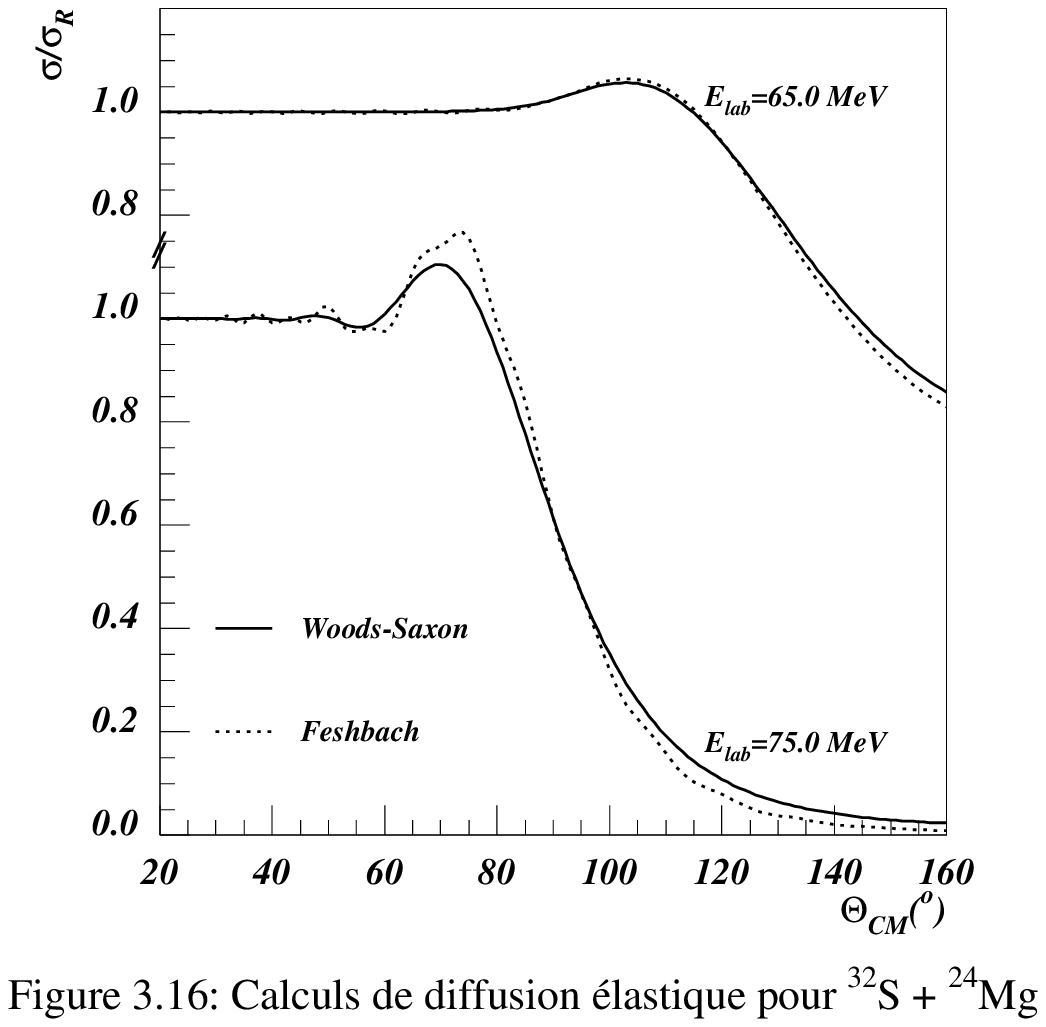}}\end{center}
\end{figure}
%&&&&&&&&&&&&&&&&&&&&&&&&&&&&&&&&&&&&&&&&&&&&&&&&&&&&&&&&&&&&&&&&&&&&&&&&&&&&&

Dans la figure 3.16 nous comparons les ajustements ph\'enom\'enologiques 
pour le syst\`eme \smg \`a 65.0 et \`a 75.0 MeV (ligne continue) 
aux r\'esultats des calculs de 
diffusion \'elastique obtenus avec les potentiels de Feshbach (ligne 
pointill\'ee) 
inclus dans la 
figure 3.15.a-b. \'Etant donn\'e que nous avons utilis\'e le m\^eme potentiel 
r\'eel que pour les analyses semiph\'enom\'enologiques, nos calculs 
repr\'esentent, en fait, une comparaison entre les termes d'absorption du 
potentiel. Le bon accord existant \`a 65.0 MeV indique l'\'equivalence entre 
les deux termes d'absorption. Par contre, \`a 75.0 MeV nous observons que la 
diffusion \'elastique calcul\'ee avec les potentiels de Feshbach ne reproduit 
pas correctement les donn\'ees exp\'erimentales. La section efficace de 
reaction ($\sigma_R^{in^{*}}$ du tableau 3.15) 
s'av\`ere \^etre inf\'erieure \`a celle obtenue par ajustement des 
donn\'ees ($\sigma_R$ du tableau 3.4).
 Ce d\'eficit en absorption indique que dans les calculs du potentiel 
nous n'avons pas pris en consid\'eration tous les canaux qui participent \`a 
l'absorption. En effet, dans la figure 3.15.b nous observons qu'\`a 75.0 MeV 
il existe un \'ecart entre le potentiel ph\'enom\'enologique et celui de Feshbach, 
ce qui sugg\`ere la necessit\'e d'inclure la contribution \`a l'absorption 
totale de processus plus p\'eriph\'eriques tels que les transferts d'un ou de 
plusieurs nucl\'eons.

Dans les figures 3.17.a \`a 3.20.a nous comparons, pour tous les autres 
syst\`emes \'etudi\'es,

%&&&&&&&&&&&&&&&&&&&&&&&&&&&&&&&&&&&&&&&&&&&&&&&&&&&&&&&&&&&&&&&&&&&&&&&&&&&&&
\begin{figure}
\begin{center}\mbox{\epsfig{file=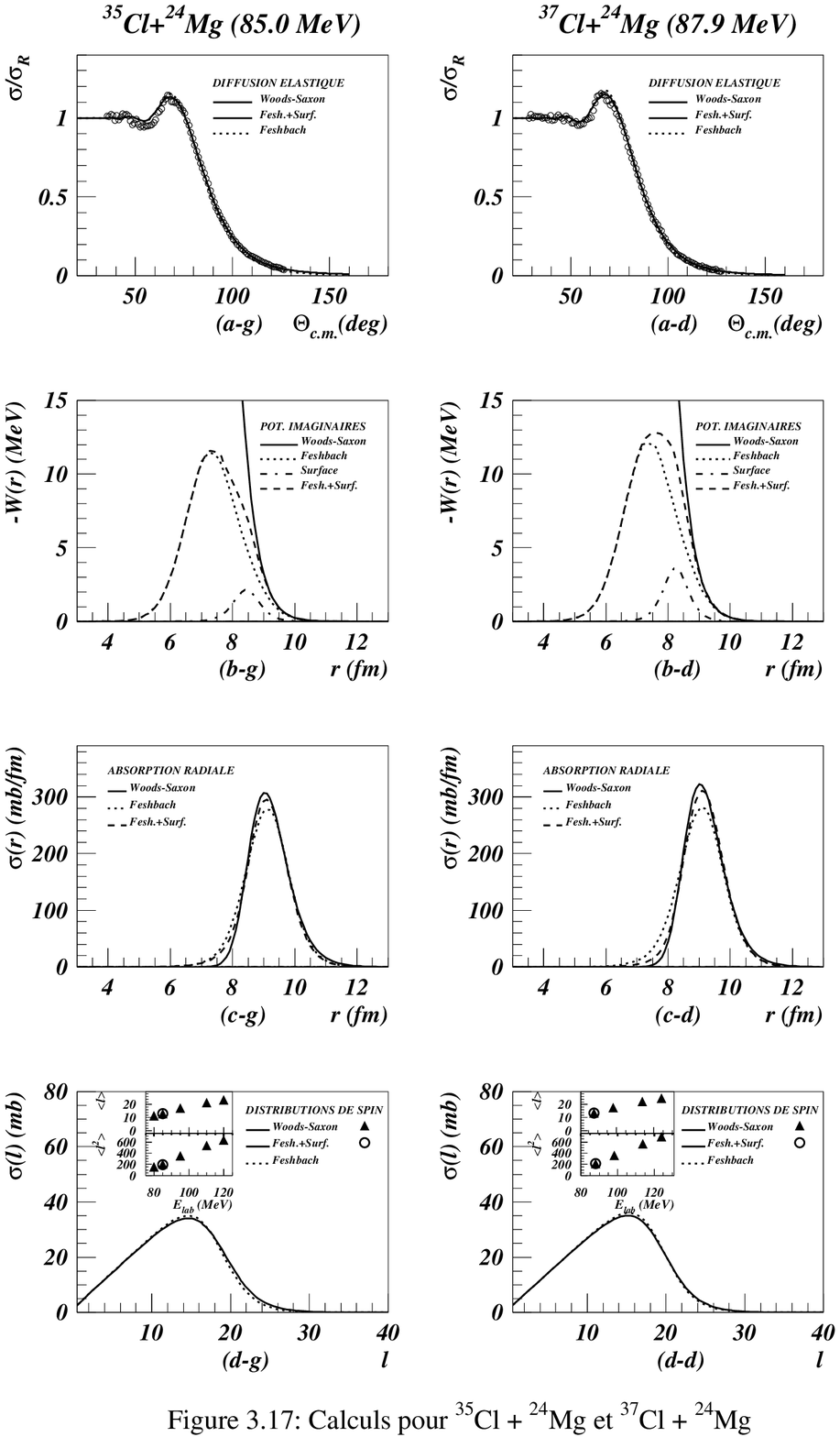}}\end{center}
\end{figure}
%&&&&&&&&&&&&&&&&&&&&&&&&&&&&&&&&&&&&&&&&&&&&&&&&&&&&&&&&&&&&&&&&&&&&&&&&&&&&&

%&&&&&&&&&&&&&&&&&&&&&&&&&&&&&&&&&&&&&&&&&&&&&&&&&&&&&&&&&&&&&&&&&&&&&&&&&&&&&
\begin{figure}
\begin{center}\mbox{\epsfig{file=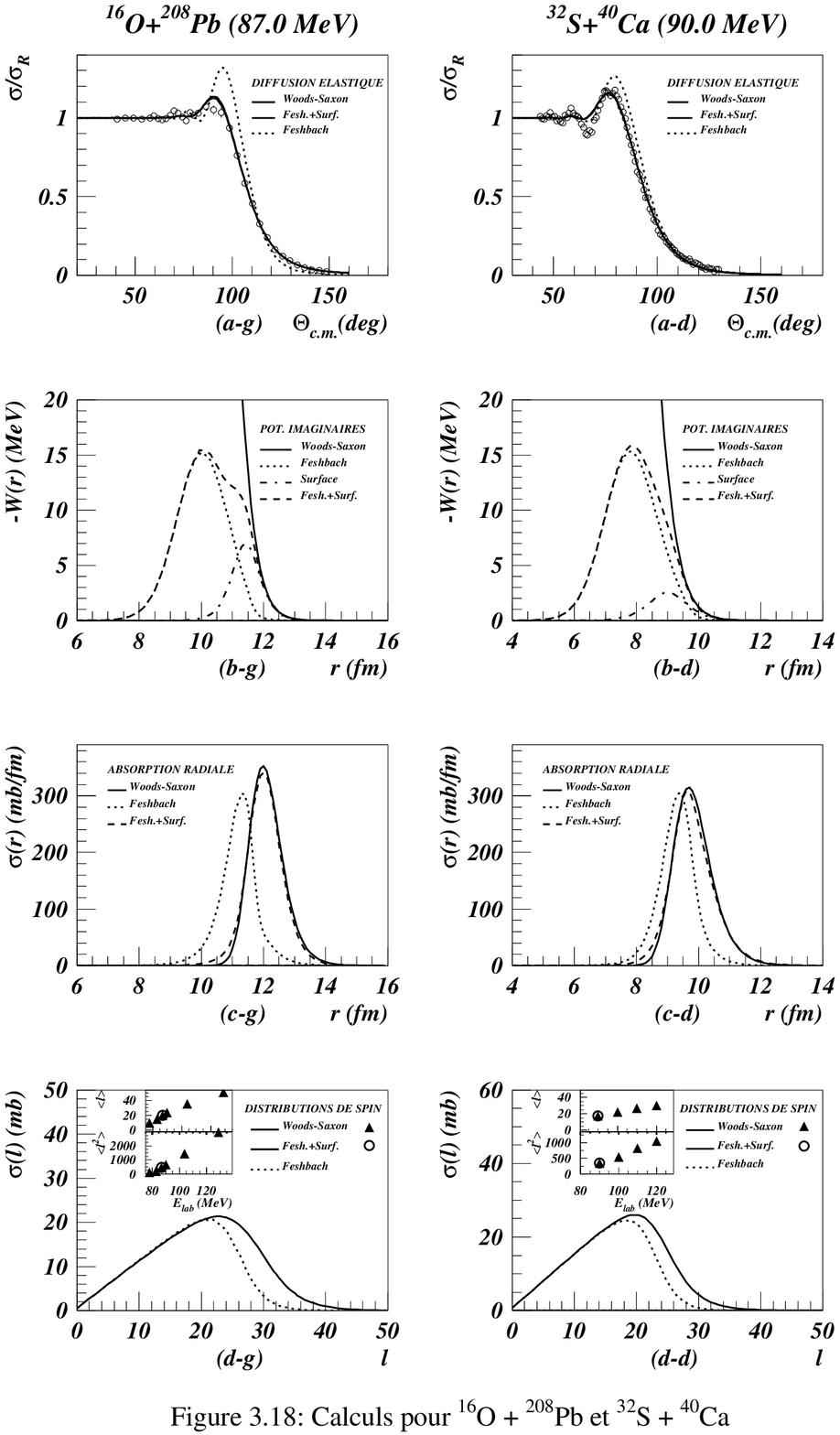}}\end{center}
\end{figure}
%&&&&&&&&&&&&&&&&&&&&&&&&&&&&&&&&&&&&&&&&&&&&&&&&&&&&&&&&&&&&&&&&&&&&&&&&&&&&&
%&&&&&&&&&&&&&&&&&&&&&&&&&&&&&&&&&&&&&&&&&&&&&&&&&&&&&&&&&&&&&&&&&&&&&&&&&&&&&
\begin{figure}
\begin{center}\mbox{\epsfig{file=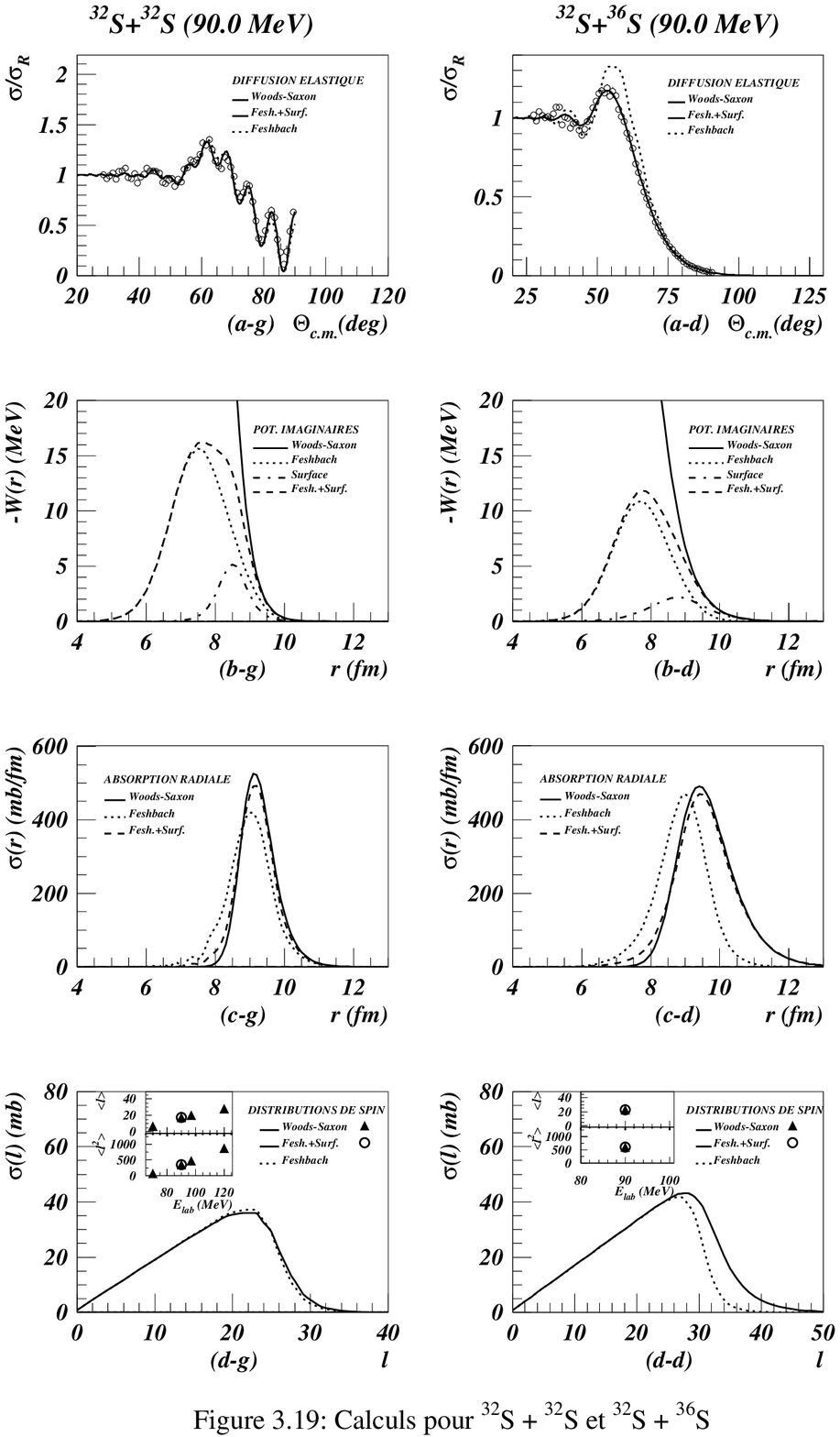}}\end{center}
\end{figure}
%&&&&&&&&&&&&&&&&&&&&&&&&&&&&&&&&&&&&&&&&&&&&&&&&&&&&&&&&&&&&&&&&&&&&&&&&&&&&&
%&&&&&&&&&&&&&&&&&&&&&&&&&&&&&&&&&&&&&&&&&&&&&&&&&&&&&&&&&&&&&&&&&&&&&&&&&&&&&
\begin{figure}
\begin{center}\mbox{\epsfig{file=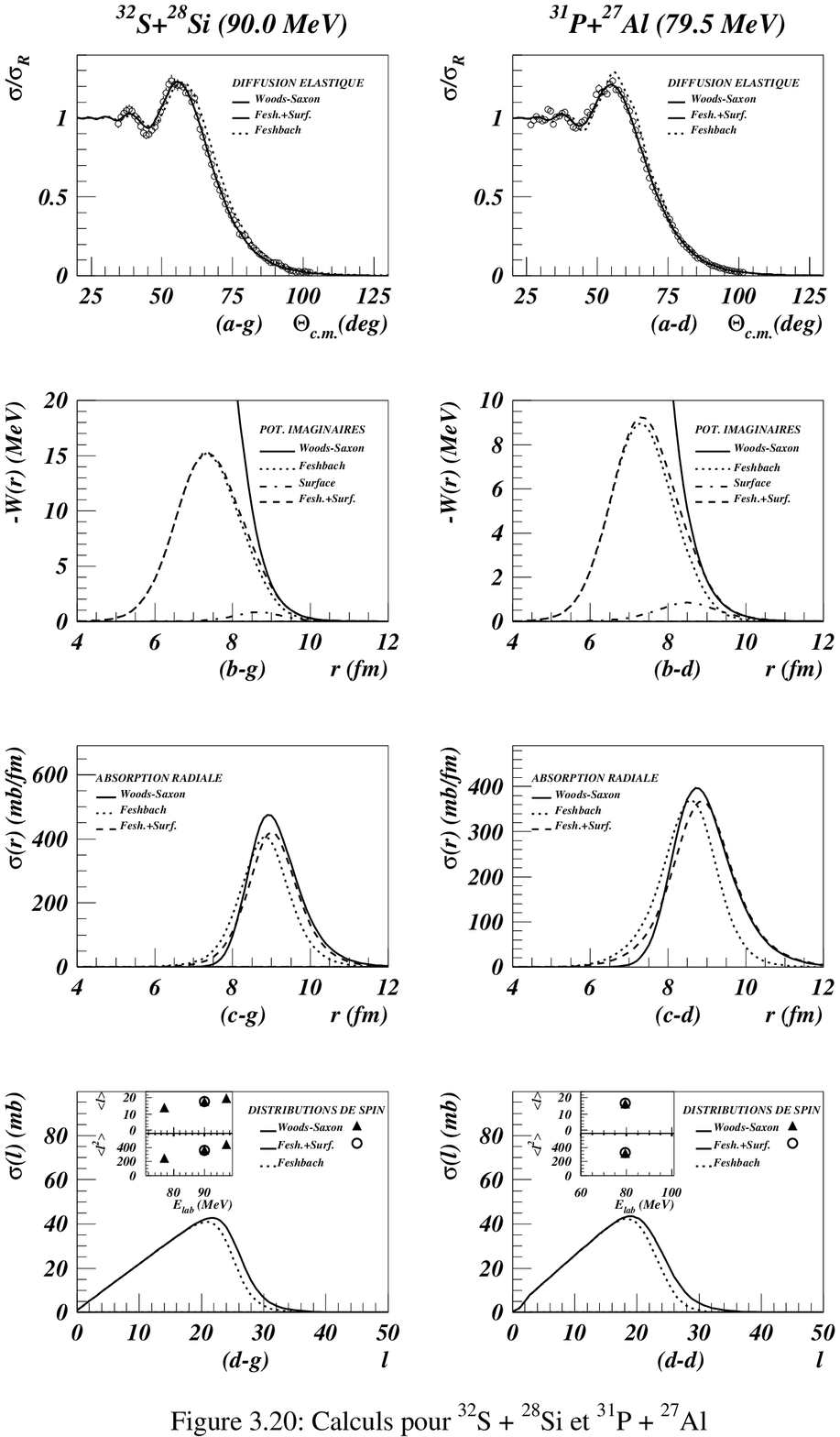}}\end{center}
\end{figure}
%&&&&&&&&&&&&&&&&&&&&&&&&&&&&&&&&&&&&&&&&&&&&&&&&&&&&&&&&&&&&&&&&&&&&&&&&&&&&&
\noindent les ajustements de la diffusion \'elastique obtenus avec les 
potentiels semiph\'enom\'enologiques (ligne continue) aux calculs 
microscopiques (ligne pointill\'ee). 

Les figures 3.17.b \`a 3.20.b permettent de comparer les potentiels 
imaginaires ph\'enom\'enologiques correspondants (ligne continue) \`a ceux de 
Feshbach (ligne pointill\'ee).

Une \'etude de ces figures montre que seules les donn\'ees de la diffusion
\'elastique de $^{35,37}$Cl + $^{24}$Mg peuvent \^etre raisonnablement 
reproduites en utilisant  les  potentiels microscopiques qui n'incluent que 
la contribution \`a l'absorption due aux \'etats in\'elastiques les plus bas. 
La caract\'eristique commune de ces syst\`emes est la pr\'esence du $^{24}$Mg 
qui est un noyau fortement d\'eform\'e, responsable d'une grande absorption 
comme nous avons pu le constater dans l'\'etude du syst\`eme \smg. Ceci 
implique une pr\'edominance de la contribution de canaux in\'elastiques dans 
l'absorption aux \'energies proches de la barri\`ere de Coulomb. De plus les 
sections efficaces des r\'eactions de transfert sont faibles pour ces 
syst\`emes {\bf(BA91)} et {\bf(BA92)}.
\begin{center}
\begin{tabular}{||c|c|c|c|c|c||} \hline \hline
Syst\`eme&$E_{lab}$&$\sigma^{tr}$&$\sigma^{in}$&$\sigma^{fis-fus}$
&$\sigma_R^{tot}$\\
 &(MeV)&(mb)&(mb)& (mb)& (mb)  \\ \hline \hline
$^{35}Cl+^{24}Mg$& 73.6 & 0.13$\pm$0.06 & 128$\pm$8 &- & 257$\pm$20$^a$ \\
                 & 100.0 & 10$\pm$2 & 193$\pm$13 &- & 870$\pm$17$^a$ \\
$^{37}Cl+^{24}Mg$& 76.1 & 0.39$\pm$0.15 & 129$\pm$11 &- & 316$\pm$12$^a$ \\
                 & 103.5 & 25$\pm$3 & 184$\pm$23 &- & 984$\pm$20$^a$ \\
$^{16}O+^{208}Pb$& 86.0 & 81.4$\pm$10.4 & 87.5$\pm$26.2 & 218$\pm$11 & 
384$\pm$48 \\
                 & 88.0 & 105.8$\pm$6.2 & 117$\pm$25 &350$\pm$40 
& 572.8$\pm$71.2 \\
\hline\hline
\end{tabular}
\vskip 0.1mm
a) Valeurs calcul\'ees par ajustement des donn\'ees exp\'erimentales de 
diffusion \'elastique.
%\vskip 4mm
Tableau 3.14 : Sections efficaces mesur\'ees pour \opb et \clmg.
\end{center}
\vskip 2mm
Dans le tableau 3.14 nous indiquons, pour les syst\`emes \clmg, les 
sections efficaces de r\'eaction mesur\'ees exp\'erimentalement pour les 
r\'eactions de transfert. Ces mesures ont \'et\'e r\'ealis\'ees, pour chaque 
syst\`eme, \`a deux \'energies entre lesquelles se trouvent celles correspondant 
aux diffusions \'etudi\'ees dans ce travail. Vu les faibles valeurs des 
sections efficaces de transfert en comparaison avec celles des sections 
efficaces totales de r\'eaction, il est ais\'e de comprendre qu'en ne 
prenant en compte que la contribution des voies in\'elastiques de plus basses 
\'energies d'excitation, nous pouvons reproduire raisonablement bien les 
donn\'ees exp\'erimentales. Nous comprenons \'egalement pourquoi l'accord entre 
les calculs et les mesures est meilleur pour le $^{35}Cl+^{24}Mg$ que pour le 
$^{37}Cl+^{24}Mg$ \'etant donn\'e que pour ce dernier syst\`eme, la 
contribution des processus de transfert, bien que petite, est beaucoup plus 
grande que pour le $^{35}Cl+^{24}Mg$.
 
Pour les autres syst\`emes \'etudi\'es, il existe toujours une diff\'erence 
entre le 
potentiel imaginaire microscopique et le potentiel ph\'enom\'enologique. Il 
en r\'esulte une mauvaise reproduction des donn\'ees de diffusion \'elastique 
avec les potentiels que nous avons calcul\'es. Cependant, pour ces syst\`emes 
et contrairement \`a ceux incluant le ${}^{24}Mg$, il a toujours \'et\'e 
observ\'e exp\'erimentalement  une importante contribution  des processus de 
transfert, et cel\`a, m\^eme \`a des \'energies proches de la barri\`ere 
{\bf(LIL)} et {\bf(VI77)}.
En effet, dans le tableau 3.14 nous indiquons, pour le syst\`eme \opb les 
sections efficaces de r\'eaction mesur\'ees exp\'erimentalement pour 
diff\'erents processus de r\'eaction \`a des \'energies tr\`es proches de celle 
\'etudi\'ee dans cette th\`ese. Nous observons que les sections efficaces de 
transfert sont du m\^eme ordre de grandeur que celles correspondant aux 
processus in\'elastiques et qu'elles repr\'esentent une partie importante des 
sections efficaces totales de r\'eaction (20 \`a 25 \%). Il n'est donc pas 
surprenant que l'on ne puisse pas reproduire les mesures de diffusion 
\'elastique en ne consid\'erant que l'absorption due aux canaux in\'elastiques.

Afin de corriger le fait de ne pas avoir pris en compte les processus de 
transfert dans nos calculs microscopiques, nous introduisons cette contribution 
de mani\`ere ph\'enom\'enologique. En nous basant sur les facteurs de forme 
semi-classiques des transferts ({\bf(BRO81)}) nous avons assum\'e un terme 
d'absorption de surface de la forme: 
\be 	W^{trans.}_{ph\acute{e}n.}(R) = -4\,W_sa_s {{d f(R)}\over {d r}}  \ee
avec 
\be 	f(r) ={1\over{1+e^{{(r-R_s)}\over {a_s}}}}	\ee
o\`u $W_S$, $R_S$ et $a_S$ sont, respectivement, la profondeur, le rayon et le 
param\`etre de diffusivit\'e du potentiel qui ont \'et\'e d\'etermin\'es par 
ajustement aux donn\'ees exp\'erimentales.  De cette mani\`ere, nous utilisons 
dans nos calculs un potentiel total imaginaire donn\'e par :
 \be	W_L^T(R) = W_L^{in}(R) + W^{trans.}_{ph\acute{e}n.}(R)	\ee

\'Etant donn\'e que le terme $W^{trans.}_{ph\acute{e}n.}(R)$ a \'et\'e 
d\'etermin\'e par ajustement des donn\'ees avec le potentiel imaginaire de 
l'\'equation 3.15, il doit contenir bien plus que les processus de transfert de 
nucl\'eons. Entre autres, il contient des corrections au terme $W_L^{in}(R)$. 
Cependant, comme $W_L^{in}(R)$ a \'et\'e obtenu en incluant dans les calculs du 
potentiel de Feshbach tous les canaux ouverts donnant une contribution 
sup\'erieure \`a 1\% \`a la section efficace de r\'eaction, les corrections au 
terme $W_L^{in}(R)$ doivent \^etre tr\`es peu significatives et 
$W^{trans.}_{ph\acute{e}n.}(R)$ correspond principalement aux processus de 
transfert.

 Le tableau 3.15 montre, pour chaque syst\`eme, le nombre de canaux 
in\'elastiques inclus dans les calculs du potentiel de Feshbach ainsi que les 
valeurs des param\`etres $W_S$, $R_S$ et $a_S$ d\'eduites de l'ajustement aux 
donn\'ees de diffusion \'elastique en utilisant le potentiel imaginaire total 
de l'\'equation 3.15.
\begin{center}
\begin{tabular}{||c|c|c|c|c|c|c|c|c|c||} \hline \hline
Syst\`eme&$E_{lab}$&Nombre&$\sigma_R^{in^{*}}$&$W_s$&$R_s$&$a_s$&$\chi^2/n$&
$\sigma_R^{in^{*}+tr}$&$\sigma_R^{tr}$  \\
 &(MeV)&d'\'etats&(mb)&(MeV) & (fm) & (fm) &  & (mb) &(mb) \\ \hline \hline
$^{16}O+^{208}Pb$&87.0&4&373&7.00&11.43&0.294&1.00&460&87 \\
$^{31}P+^{27}Al$ &79.5&6&662&0.85&8.500&0.519&5.72&748&86 \\
$^{32}S+^{24}Mg$ &65.0&2&57.0&-- &-- &-- &3.9&--  &--  \\
$^{32}S+^{24}Mg$ &75.0&2&422&2.75&8.467&0.324&3.42&440&18 \\
$^{32}S+^{28}Si$ &90.0&4&683&1.29&8.500&0.413&1.02&742&59 \\
$^{32}S+^{32}S$  &90.0&3&629&5.15&8.500&0.271&2.79&632& 3 \\
$^{32}S+^{36}S$  &90.0&4&790&2.19&8.783&0.529&2.29&942&152 \\
$^{32}S+^{40}Ca$ &90.0&5&394&2.53&9.000&0.413&3.29&472&78 \\
$^{35}Cl+^{24}Mg$&85.0&5&484&2.25&8.433&0.277&1.58&485& 1 \\
$^{37}Cl+^{24}Mg$&87.9&5&505&3.64&8.244&0.264&0.95&508& 3 \\  
\hline\hline
\end{tabular}
\vskip 4mm
Tableau 3.15 : Meilleurs param\`etres des potentiels de surface.
\end{center}
\vskip 2mm

 Nous incluons \'egalement dans ce tableau les sections 
efficaces de r\'eaction $\sigma_R^{in^{*}}$ et $\sigma_R^{in^{*}+tr}$ calcul\'ees avec 
le potentiel imaginaire de Feshbach et celui de l'\'equation 3.15, 
respectivement. Les augmentations de la 
section efficace de r\'eaction obtenues apr\`es avoir inclus le terme 
ph\'enom\'enologique de transfert ($\sigma_R^{in^{*}+tr} - \sigma_R^{in^{*}}$) 
sont compatibles avec les sections efficaces 
de transfert dans le cas des syst\`emes pour lesquels existent des mesures 
exp\'erimentales (tableau 3.14). Il est important de signaler que le fait 
d'avoir utilis\'e un propagateur WKB complet dans la mod\'elisation du potentiel de 
Feshbach (\'eq. 2.52), nos calculs permettent d'obtenir l'absorption due tant 
\`a la contribution des processus directs qu'\`a ceux de multi-\'etapes qui ont 
comme "porte d'entr\'ee" l'excitation des \'etats in\'elastiques explicitement 
inclus. \'Etant donn\'e que dans un travail ant\'erieur {\bf(VI93)} il a \'et\'e 
utilis\'e avec succes l'hypoth\`ese que ce sont les processus de multi-\'etapes 
qui conduisent \`a la fusion, les valeurs de $\sigma_R^{in^{*}}$) obtenues dans 
nos calculs doivent inclure l'absorption due aux processus directs et de 
fusion.

Pour le syst\`eme \opb, les sections efficaces d'absorption 
correspondant \`a toutes les voies ouvertes \`a des \'energies tr\`es proches de la  
notre ont \'et\'e mesur\'ees {\bf(LIL)} et {\bf(VI77)}. 
En accord avec notre id\'ee, la valeur de 
$\sigma_R^{in^{*}}$ calcul\'ee \`a 87 MeV 
s'av\`ere \^etre compatible avec la valeur de $\sigma^{in} + \sigma^{fis-fus}$ 
d\'eduite des valeurs mesur\'ees \`a 86 et \`a 88 MeV (tableau 3.14) ce qui 
n'exclut pas que les processus de transfert soient \'egalement des portes 
d'entr\'ees conduisant \`a la fusion.

La contribution de $W^{trans.}_{ph\acute{e}n.}(R)$ (ligne tiret-pointill\'ee)
 ainsi que le potentiel 
imaginaire total de l'\'equation 3.15 (tirets)
sont repr\'esent\'es sur la figure 3.21.b, pour le \smg et sur 
les figures 3.17.b \`a 3.20.b pour tous les autres syst\`emes. 
Nous pouvons observer qu'il existe un large domaine radial dans 
lequel le potentiel total imaginaire (Eq. 3.15) et celui de Woods-Saxon s'accordent 
parfaitement.
%&&&&&&&&&&&&&&&&&&&&&&&&&&&&&&&&&&&&&&&&&&&&&&&&&&&&&&&&&&&&&&&&&&&&&&&&&&&&&
\begin{figure}
\vspace{-1cm}
\begin{center}\mbox{\epsfig{file=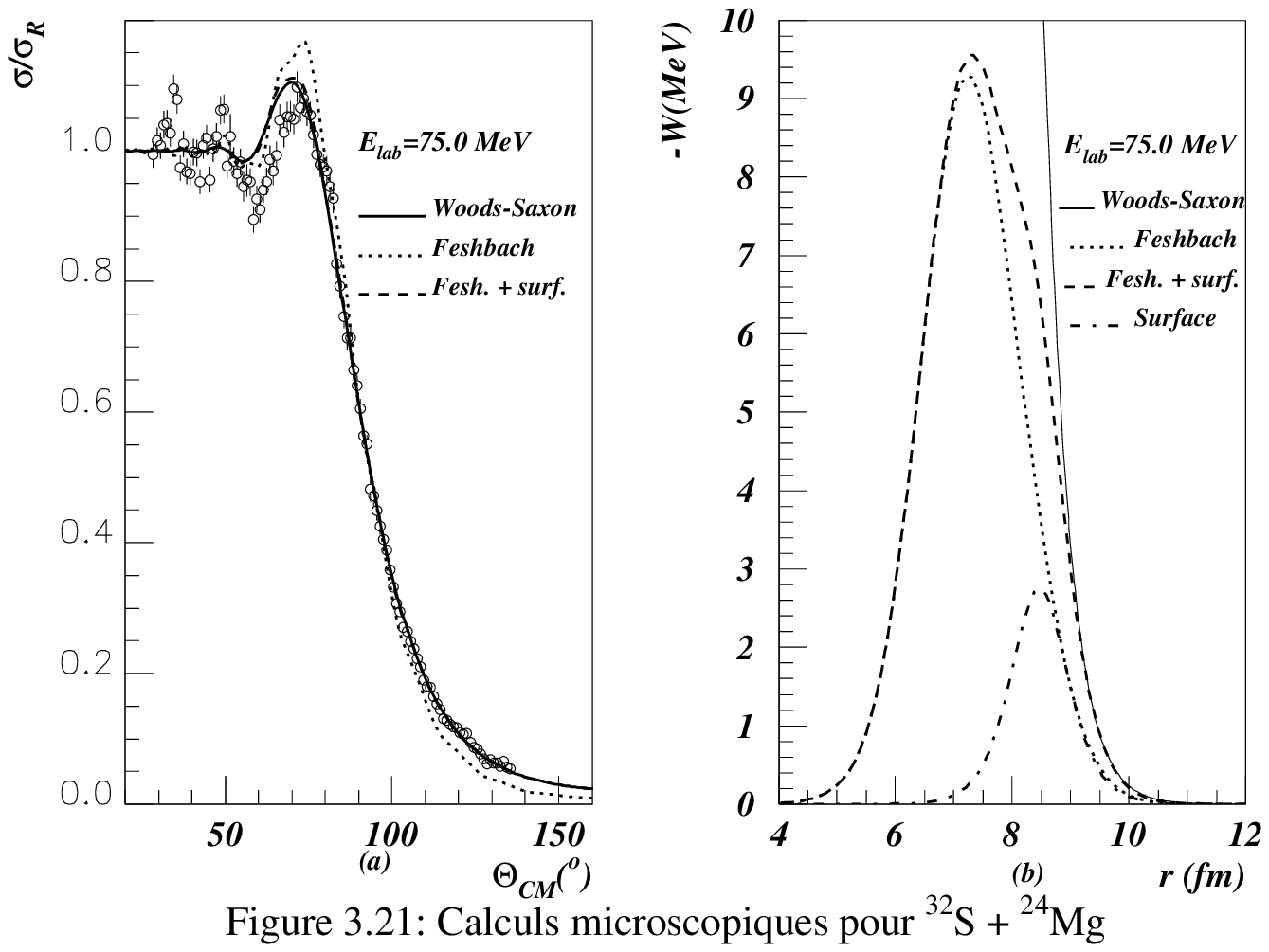}}\end{center}
\vspace{-0.8cm}
\end{figure}
%&&&&&&&&&&&&&&&&&&&&&&&&&&&&&&&&&&&&&&&&&&&&&&&&&&&&&&&&&&&&&&&&&&&&&&&&&&&&&
En introduisant un potentiel de surface ph\'enom\'enologique "plus 
p\'eriph\'erique que celui de Feshbach", pour d\'ecrire les processus de 
transfert, nous avons \'et\'e capables de 
reproduire les donn\'ees de diffusion \'elastique avec la m\^eme pr\'ecision 
que celle obtenue lors de l'analyse semiph\'enom\'enologique d\'ecrite \`a la 
section pr\'ec\'edente. En effet, sur les figures 3.17.a \`a 3.21.a, il n'est 
pas possible de distinguer les sections efficaces de diffusion \'elastique 
calcul\'ees avec l'un (tableau 3.12) ou l'autre (eq. 3.15) des potentiels 
imaginaires (lignes continues). Les lignes pointill\'ees repr\'esentent les 
calculs effectu\'es avec les potentiels imaginaires de Feshbach.

\subsection{Distributions radiales de l'absorption}
                                                         
Suivant le proc\'ed\'e d\'ecrit dans la section 3.2.2 nous avons calcul\'e les 
fonctions d'onde du mouvement relatif, utilisant le potentiel de Feshbach et 
celui donn\'e par l'\'equation 3.15. L'\'equation 3.10 nous a permis de tracer
les distributions radiales de l'absorption.
Les courbes 
correspondantes sont 
report\'ees dans les figures 3.22 pour le \smg. En particulier, dans la figure 
3.22.a, est indiqu\'ee l'absorption pour le syst\`eme \smg \`a 65 MeV qui est 
l'unique \'energie, parmis celles \'etudi\'ees, qui corresponde  \`a une 
collision en dessous de la barri\`ere de Coulomb.  Comme il a \'et\'e justifi\'e 
dans la section 2.2.3, la th\'eorie nous montre qu'il n'existe pas d'absorption 
en dessous de la barri\`ere.  Les calculs effectu\'es avec le potentiel de 
Feshbach (ligne pointill\'ee) montrent une discontinuit\'e  de la distribution 
de l'absorption correspondante entre 9 et 11 fm.  Cette discontinuit\'e qui 
concorde avec le domaine de distances situ\'e en dessous de la barri\`ere, 
indique, de par la valeur nulle du potentiel imaginaire, qu'il n'y a pas 
d'absorption (voir figure 3.15.a-ligne continue A).  Par contre, les calculs
effectu\'es avec des potentiels ph\'enom\'enologiques (ligne continue) non
nuls dans tout le domaine radial (voir figure 3.15.b-a continue B), simulent
ce manque d'absorption pr\'esentant une diffusivit\'e tr\`es petite (table 3.4).
Cependant, malgr\'e les diff\'erentes formes des deux distributions, les 
positions des maximums d'absorption obtenus ainsi que les valeurs de l'absorption 
totale (mesur\'ee par l'aire sous la courbe) sont en excellent accord (voir
les tableaux 3.4 et 3.15).
  
Les figures 3.17.c \`a 3.20.c montrent les 
distributions radiales de l'absorption calcul\'ees avec le potentiel de Feshbach 
(lignes pointill\'ees) et avec celui de l'\'eq. 3.15 (tirets discontinus), 
pour les autres syst\`emes analys\'es. Elles sont compar\'ees \`a celles obtenues avec les potentiels 
semiph\'enom\'enologiques (lignes continues).
%&&&&&&&&&&&&&&&&&&&&&&&&&&&&&&&&&&&&&&&&&&&&&&&&&&&&&&&&&&&&&&&&&&&&&&&&&&&&&
\begin{figure}
\vspace{-2.0cm}
\begin{center}\mbox{\epsfig{file=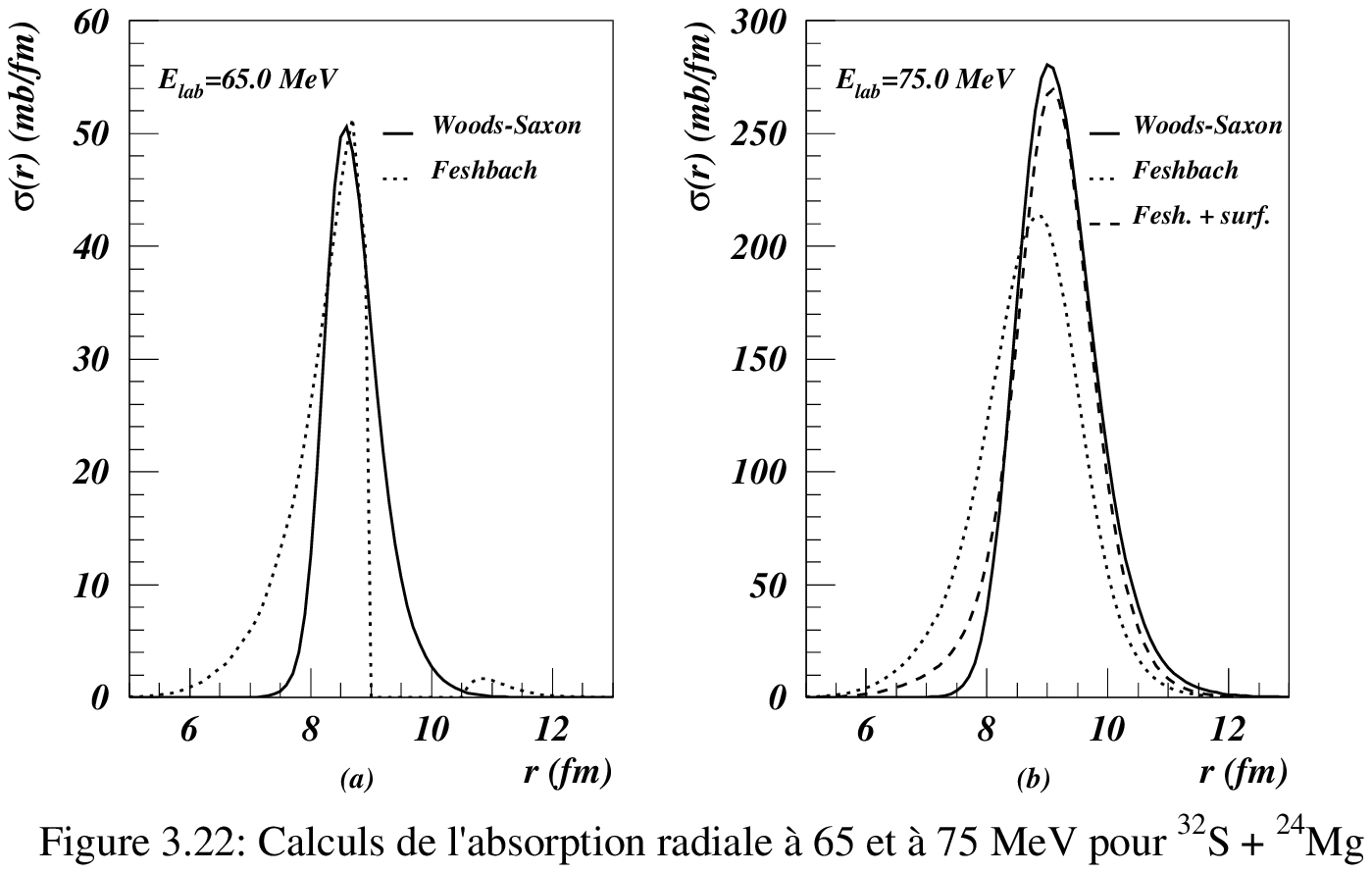}}\end{center}
\vspace{-0.6cm}
\end{figure}
%&&&&&&&&&&&&&&&&&&&&&&&&&&&&&&&&&&&&&&&&&&&&&&&&&&&&&&&&&&&&&&&&&&&&&&&&&&&&&
Comme nous l'avons expliqu\'e pr\'ec\'edement, \'etant 
donn\'e que nous avons assum\'e le m\^eme potentiel r\'eel dans tous les 
calculs, nous obtenons, en fait, une comparaison directe entre les parties 
absorbantes des diff\'erents~potentiels. 
Nous observons que les sections efficaces totales d'absorption calcul\'ees sont 
approximativement les m\^emes pour le potentiel ph\'enom\'enologique et celui 
de l'\'equation 3.15. 
Il apparait donc que pr\`es de la barri\`ere de Coulomb, 
l'accord entre les pr\'edictions th\'eoriques et les calculs effectu\'es avec 
les potentiels ph\'enom\'enologiques d\'eduits de l'ajustement des 
exp\'eriences 
est bon et que seuls quelques canaux in\'elastiques et les processus de 
transfert de nucl\'eons jouent un r\^ole pr\'edominant dans l'absorption.

\subsection{Distributions de spin de l'absorption}

Suivant le proc\'ed\'e largement d\'ecrit dans la section 3.2.3, nous avons 
utilis\'e l'\'equation 3.12 pour calculer 
la contribution de chaque onde 
partielle \`a l'absorption totale, $\sigma_R (l)$, lorsque l'on utilise le 
potentiel de Feshbach ou celui donn\'e par l'\'equation 3.15 pour d\'ecrire le 
terme d'absorption.
%&&&&&&&&&&&&&&&&&&&&&&&&&&&&&&&&&&&&&&&&&&&&&&&&&&&&&&&&&&&&&&&&&&&&&&&&&&&&&
\begin{figure}
\vspace{-0.5cm}
\begin{center}\mbox{\epsfig{file=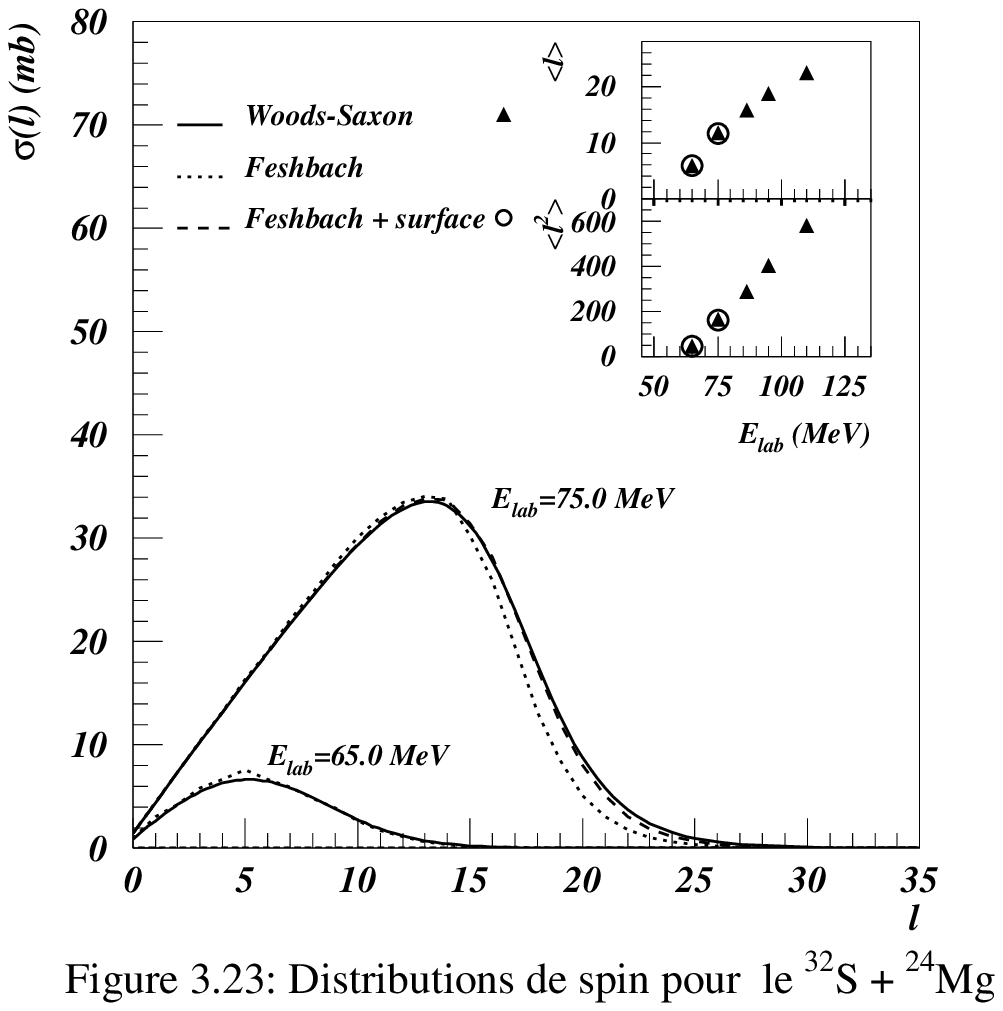}}\end{center}
\vspace{-1.0cm}
\end{figure}
%&&&&&&&&&&&&&&&&&&&&&&&&&&&&&&&&&&&&&&&&&&&&&&&&&&&&&&&&&&&&&&&&&&&&&&&&&&&&&
  Nous comparons, dans la figure 3.23 pour le 
\smg et dans les figures 3.17.d \`a 3.20.d pour les autres syst\`emes, les 
distributions de spin obtenues avec le potentiel de Feshbach (lignes 
pointill\'ees), avec le potentiel ph\'enom\'enologique de Woods-Saxon (lignes 
continues) et celui de l'\'equation 3.15 (lignes discontinues). Les 
distributions obtenues avec les deux derniers potentiels sont en parfait 
accord au point d'\^etre indistingables. 
Nous comparons 
aussi les valeurs moyennes $<~l~>$ et $<~l^2~>$ calcul\'ees avec les 
potentiels semiph\'enom\'enologiques (triangles) \`a celles calcul\'ees 
\`a une certaine \'energie avec notre potentiel incluant les contributions 
des canaux in\'elastiques les plus bas et ceux des transferts (cercles). 
  
Dans le tableau 3.16, sont indiqu\'ees les valeurs de $< l >$, $< l^2 >$ et 
$\sigma_R$ (en mb) calcul\'ees avec les potentiels semiph\'enom\'enologiques 
et avec celui de 
l'\'equation 3.15. Les valeurs de $\chi^2/n$ obtenues lors de l'ajustement 
aux donn\'ees de diffusion \'elastique avec les deux potentiels indiqu\'es 
ci-dessus, sont \'egalement incluses, et montrent l'excellent accord obtenu 
entre les mesures exp\'erimentales et les calculs r\'ealis\'es avec ces 
deux types de potentiels.
\hspace{1.5cm}
\begin{center}
\begin{tabular}{||c|c|c|c||} \hline \hline
Syst\`eme & $E_{lab}$  & Calculs semiph\'enom. & Feshbach + Surf. (ph\'enom.)\\
  & (MeV) & $<l>  \,\,\,\,\,\,\ <l^2>  \ \ \sigma_R^T \,\, \ \chi^2/n$ &  
 $<l>  \,\,\,\,\,\,\ <l^2>  \ \ \sigma_R^{in^{*}+tr} \,\, \ \chi^2/n$\\ \hline \hline
${}^{16}O + {}^{208}Pb$ & 87.0 & 20.15\,\,\,\,\,\,\,\, 475.00\,\,\,\,\,\,\, 
474\,\,\,\,\,\,\,\,\, 1.08 
	& 19.95\,\,\,\,\,\,\,\, 464.38\,\,\,\,\,\,\, 460\,\,\,\,\,\,\,\,\, 
1.00 \\
${}^{31}P + {}^{27}Al$ & 79.5 & 16.57\,\,\,\,\,\,\,\, 318.12\,\,\,\,\,\,\, 
737\,\,\,\,\,\,\,\,\, 1.80 
	& 16.70\,\,\,\,\,\,\,\, 323.20\,\,\,\,\,\,\, 748\,\,\,\,\,\,\,\,\, 
5.72 \\
${}^{32}S + {}^{24}Mg$ & 75.0 & 11.80\,\,\,\,\,\,\,\, 165.00\,\,\,\,\,\,\, 
445\,\,\,\,\,\,\,\,\, 3.40 
	& 11.70\,\,\,\,\,\,\,\, 160.84\,\,\,\,\,\,\, 440\,\,\,\,\,\,\,\,\, 
3.42 \\
${}^{32}S + {}^{28}Si$ & 90.0 & 17.78\,\,\,\,\,\,\,\, 364.35\,\,\,\,\,\,\, 
749\,\,\,\,\,\,\,\,\, 1.15 
	& 17.69\,\,\,\,\,\,\,\, 359.99\,\,\,\,\,\,\, 745\,\,\,\,\,\,\,\,\, 
1.02 \\
${}^{32}S + {}^{32}S$  & 90.0 & 17.51\,\,\,\,\,\,\,\, 352.68\,\,\,\,\,\,\, 
631\,\,\,\,\,\,\,\,\, 2.90 
	& 17.47\,\,\,\,\,\,\,\, 350.71\,\,\,\,\,\,\, 632\,\,\,\,\,\,\,\,\, 
2.79 \\
${}^{32}S + {}^{36}S$  & 90.0 & 22.76\,\,\,\,\,\,\,\, 597.03\,\,\,\,\,\,\, 
943\,\,\,\,\,\,\,\,\, 1.75 
	& 22.77\,\,\,\,\,\,\,\, 597.72\,\,\,\,\,\,\, 942\,\,\,\,\,\,\,\,\, 
2.29 \\
${}^{32}S + {}^{40}Ca$  & 90.0 & 17.23\,\,\,\,\,\,\,\, 347.16\,\,\,\,\,\,\, 
473\,\,\,\,\,\,\,\,\, 3.59 
	& 17.24\,\,\,\,\,\,\,\, 348.00\,\,\,\,\,\,\, 472\,\,\,\,\,\,\,\,\, 
3.29 \\
${}^{35}Cl + {}^{24}Mg$ & 85.0 & 13.13\,\,\,\,\,\,\,\, 202.32\,\,\,\,\,\,\, 
489\,\,\,\,\,\,\,\,\, 1.60 
	& 13.00\,\,\,\,\,\,\,\, 197.41\,\,\,\,\,\,\, 485\,\,\,\,\,\,\,\,\, 
1.58 \\
${}^{37}Cl + {}^{24}Mg$ & 87.9 & 13.45\,\,\,\,\,\,\,\, 211.41\,\,\,\,\,\,\, 
508\,\,\,\,\,\,\,\,\, 0.90 
	& 13.39\,\,\,\,\,\,\,\, 209.88\,\,\,\,\,\,\, 508\,\,\,\,\,\,\,\,\, 
0.95 \\  
\hline\hline
\end{tabular}
\vskip 3mm
Tableau 3.16 : Sections efficaces totales de r\'eaction et valeurs 
moyennes des moments angulaires.
\end{center}

\newpage
$ $
\newpage
\setcounter{chapter}{3}
\chapter{{\Large{\bf{EFFETS COLLECTIFS AUX HAUTES \'ENERGIES.}}}}
Dans le chapitre pr\'ec\'edent nous avons \'etudi\'e l'absorption au voisinage
de la barri\`ere de Coulomb. 
Nous avons montr\'e qu'il \'etait possible de d\'eterminer des potentiels
microscopiques,  calcul\'es
dans le formalisme de Feshbach avec un nombre limit\'e d' \'etats excit\'es, 
qui permettent de reproduire les donn\'ees exp\'erimentales pour les
syst\`emes tr\`es d\'eform\'es. Pour les autres syst\`emes \'etudi\'es, 
sph\'eriques ou peu d\'eform\'es, 
\`a ce potentiel 
devait \^etre ajout\'e un terme plus externe qui rende compte de l'absorption
due aux r\'eactions de tranferts.

Nous avons donc \`a notre disposition un outil qui nous permet de calculer 
l'absorption aux \'energies
proches de la barri\`ere de Coulomb, dans un domaine o\`u le mod\`ele de
fermeture est inadapt\'e. 
Nous savons, par ailleurs, que ce mod\`ele est tr\`es efficace pour d\'ecrire
l'absorption aux \'energies plus \'elev\'ees, lorsque toutes les voies 
d'absorption, ou du moins les plus importantes, sont ouvertes. En particulier, 
le syst\`eme \opb a \'et\'e largement \'etudi\'e  {\bf (AN90)}, {\bf (VI91)} 
et {\bf (VI93)}. Du point de vue th\'eorique, c'est un 
syst\`eme tr\`es interessant car les deux noyaux en interaction sont 
sph\'eriques et \`a couches ferm\'ees. Des r\'esultats similaires ont \'et\'e 
obtenus pour des syst\`emes peu d\'eform\'es comme le \sca {\bf (VI91)}. 
Cependant les \'etudes
r\'ealis\'ees jusqu'\`a pr\'esent pour le syst\`eme \opb ont \'et\'e 
limit\'ees aux \'energies comprises
entre 78 MeV et 192 MeV ce qui correspond \`a une variation de 4.75 \`a 12 
MeV/Nucleon. La question de la limite de 
validit\'e du mod\`ele aux grandes \'energies  reste ouverte. Pour y r\'epondre
 nous exposons dans ce
chapitre l'\'etude effectu\'ee \`a l'aide du syst\`eme \opb  \`a 793 MeV et 
1503 MeV; soit, respectivement, \`a des \'energies de 49.6 MeV/Nucleon et 94 MeV/Nucleon, 
respectivement. 
Ainsi, l'int\'er\^et de cette \'etude est double : Il s'agit, d'abord de 
d\'eterminer les limites de validit\'e du mod\`ele en termes de variation de 
l'\'energie de la collision, et ensuite, d'obtenir des informations relatives 
\`a l'importance des effets collectifs sur l'absorption \`a des \'energies 
suffisamment \'elev\'ees pour lesquelles les processus de diffusion 
nucl\'eon-nucl\'eon peuvent avoir une contribution significative. 
Comme nous l'avons pr\'ec\'edemment indiqu\'e, et 
\`a cause principalement de l'utilisation de certaines approximations dans 
l'\'evaluation du potentiel noyau-noyau, comme par exemple l'emploi d'une 
interaction nucl\'eon-nucl\'eon de port\'ee nulle, le mod\`ele de 
l'approximation de fermeture est mieux adapt\'e pour le calcul de la partie 
d'absorption du potentiel que pour la determination du potentiel r\'eel qui, 
comme d'habitude {\bf (VIN93)}, est remplac\'e dans l'\'evaluation du 
propagateur par un potentiel d\'eduit de l'ajustement aux donn\'ees de 
diffusion \'elastique ou par un autre qui lui est \'equivalent. En 
particulier, nous suivrons la m\^eme proc\'edure d\'ecrite dans la 
r\'ef\'erence {\bf (VIN93)} et nous d\'eterminerons la d\'ependance 
\'energ\'etique du potentiel r\'eel obtenu par double convolution des 
densit\'es des noyaux cible et projectile avec la force effective M3Y, faisant 
usage du pouvoir pr\'edictif de la relation de dispersion que nous construirons 
en suivant la proc\'edure d\'ecrite dans les r\'ef\'erences {\bf (LI85)}, 
{\bf (NA85)} et {\bf (MA86)}.

\vskip 20mm

\section{\large {\bf ANALYSE DE LA DIFFUSION \'ELASTIQUE \opb:}}

%\label{sec:anaphen}
Dans l'analyse des donn\'ees de la diffusion \'elastique \opb \`a 793 MeV 
{\bf (LI)} et \`a 1503 MeV {\bf (RO)}, nous avons utilis\'e la 
param\'etrisation ph\'enom\'enologique du potentiel optique donn\'ee par 
l'\'equation 2.18 
 \be V_{opt}=-V_0f_R(r)-iW_0f_I(r) \ee
f(r) \'etant le facteur de forme habituel de Woods-Saxon d\'efini par 
l'\'equation 2.16
 \be f(r)=\frac {1}{1+e^{\frac{r-R}{a}}}\ee 
o\`u $a$ et $R=r_0(A_P^{\frac{1}{3}}+A_C^{\frac{1}{3}})$ repr\'esentent 
respectivement, la diffusivit\'e et le rayon du potentiel noyau-noyau. 
Dans cette premi\`ere analyse nous avons adopt\'e les param\`etres du potentiel 
optique inclus dans le tableau 4.1, qui furent obtenus par Mermaz et al. 
{\bf (ME87)} \`a 793 MeV et par Roussel et al. {\bf (RO86)} \`a 1503 MeV par 
ajustement aux donn\'ees de la diffusion \'elastique \opb correspondantes.
%&&&&&&&&&&&&&&&&&&&&&&&&&&&&&&&&&&&&&&&&&&&&&&&&&&&&&&&&&&&&&&&&&&&&&&&&&&&&&
\begin{center}
\begin{tabular}{||c|c|c|c|c|c|c|c|c|c||} \hline \hline
$E_{lab}$&$V_0$&$R_R$&$a_R$&$W_0$&$R_W$&$a_W$&R\'ef\'erences&$\chi^2/n$&$\sigma_R$ 
  \\
 (MeV) &   (MeV) & (fm) & (fm) &(MeV) & (fm)& (fm)& &  & (mb) \\ \hline \hline
793.&50.&1.083&0.755&42.2&1.083&0.755& ME87& 0.75& 3600\\
1503.&80.&1.072&0.718&51.6&1.072&0.718& RO86,RO88& 54.& 3586\\
1503.&44.&1.072&0.718&26.1&1.072&0.718& pr\'esent travail & 2.18& 3209\\
\hline\hline
\end{tabular}
\vskip 6mm
Tableau 4.1 :  Param\`etres du potentiel optique ph\'enom\'enologique.
\end{center}
%&&&&&&&&&&&&&&&&&&&&&&&&&&&&&&&&&&&&&&&&&&&&&&&&&&&&&&&&&&&&&&&&&&&&&&&&&&&&&
\vskip 4mm
Dans les deux derni\`eres colonnes, nous avons inclus les valeurs du 
param\`etre $\chi^2/n$, qui nous indique la qualit\'e de l'ajustement obtenu 
avec les param\`etres optiques correspondants, ainsi que la valeur de la 
section efficace de r\'eaction calcul\'ee pour chaque cas.
%&&&&&&&&&&&&&&&&&&&&&&&&&&&&&&&&&&&&&&&&&&&&&&&&&&&&&&&&&&&&&&&&&&&&&&&&&&&&&
\begin{figure}
\vspace {-0.8cm}
\begin{center}\mbox{\epsfig{file=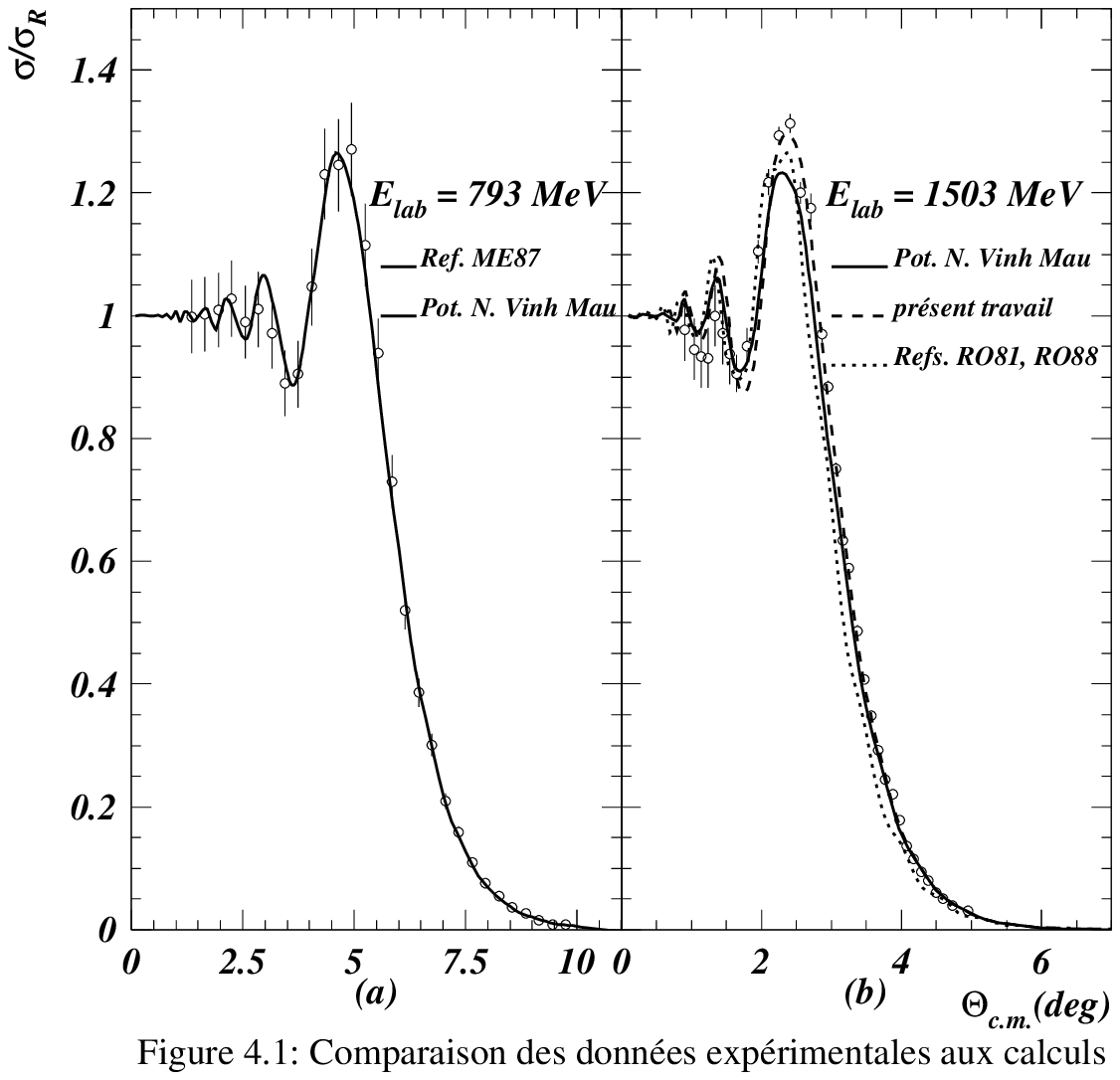}}\end{center}
\vspace {-0.8cm}
\end{figure}
%&&&&&&&&&&&&&&&&&&&&&&&&&&&&&&&&&&&&&&&&&&&&&&&&&&&&&&&&&&&&&&&&&&&&&&&&&&&&&
La figure 4.1.a indique l'excellent accord obtenu \`a 793 MeV entre 
l'exp\'erience et les calculs r\'ealis\'es, tandis que la figure 4.1.b met en 
\'evidence le d\'esaccord existant entre les donn\'ees de diffusion 
\'elastique mesur\'ees \`a 1503 MeV et le r\'esultat des calculs (ligne 
pointill\'ee) utilisant les param\`etres de Roussel et al. 

 Devant ce d\'esaccord, 
nous avons r\'ealis\'e un nouvel ajustement aux donn\'ees exp\'erimentales, en 
maintenant la m\^eme g\'eom\'etrie et en recherchant les valeurs des 
profondeurs de la partie r\'eelle $V_0$ et imaginaire $W_0$ du potentiel 
optique, qui conduisent \`a une valeur minimale du param\`etre $\chi^2/n$. Les 
r\'esultats sont inclus dans le tableau 4.1 et la distribution angulaire 
obtenue est repr\'esent\'ee dans la figure 4.1.b par une ligne discontinue. 
L'ajustement aux donn\'ees est, maintenant, excellent. 

Les potentiels obtenus par cette proc\'edure standard ainsi que les sections 
efficaces de r\'eaction calcul\'ees \`a partir de ces potentiels, seront 
utilis\'es par la suite comme r\'ef\'erence pour tous les calculs que nous 
r\'ealiserons en utilisant des potentiels de diff\'erente nature.
  
Les donn\'ees de diffusion \'elastique mesur\'ees aux \'energies indiqu\'ees 
dans le tableau 4.2, ont \'et\'e prises de la r\'ef\'erence {\bf (LIL)} et 
analys\'ees selon la m\^eme m\'ethode semi-ph\'enom\'enologique indiqu\'ee dans la 
r\'ef\'erence {\bf (LI85)}. En accord avec cette m\'ethode, la partie 
d'absorption du potentiel est d\'ecrite par un terme de Woods-Saxon 
d\'ependant de l'\'energie, dont les param\`etres sont pris de la 
r\'ef\'erence {\bf (NAG)} et inclus dans le tableau 4.2. La partie r\'eelle du 
potentiel est d\'etermin\'ee comme un potentiel de double convolution dans 
lequel l'interaction entre les noyaux est d\'ecrite par une force effective 
locale M3Y qui a \'et\'e amplement utilis\'ee avec beaucoup de succ\`es 
{\bf (SA79)}. L'unique diff\'erence entre les calculs que nous r\'esumons dans 
le tableau 4.2 et ceux des r\'ef\'erences {\bf (LI85)} et {\bf (NAG)} r\'eside 
dans la forme adopt\'ee pour la densit\'e de l'\'etat fondamental des noyaux 
cible et projectile que nous avons consid\'er\'e dans nos calculs comme une 
distribution de Fermi (Cf. \'equation 3.4). Comme nous l'indiquions dans le 
chapitre 3, la densit\'e du $^{208}Pb$ a \'et\'e obtenue par une m\'ethode 
variationelle semi-classique, d\'eterminant s\'epar\'ement les param\`etres 
pour la densit\'e de protons et de neutrons indiqu\'es dans le tableau 3.2. 
Cependant, dans le cas du noyau $^{16}O$, vu que sa densit\'e a \'et\'e 
d\'eduite d'exp\'eriences de diffusion d'\'el\'ectrons, nous avons effectu\'e 
les corrections ad\'equates dues \`a la distribution de charge non nulle du 
neutron et \`a la dimension finie du proton {\bf (SA79)}. Les potentiels 
folding calcul\'es dans ces conditions sont des potentiels nus dont la 
d\'ependance avec l'\'energie s'obtient en les renormalisant au moyen 
d'ajustement aux distributions angulaires correspondant \`a chaque \'energie. 
Les facteurs de renormalisation obtenus, d\'ependant de l'\'energie, sont 
inclus dans le tableau 4.2, o\`u nous indiquons \'egalement les valeurs des 
parties r\'eelle et imaginaire du potentiel optique calcul\'ees \`a 12.4 fm, 
ainsi que les sections efficaces de r\'eaction calcul\'ees avec les potentiels 
qui conduisent \`a un ajustement optimum donn\'e par la valeur du param\`etre 
$\chi^2/n$.
%&&&&&&&&&&&&&&&&&&&&&&&&&&&&&&&&&&&&&&&&&&&&&&&&&&&&&&&&&&&&&&&&&&&&&&&&&&&&&
\begin{center}
\begin{tabular}{||l|c|c|c|c|c|c|r|c||}  \hline \hline
E${}_{lab}\;\;$ & N(E) & W${}_0$ & r${}_W$ & a${}_W$ & -V(12.4 fm) 
& -W(12.4 fm) & $\sigma_R\;\;$ & ${\chi^2}$/n \\ 
(MeV) & & (MeV) & (fm) & (fm) & (MeV) & (MeV) & (mb) &  \\ \hline \hline 
 78.00 & 1.726 & 28.350 & 1.2635 & 0.4206 & 2.859 & 0.456 &   40.30& 0.7\\ 
 80.00 & 1.479 & 29.590 & 1.2695 & 0.4408 & 2.450 & 0.641 &  109.30 & 1.2\\ 
 81.00 & 1.581 & 32.280 & 1.2666 & 0.4201 & 2.618 & 0.549 &  140.20 & 1.3\\ 
 82.00 & 1.635 & 29.809 & 1.3482 & 0.3077 & 2.708 & 1.063 &  205.50 & 0.4\\ 
 83.00 & 1.606 & 26.760 & 1.3428 & 0.3211 & 2.659 & 0.950 &  253.80 & 1.0\\ 
 85.95 & 1.414 & 27.436 & 1.3284 & 0.3847 & 2.342 & 1.214 &  451.90 & 1.0\\ 
 87.00 & 1.429 & 29.380 & 1.3251 & 0.3730 & 2.366 & 1.104 &  475.60 & 0.9\\ 
 88.00 & 1.380 & 25.476 & 1.3208 & 0.3681 & 2.285 & 0.835 &  495.80 & 1.3\\ 
 90.00 & 1.266 & 32.953 & 1.2924 & 0.4339 & 2.097 & 1.039 &  626.80 & 0.7\\ 
 94.00 & 1.112 & 35.300 & 1.2787 & 0.4840 & 1.842 & 1.245 &  842.30 & 1.7\\ 
 96.00 & 1.259 & 32.240 & 1.2900 & 0.4424 & 2.085 & 1.037 &  894.30 & 0.5\\ 
102.00 & 1.000 & 36.150 & 1.2286 & 0.5687 & 1.656 & 0.999 & 1153.00 & 0.7\\ 
104.00 & 1.311 & 32.800 & 1.3099 & 0.4088 & 2.171 & 1.197 & 1208.00 & 0.2\\ 
129.50 & 1.100 & 35.578 & 1.2370 & 0.6271 & 1.822 & 1.511 & 2079.00 & 0.6\\ 
138.50 & 0.958 & 40.203 & 1.2155 & 0.6029 & 1.586 & 1.132 & 2152.00 & 3.1\\ 
192.00 & 0.952 & 42.223 & 1.2098 & 0.6196 & 1.577 & 1.209 & 2893.00 & 1.1\\ 
216.60 & 0.584 & 38.092 & 1.1595 & 0.7363 & 0.967 & 1.071 & 3069.00 & 1.5\\ 
312.60 & 0.868 & 23.783 & 1.2178 & 0.6478 & 1.437 & 0.874 & 3419.00 & 0.1\\ 
\hline \hline 
\end{tabular}
\end{center}
\begin{center}
Tableau 4.2 : Param\`etres du mod\`ele optique semi-ph\'enom\'enologique 
pour le syst\`eme $^{16}O+^{208}Pb$ .
\end{center}
%&&&&&&&&&&&&&&&&&&&&&&&&&&&&&&&&&&&&&&&&&&&&&&&&&&&&&&&&&&&&&&&&&&&&&&&&&&&&&
Dans le chapitre 3 nous avons montr\'e que pour des \'energies proches de la 
barri\`ere de Coulomb, l'absorption est control\'ee par des processus 
p\'eriph\'eriques. Nous avons aussi montr\'e qu'il n'y a de sensibilit\'e que dans une r\'egion tr\`es 
\'etroite de la surface du potentiel. Pour cette raison nous avons 
consid\'er\'e, dans les analyses ph\'enom\'enologiques r\'ealis\'ees, une 
profondeur constante de 60 MeV pour la partie d'absorption du potentiel. 
Cependant, la philosophie adopt\'ee pour la pr\'esente analyse a \'et\'e 
diff\'erente car nous avons consid\'er\'e les param\`etres du potentiel 
imaginaire de la r\'ef\'erence {\bf (NAG)} qui utilise diff\'erentes 
profondeurs pour le terme d'absorption, et nous n'avons recherch\'e que la 
nouvelle renormalisation du potentiel r\'eel dans le but de corriger les 
petites diff\'erences dues aux densit\'es distinctes que nous avons utilis\'ees 
pour les noyaux cible et projectile dans le calcul du potentiel folding nu.

Deux raisons nous ont pouss\'es \`a utiliser cette proc\'edure : \\
En premier lieu, nous savons que le domaine de distance dans lequel se produit 
l'absorption augmente avec l'\'energie et que par cons\'equent, la sensibilit\'e 
aux d\'etails du potentiel aux petites distances doit cro\^{\i}tre. Notre 
hypoth\`ese consistant \`a maintenir constante la profondeur du terme 
d'absorption pourrait alors ne pas \^etre ad\'equate aux \'energies plus 
\'elev\'ees; cependant, dans les calculs inclus dans le tableau 3.6 nous avons 
maintenu constante la profondeur du terme imaginaire jusqu'\`a une \'energie 
de 192 MeV et nous avons obtenu des valeurs du facteur de renormalisation du 
potentiel r\'eel compatibles avec celles inclues dans le tableau 4.2.\\
En second lieu, nous d\'esirons reproduire de la forme la plus pr\'ecise 
possible les calculs qui ont permis d'obtenir l'\'evolution du potentiel avec 
l'\'energie indiqu\'ee dans la r\'ef\'erence {\bf (LI85)}, pour une distance 
de 12.4 fm adopt\'ee comme rayon de sensibilit\'e.
  
\vskip 20mm 

\section{\large {\bf D\'EPENDANCE DE L'\'ENERGIE DU POTENTIEL R\'EEL:}}

Comme il a \'et\'e indiqu\'e ant\'erieurement, la d\'ependance \'energ\'etique du potentiel r\'eel fut observ\'ee 
exp\'erimentalement {\bf (BA84)}, {\bf (LI85)}, {\bf (NA85)} et {\bf (DI89)}. 
Elle a \'et\'e l'objet de nombreux travaux th\'eoriques qui ont permis de bien la 
comprendre {\bf (MA86)}, {\bf (VI91)} et {\bf (PA91)}. Ce ph\'enom\`ene est 
d\^u au couplage des canaux non \'elastiques \`a la voie \'elastique et se 
manifeste par une augmentation du potentiel r\'eel lorsque diminue l'\'energie 
de la collision des deux noyaux en interaction. Cette augmentation du potentiel 
est d'autant plus grande que le couplage est intense. Elle est aussi 
directement li\'ee, au moyen d'une relation de dispersion, \`a la variation de 
l'absorption avec l'\'energie. Cette relation de dispersion, donn\'ee par 
l'\'equation 2.15 peut \^etre \'ecrite de la fa\c{c}on suivante :
\be	{\cal{R}}e \Delta V_{E_S}(r,E) = {E-E_S\over \pi} {\cal{P}}
	 \int_{E_0}^{\infty} {{\cal{I}}m \Delta V(r,{E'})\over 
	({E'}-E_S)({E'}-E)} d{E'}   \ee
o\`u $\cal{P}$ repr\'esente la valeur principale de l'int\'egrale et 
${\cal{I}}m \Delta V(r,{E'})$ le potentiel imaginaire que nous appellerons, 
par souci de simplicit\'e dans ce qui suit, $W(r,E')$. E$_0$ est l'\'energie 
seuil \`a laquelle le potentiel imaginaire s'annule, $E_S$ est l'\'energie de 
r\'ef\'erence et ${\cal{R}}e \Delta V_{E_S}(r,E)$ est donn\'e par :
\be	{\cal{R}}e \Delta V_{E_S}(r,E) = V(r,E) - V(r,E_S)	\ee
o\`u V(r,$E_S$) est la valeur du potentiel r\'eel total \`a l'\'energie de 
r\'ef\'erence. 

L'\'equation 4.3 peut \^etre \'evalu\'ee, pour chaque valeur de r, en faisant 
usage d'un mod\`ele lin\'eaire tr\`es simple {\bf (MA86)} dans lequel nous 
supposons que l'\'evolution de $W(r,E')$ avec l'\'energie peut \^etre 
repr\'esent\'ee par des segments de droites qui unissent ses valeurs \`a 
diff\'erentes \'energies. Ainsi, pour un rayon donn\'e, nous pouvons obtenir :
\be	{\cal{R}}e \Delta V_{E_S}(r,E) = {1 \over \pi} Ln  
	{{\ds \prod_{i=0}^{n-1}}\left| {E_{i+1}-E \over E_i-E} \right|^{W_i(E)}
	 \left| {1 \over E_n-E} \right|^{W_n} \over 
	{\ds \prod_{i=0}^{n-1}} \left| {E_{i+1} - E_S \over E_i - E_S} 
	\right|^{W_i({E_S})} \left| {1 \over E_n-E_S} \right|^{W_n}}	\ee
o\`u   
\be	W_i(E) = {W_{i+1} - W_i \over E_{i+1} - E_i} (E-E_i) + W_i	\ee
$E_n$ \'etant une valeur tr\`es grande de l'\'energie \`a laquelle nous 
supposons que le potentiel imaginaire $W_n$ est approximativement constant. La 
relation 4.5 nous permet d'obtenir, \`a chaque \'energie, l'augmentation du 
potentiel r\'eel par rapport \`a sa valeur \`a l'\'energie de r\'ef\'erence 
$E_S$. Le potentiel r\'eel total ainsi obtenu peut \^etre exprim\'e comme 
produit du potentiel nu par un facteur de renormalisation, N(E), 
d\'ependant de l'\'energie :
 \be V_{M3Y}+\Delta V(E)=N(E)V_{M3Y}\ee
Nos calculs ont \'et\'e r\'ealis\'es dans les conditions suivantes :

a)- Nous avons pris comme \'energie de r\'ef\'erence $E_S = 138.5 MeV$ car le 
facteur de renormalisation du potentiel r\'eel obtenu par ajustement aux 
donn\'ees exp\'erimentales a une valeur tr\`es proche de l'unit\'e.
  
b)- L'\'equation 4.5 a \'et\'e \'evalu\'ee au rayon de sensibilit\'e 
$R_S = 12.4 fm$ dont la valeur a \'et\'e prise des r\'ef\'erences {\bf (TH85)}, 
{\bf (LI85)} et {\bf (NA85)} dans lesquelles il a \'et\'e calcul\'e \`a partir 
de l'ajustement aux donn\'ees de diffusion \'elastique inclues dans le 
tableau 4.2.
  
c)- Nous avons choisi quatre segments qui connectent les valeurs du potentiel 
imaginaire $-W(r = 12.4 fm, E_{lab})$ = 0.0, 1.21, 0.84, 0.56 et 0.24 MeV  qui 
correspondent, respectivement, aux \'energies $E_{lab}$=74, 85.95, 400, 793 et 
1503 MeV. Les valeurs du potentiel imaginaire \`a ces \'energies ont \'et\'e 
calcul\'ees \`a partir des param\`etres optiques trouv\'es lors de 
l'ajustement aux donn\'ees de diffusion \'elastique inclus dans les 
tableaux 4.1 et 4.2.
  
Dans la partie inf\'erieure de la figure 4.2, nous comparons notre hypoth\`ese 
aux valeurs de $W(E)$ obtenues \`a 12.4 fm \`a partir des param\`etres 
optiques qui s'ajustent aux donn\'ees. La partie sup\'erieure de cette m\^eme 
figure indique, \`a la m\^eme distance, une comparaison similaire entre les 
valeurs du potentiel r\'eel total d\'eduit de l'ajustement aux donn\'ees et 
%&&&&&&&&&&&&&&&&&&&&&&&&&&&&&&&&&&&&&&&&&&&&&&&&&&&&&&&&&&&&&&&&&&&&&&&&&&&&&
\begin{figure}
\vspace {-3.0cm}
\begin{center}\mbox{\epsfig{file=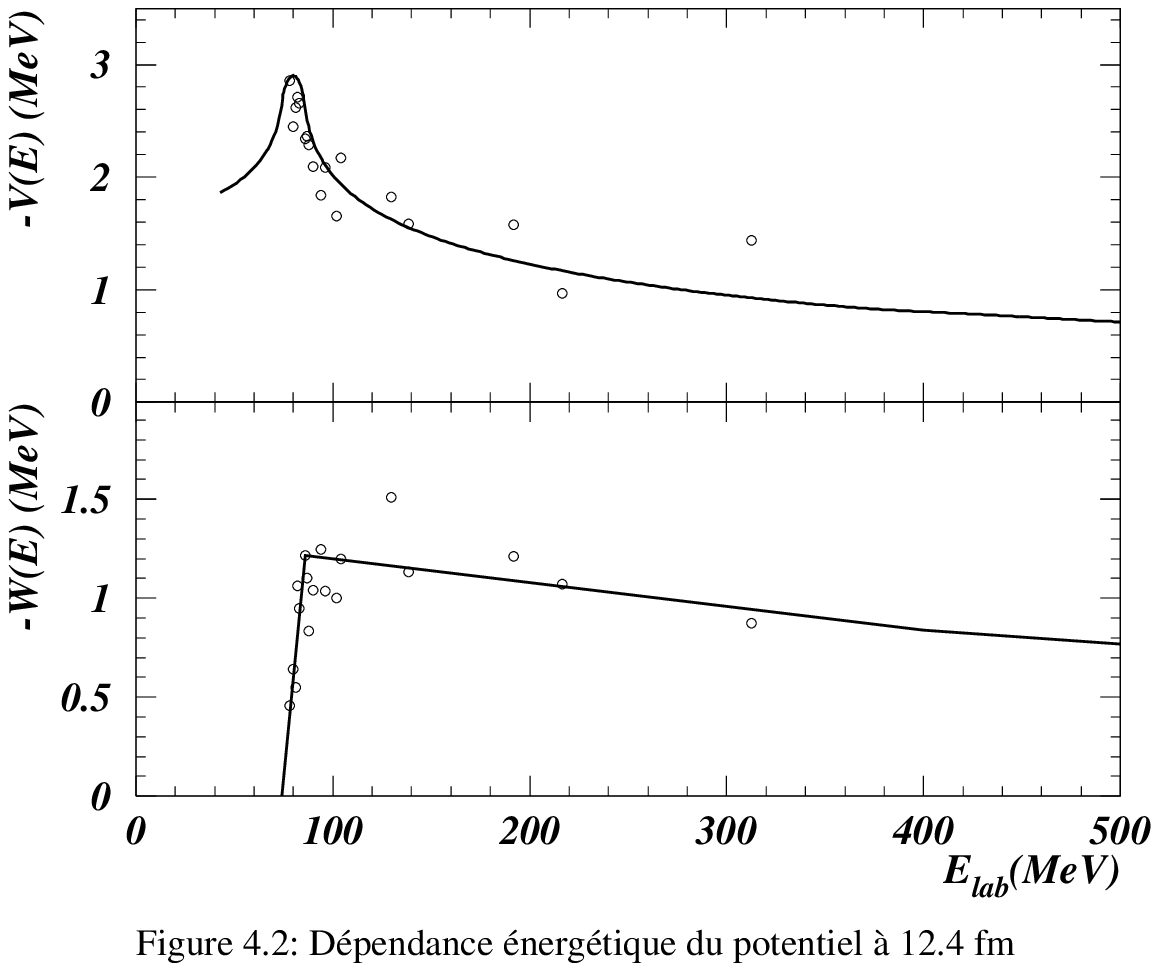}}\end{center}
\vspace {-0.8cm}
\end{figure}
%&&&&&&&&&&&&&&&&&&&&&&&&&&&&&&&&&&&&&&&&&&&&&&&&&&&&&&&&&&&&&&&&&&&&&&&&&&&&&
celui calcul\'e \`a partir du potentiel nu et du terme de polarisation donn\'e 
par l'\'equation 4.3 param\'etris\'ee de la forme indiqu\'ee par l'\'equation 
4.5. L'accord est excellent et tout \`a fait similaire \`a celui indiqu\'e dans 
la r\'ef\'erence {\bf (MA86)}.
%&&&&&&&&&&&&&&&&&&&&&&&&&&&&&&&&&&&&&&&&&&&&&&&&&&&&&&&&&&&&&&&&&&&&&&&&&&&&&
\begin{center}
\vskip 6mm
\begin{tabular}{||c|c|c|c|c|c|c|c||} \hline \hline
 $E_{lab}$  & N(E) &W& $R_W$ & $a_W$  &  -W(12.4fm) & $\sigma_R$ &
 $\chi^2/n$ \\
 (MeV) & &MeV& (fm) & (fm) & (MeV) &  (mb) & \\ \hline \hline
793 & 0.3572 & 42.2 & 1.083 & 0.755&  0.56 & 3600 & 0.78 \\
1503 & 0.1957 & 26.1 & 1.072 & 0.718&  0.24  & 3210 & 2.76 \\
\hline\hline
\end{tabular}
\vskip 6mm
Tableau 4.3 : Analyse semi-ph\'enom\'enologique.
\end{center}
%&&&&&&&&&&&&&&&&&&&&&&&&&&&&&&&&&&&&&&&&&&&&&&&&&&&&&&&&&&&&&&&&&&&&&&&&&&&&&
\vskip 4mm
Ainsi, en utilisant le pouvoir pr\'edictif de la relation de dispersion, nous 
avons calcul\'e le facteur de renormalisation du potentiel folding nu. Les 
valeurs obtenues \`a 793 et \`a 1503 MeV sont indiqu\'ees dans le tableau 4.3. 
On pourrait s'interroger sur la validit\'e, aux \'energies plus \'elev\'ees, de la 
relation de dispersion construite au rayon de sensibilit\'e obtenu pour des 
\'energies plus basses, car il est bien connu que le domaine de sensibilit\'e du 
potentiel imaginaire augmente avec l'\'energie, tandis que le rayon de 
sensibilit\'e diminue. Il faudrait esp\'erer que la valeur du rayon de sensibilit\'e 
soit inf\'erieure \`a 12.4 fm {\bf (RO86)} pour les \'energies plus \'elev\'ees et que 
le domaine de sensibilit\'e du potentiel soit diff\'erent de celui observ\'e aux 
\'energies plus basses  qui, nous le savons, est tr\`es \'etroit {\bf (PA95)}. 
Nous aurions pu construire la relation de dispersion \`a une valeur du rayon 
correspondant 
\newpage
\noindent
au rayon de sensibilit\'e obtenu pour les \'energies plus \'elev\'ees, 
mais une telle relation de dispersion n'aurait pas de sens physique. En effet, 
la d\'ependance \'energ\'etique du potentiel r\'eel est li\'ee \`a la d\'ependance 
\'energ\'etique du potentiel imaginaire qui varie tr\`es rapidement \`a proximit\'e 
de la barri\`ere de Coulomb o\`u le domaine de sensibilit\'e du potentiel 
imaginaire est tr\`es \'etroit et se trouve localis\'e autour de 12.4 fm. En 
consid\'erant, \`a ces \'energies, un rayon de sensibilit\'e plus petit que 
12.4 fm correspondant aux \'energies plus \'elev\'ees, nous nous situerions 
hors du domaine 
de sensibilit\'e o\`u se produit l'absorption et, par cons\'equent, la valeur du 
potentiel imaginaire n'aurait aucun sens physique. Bien que notre relation de 
dispersion ait \'et\'e bien construite pour les \'energies plus basses, la 
question demeure de savoir si le potentiel calcul\'e pour les \'energies plus hautes est un 
potentiel correct. Afin de tester sa validit\'e, nous avons calcul\'e les 
distributions angulaires de la diffusion \'elastique \opb \`a 793 et \`a 1503 
MeV, utilisant les param\`etres optiques inclus dans le tableau 4.3, dans 
lequel nous pouvons observer que nous avons maintenu la m\^eme partie 
imaginaire du potentiel que pour les calculs ph\'enom\'enologiques qui conduisaient 
au meilleur ajustement aux donn\'ees, tandis que nous avons remplac\'e les 
potentiels r\'eels ph\'enom\'enologiques, inclus dans le tableau 4.1, par ceux d\'eduits 
de la relation de dispersion. \'Etant donn\'e que dans ces calculs nous avons 
maintenu les m\^emes potentiels imaginaires que dans les calculs 
ph\'enom\'enologiques de r\'ef\'erence, ils constituent de fait une comparaison entre 
les potentiels r\'eels.

Dans le tableau 4.3 nous trouvons les valeurs des sections efficaces de 
r\'eaction calcul\'ees avec les potentiels indiqu\'es, ainsi que les valeurs du 
param\`etre $\chi^2/n$. Ces valeurs sont pratiquement identiques \`a celles 
obtenues avec les potentiels du tableau 4.1. Dans la figure 4.1 nous montrons 
les r\'esultats de ces calculs que nous ne pouvons pas distinguer des 
ajustements obtenus dans les calculs avec les potentiels ph\'enom\'enologiques 
du tableau 4.1. 
L'excellent accord obtenu nous permet d'affirmer que le potentiel r\'eel 
ph\'enom\'enologique et celui calcul\'e  avec la relation de dispersion sont 
totalement \'equivalents tant \`a 793 MeV qu'\`a 1503 MeV.
%&&&&&&&&&&&&&&&&&&&&&&&&&&&&&&&&&&&&&&&&&&&&&&&&&&&&&&&&&&&&&&&&&&&&&&&&&&&&&
\begin{figure}
\vspace {-1.0cm}
\begin{center}\mbox{\epsfig{file=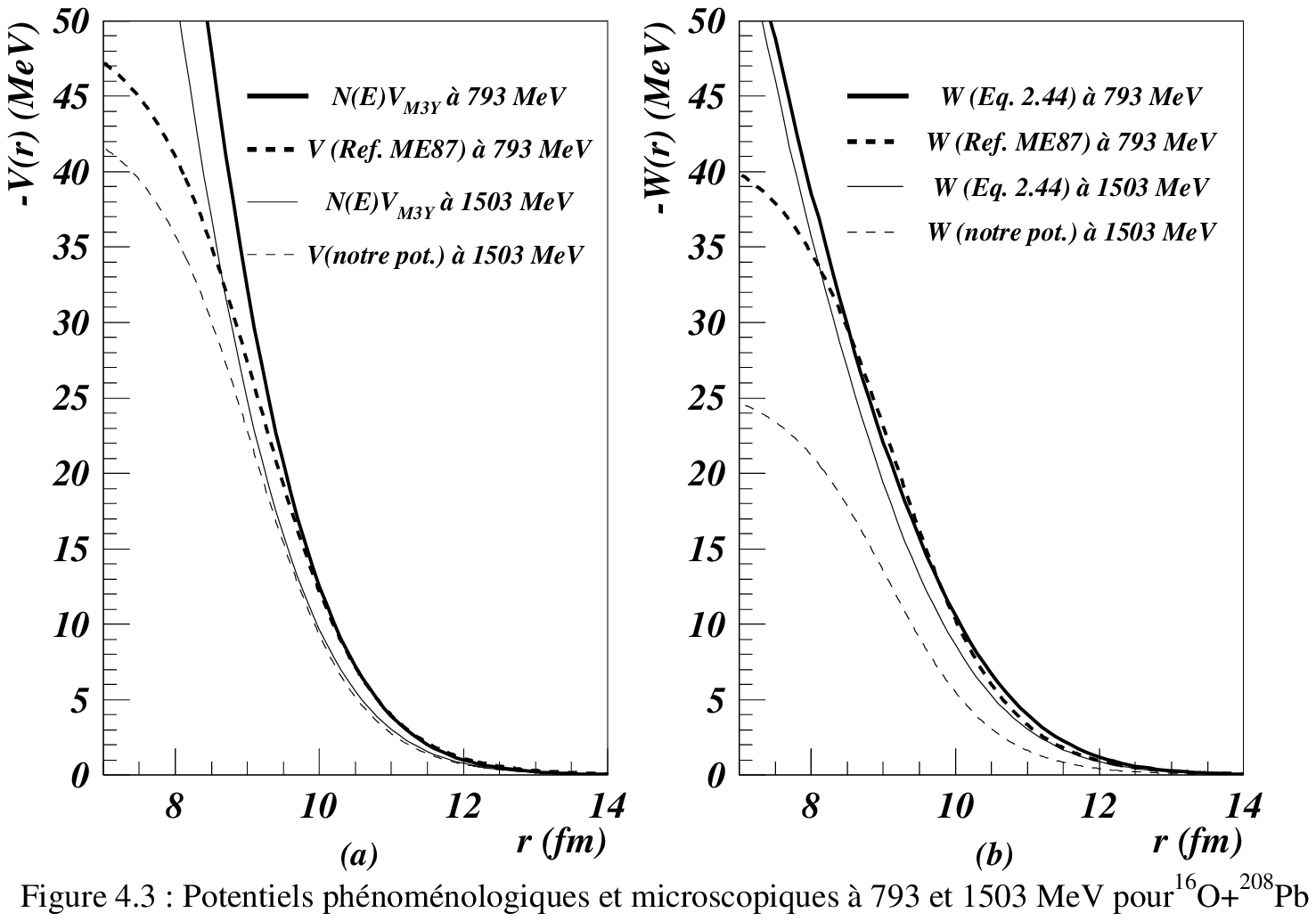}}\end{center}
\vspace {-0.6cm}
\end{figure}
%&&&&&&&&&&&&&&&&&&&&&&&&&&&&&&&&&&&&&&&&&&&&&&&&&&&&&&&&&&&&&&&&&&&&&&&&&&&&&
Dans la figure 4.3.a nous comparons le potentiel r\'eel ph\'enom\'enologique 
(ligne prononc\'ee discontinue) \`a celui d\'eduit de la relation de 
dispersion (ligne prononc\'ee continue) \`a 793 MeV et nous observons qu'ils 
co\"{\i}ncident \`a partir de 10 fm. De mani\`ere analogue, \`a 1503 MeV, 
l'accord entre le potentiel r\'eel ph\'enom\'enologique (ligne fine 
discontinue) et celui calcul\'e avec la relation de dispersion (ligne fine 
continue) est excellent \`a partir de 9.5 fm.

Tous ces r\'esultats nous permettent de conclure que, tant \`a 793 MeV qu'\`a 
1503 MeV :
  
1) Les potentiels r\'eels, ph\'enom\'enologique et calcul\'e avec les 
\'equations 4.3 et 4.5, sont totalement \'equivalents et conduisent \`a des 
r\'esultats pratiquement indiscernables.
  
2) \'Etant donn\'e que les deux potentiels r\'eels sont tr\`es diff\'erents aux 
petites distances et ont des valeurs tr\`es proches \`a partir d'environ 
10 fm \`a 793 MeV et 9.5 fm \`a 1503 MeV, nous pouvons affirmer que la r\'egion 
de sensibilit\'e du potentiel r\'eel est inclue dans ce domaine et sa valeur aux 
distances inf\'erieures n'a aucune incidence.

3) \'Etant donn\'e que le rayon de sensibilit\'e calcul\'e aux \'energies plus \'elev\'ees  
{\bf (RO85)} est situ\'e dans la r\'egion o\`u co\"{\i}ncident les potentiels r\'eels 
ph\'enom\'enologique et calcul\'e, nous pouvons dire que nous avons adopt\'e une 
mod\'elisation correcte de l'\'equation 4.3 et nous pouvons admettre que, malgr\'e 
leur grande diff\'erence aux distances inf\'erieures \`a 9.5 fm, les potentiels 
sont totalement \'equivalents.  
  
4) La d\'ependance \'energ\'etique observ\'ee dans le domaine de sensibilit\'e est la 
m\^eme pour les deux potentiels r\'eels.

Toutes ces conclusions justifient l'utilisation de potentiels r\'eels d\'eduits de 
la relation de dispersion pour les calculs microscopiques du potentiel 
imaginaire.

\vskip 20mm

\section{\large {\bf{{CALCULS MICROSCOPIQUES :}}}}

Les potentiels ph\'enom\'enologiques du tableau 4.1 nous ont permis de 
reproduire de fa\c{c}on tr\`es satisfaisante les distributions angulaires \opb 
\`a 793 MeV et \`a 1503 MeV, tout au moins dans l'intervalle angulaire o\`u ont 
\'et\'e r\'ealis\'ees les mesures. Ils nous ont \'egalement permis d'obtenir les valeurs 
correspondantes des sections efficaces de r\'eaction. Cependant, ils ne nous 
apportent aucune information sur les processus de r\'eaction qui ont une importance 
significative dans l'absorption totale, ni sur la r\'egion du potentiel, 
superficielle ou profonde, o\`u l'absorption se produit. C'est pour cette 
raison que nous utilisons des potentiels construits sur des bases 
microscopiques qui nous permettent de savoir quelles sont les voies 
d'absorption inclues dans les calculs et, par cons\'equent, quelle est leur 
importance dans l'absorption.

\subsection {Potentiel imaginaire}

Le calcul microscopique du potentiel imaginaire \`a 793 MeV et \`a 1503 MeV 
peut \^etre effectu\'e avec le mod\`ele de l'approximation de fermeture dont les 
hypoth\`eses de base ont \'et\'e rappel\'ees au chapitre 2. Pour l'\'evaluation de ces 
termes d'absorption nous utiliserons les potentiels r\'eels d\'eduits de la 
relation de dispersion et dont l'\'equivalence avec les potentiels 
ph\'enom\'enologiques a \'et\'e mise en \'evidence dans la section 
pr\'ec\'edente. Nous n'indiquons ici que les param\`etres que nous 
avons utilis\'es pour effectuer nos calculs {\bf(VI91)}:

1) Les \'energies moyennes des noyaux cible et projectile sont, respectivement, 
$<E_C>$ = 3.5 MeV pour $^{208}Pb$ et $<E_P>$ = 6.5 MeV pour $^{16}O$. Elles 
ont \'et\'e d\'etermin\'ees comme les \'energies moyennes des \'etats 
collectifs les plus bas du projectile et de la cible ($2^+$, $3^-$ pour 
$^{16}$O et $2^+$, $3^-$, $4^+$, $5^-$ pour $^{208}Pb$).

2) Les deux param\`etres d\'efinissant l'interaction effective sont 
$V_0$ = 58.7 MeV et $\mu$ = 0.45 $fm^{-2}$ correspondant \`a une port\'ee 
$r_0$ $\simeq$ 2 fm. Ces valeurs ont \'et\'e  d\'etermin\'ees \`a l'\'energie 
de 197 MeV, au rayon de sensibilit\'e (12.4 fm) {\bf(VI86)} et {\bf(VI91)}.  

Avec ces param\`etres et les densit\'es des noyaux cible et projectile dans leurs 
\'etats fondamentaux (Cf tableau 3.2 du chapitre 3), nous avons obtenu les 
potentiels imaginaires microscopiques indiqu\'es dans la figure 4.3.b que nous 
comparons aux potentiels ph\'enom\'enologiques de r\'ef\'erence du tableau 4.3. 
A 793 MeV, les deux potentiels, ph\'enom\'enologique (ligne prononc\'ee 
discontinue) et microscopique (ligne prononc\'ee continue) sont en accord pour 
des distances sup\'erieures \`a 8 fm et pr\'esentent une port\'ee similaire. 
Cependant, \`a 1503 
MeV, le comportement des deux potentiels est compl\`etement 
diff\'erent : le potentiel ph\'enom\'enologique (ligne fine discontinue) pr\'esente 
une port\'ee et une profondeur inf\'erieures \`a celles du potentiel microscopique 
(ligne fine continue). Le d\'esaccord est manifeste dans tout le domaine de 
distances. Dans la figure 4.3.b, nous pouvons observer que la d\'ependance 
\'energ\'etique existant entre les potentiels ph\'enom\'enologiques est diff\'erente 
de celle observ\'ee entre les potentiels microscopiques.

Voici, selon nous, une explication, tout au moins qualitative de ces 
diff\'erentes d\'ependances \'energ\'etiques ainsi que de la diff\'erence de port\'ees 
observ\'ee: Le mod\`ele de l'approximation de fermeture \'evalue la 
contribution au potentiel imaginaire due aux excitations du champ moyen. En 
particulier, les \'etats collectifs de plus basses \'energies d'excitation et les 
processus de transfert de nucl\'eons, simul\'es principalement \`a travers des 
\'etats de haute \'energie d'excitation, ont une contribution de longue port\'ee au 
potentiel d'absorption dont la d\'ependance \'energ\'etique est donn\'ee uniquement par 
celle inclue dans le propagateur WKB. Nous savons cependant que, lors des 
collisions \`a hautes \'energies, les processus de diffusion nucl\'eon-nucl\'eon 
gagnent de l'importance et contribuent de mani\`ere significative \`a la partie 
de volume du potentiel imaginaire, caract\'eris\'ee par une courte port\'ee, tandis 
que la d\'ependance \'energ\'etique du champ moyen combin\'ee aux effets de recul 
tend \`a diminuer l'absorption \`a cause des processus plus p\'eriph\'eriques 
tels que  les excitations in\'elastiques de surface et les processus de 
transferts de nucl\'eons 
{\bf (SO92)}. 
Dans cette r\'ef\'erence, il est indiqu\'e pour diff\'erents syst\`emes comment, y 
compris aux \'energies de l'ordre de 50 MeV/nucl\'eon, l'absorption reste domin\'ee 
par les processus plus p\'eriph\'eriques tandis que la contribution due \`a la 
diffusion nucl\'eon-nucl\'eon \`a grands angles n'est pas encore assez 
importante. Cependant, 
\`a 150 MeV/nucl\'eon, la situation s'inverse et les processus de 
diffusion nucl\'eon-nucl\'eon sont d\'ej\`a suffisamment importants pour 
contr\^oler 
l'absorption. Nous pouvons alors facilement comprendre pourquoi notre 
potentiel microscopique calcul\'e \`a 793 MeV (49.5 MeV/nucl\'eon) co\"{\i}ncide 
presque exactement avec le potentiel ph\'enom\'enologique de r\'ef\'erence \`a partir 
de 8 fm, exhibant tous les deux la m\^eme port\'ee, alors qu'\`a 1503 MeV (93.9 
MeV/nucl\'eon) les deux potentiels ont une port\'ee diff\'erente. En effet, \`a 793 
MeV, les canaux de diffusion nucl\'eon-nucl\'eon contribuent faiblement \`a 
l'absorption qui est domin\'ee par les contributions de longue port\'ee dues aux 
excitations des \'etats d'\'energies plus basses inclus dans nos calculs. 
Cependant, 
\`a 1503 MeV la contribution \`a l'absorption due aux canaux de 
diffusion nucl\'eon-nucl\'eon ouverts est tr\`es importante, tandis que celle due 
aux \'etats d'\'energie d'excitation plus basse l'est beaucoup moins. Cette 
situation conduit \`a un potentiel d'absorption de port\'ee moindre que celle du 
potentiel calcul\'e avec le mod\`ele de l'approximation de fermeture dans lequel 
\newpage
\noindent
le potentiel imaginaire est domin\'e par les termes de longue port\'ee qui 
proviennent de la contribution des \'etats collectifs les plus bas.

\subsection {Diffusion \'elastique \opb}

Utilisant le potentiel r\'eel de double convolution, dont la d\'ependance 
\'energ\'etique a \'et\'e calcul\'ee sur la base du pouvoir de pr\'ediction de la relation 
de dispersion, et le potentiel imaginaire obtenu de la forme d\'ecrite dans la 
section pr\'ec\'edente nous avons calcul\'e la distribution angulaire de la 
diffusion \'elastique \opb \`a 793 MeV. Dans la figure 4.1.a nous comparons la 
distribution angulaire exp\'erimentale aux calculs microscopiques (ligne 
continue), qu'il est impossible de distinguer des r\'esultats obtenus avec les 
potentiels ph\'enom\'enologiques et semi-ph\'enom\'enologiques. Dans le tableau 4.4 
sont indiqu\'ees les valeurs calcul\'ees de la section efficace totale de 
r\'eaction, $\sigma_R$, et du param\`etre $\chi^2/n$, qui s'av\`erent \^etre 
tr\`es proches des valeurs obtenues avec les potentiels ph\'enom\'enologiques 
(tableau 4.1) ou semi-ph\'enom\'enologiques (tableau 4.3).
%&&&&&&&&&&&&&&&&&&&&&&&&&&&&&&&&&&&&&&&&&&&&&&&&&&&&&&&&&&&&&&&&&&&&&&&&&&&&&
\begin{center}
\vskip 6mm
\begin{tabular}{||c|c|c|c||} \hline \hline
$E_{lab} (MeV)$  & N(E) & $\sigma_R (mb)$ & $\chi^2/n$ \\
\hline \hline
793.0 & 0.3572 & 3673 & 0.68 \\
1503.0 & 0.1957 & 3527 & 11.5 \\
\hline\hline
\end{tabular}
\vskip 6mm
Tableau 4.4 : Calculs microscopiques de diffusion \'elastique.
\end{center}
%&&&&&&&&&&&&&&&&&&&&&&&&&&&&&&&&&&&&&&&&&&&&&&&&&&&&&&&&&&&&&&&&&&&&&&&&&&&&&
\vskip 4mm
L'excellent accord observ\'e entre les calculs et les r\'esultats exp\'erimentaux, 
dans tout le domaine angulaire o\`u la diffusion angulaire a \'et\'e mesur\'ee, 
montre l'\'equivalence entre le potentiel imaginaire ph\'enom\'enologique de 
Mermaz (tableau 4.1) et celui calcul\'e microscopiquement. Cet accord, et la 
co\"{\i}ncidence observ\'ee pr\'ec\'edement entre les valeurs des deux potentiels \`a 
partir de 8 fm (figure 4.3.b), ont une double signification :
  
1) Le mod\`ele de l'approximation de fermeture inclut tous les canaux qui, \`a 
793 MeV contribuent de mani\`ere significative \`a l'absorption (qui d\'ebute \`a 
partir de 8 fm environ).

2) \'Etant donn\'e que les potentiels ph\'enom\'enologiques et microscopiques 
utilis\'es dans nos calculs conduisent \`a des r\'esultats indiscernables, 
malgr\'e les diff\'erences de  
profondeurs aux distances inf\'erieures \`a 
8 fm, l'absorption doit avoir lieu aux distances sup\'erieures \`a 8 fm; c'est 
\`a dire \`a la surface du potentiel.

 Ces conclusions peuvent \^etre mises en \'evidence de mani\`ere beaucoup 
plus apparente en calculant la distribution radiale de l'absorption comme 
\`a la section 3.2.2 du chapitre 3 et en comparant le 
domaine de distances o\`u elle prend des valeurs significatives \`a la 
distribution radiale du potentiel imaginaire responsable de cette absorption. 
%&&&&&&&&&&&&&&&&&&&&&&&&&&&&&&&&&&&&&&&&&&&&&&&&&&&&&&&&&&&&&&&&&&&&&&&&&&&&&
\begin{figure}
\vspace{-2.5cm}
\begin{center}\mbox{\epsfig{file=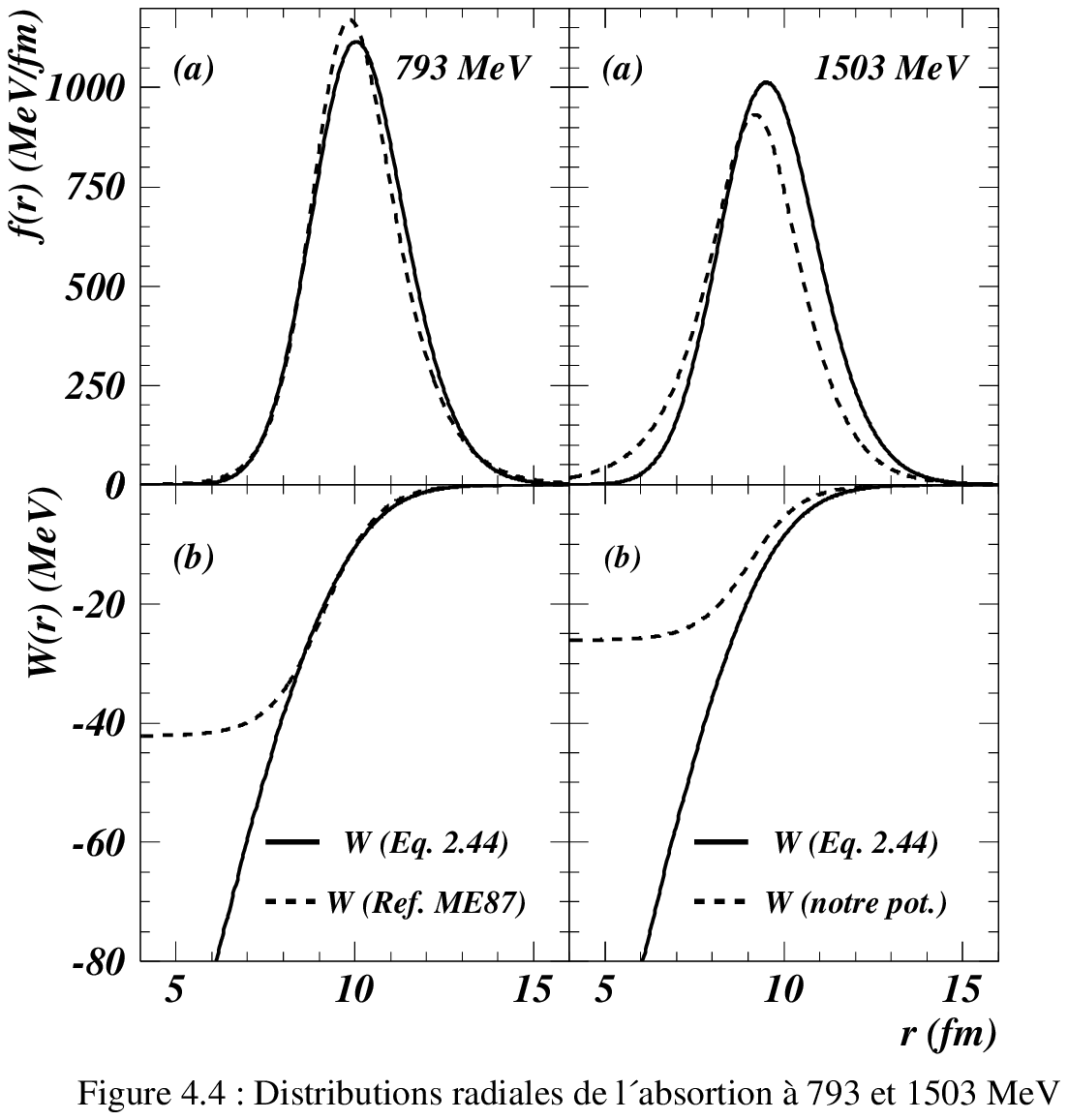}}\end{center}
\vspace{-0.8cm}
\end{figure}
%&&&&&&&&&&&&&&&&&&&&&&&&&&&&&&&&&&&&&&&&&&&&&&&&&&&&&&&&&&&&&&&&&&&&&&&&&&&&&
 
Dans la partie gauche de la figure 4.4.b sont pr\'esent\'es les potentiels 
imaginaires ph\'enom\'enologique (ligne discontinue) et microscopique (ligne 
continue) utilis\'es pour les calculs des distributions radiales de l'absorption \`a 793 MeV. 
Dans la partie gauche 
de la figure 4.4.a sont compar\'ees les distributions radiales de l'absorption 
calcul\'ees avec les potentiels imaginaires ph\'enom\'enologique
(ligne discontinue) et microscopique (ligne continue). La valeur de l'aire sous 
chaque courbe correspond \`a la section efficace totale de r\'eaction 
\`a chaque energie. Comme nous pouvons l'observer, l'accord entre 
les deux distributions radiales est excellent et les sections efficaces 
de r\'eaction correspondantes diff\`erent de moins de 2\%. Nous observons 
\'egalement que la majeure partie de l'absorption a lieu \`a des distances 
sup\'erieures \`a 8 fm.

Dans les figures 4.5.a-b-c nous comparons les valeurs de la fonction 
$(2l+1)|\chi_l(r)|^2$ calcul\'ees \`a 793 MeV avec notre potentiel d'absorption 
microscopique (ligne continue) et le potentiel ph\'enom\'enologique (ligne 
discontinue), pour diff\'erentes valeurs du moment angulaire $l$. L'accord 
observ\'e dans tous les cas est excellent. La consid\'eration simultan\'ee des 
figures 4.4.b (gauche) et 4.5.a-b-c nous facilite \'enorm\'ement la compr\'ehension 
de la forme de la distribution radiale de l'absorption dont la d\'efinition 
analytique est donn\'ee par l'\'equation 3.10.  

%&&&&&&&&&&&&&&&&&&&&&&&&&&&&&&&&&&&&&&&&&&&&&&&&&&&&&&&&&&&&&&&&&&&&&&&&&&&&&
\begin{figure}
\vspace{-25mm}
\begin{center}\mbox{\epsfig{file=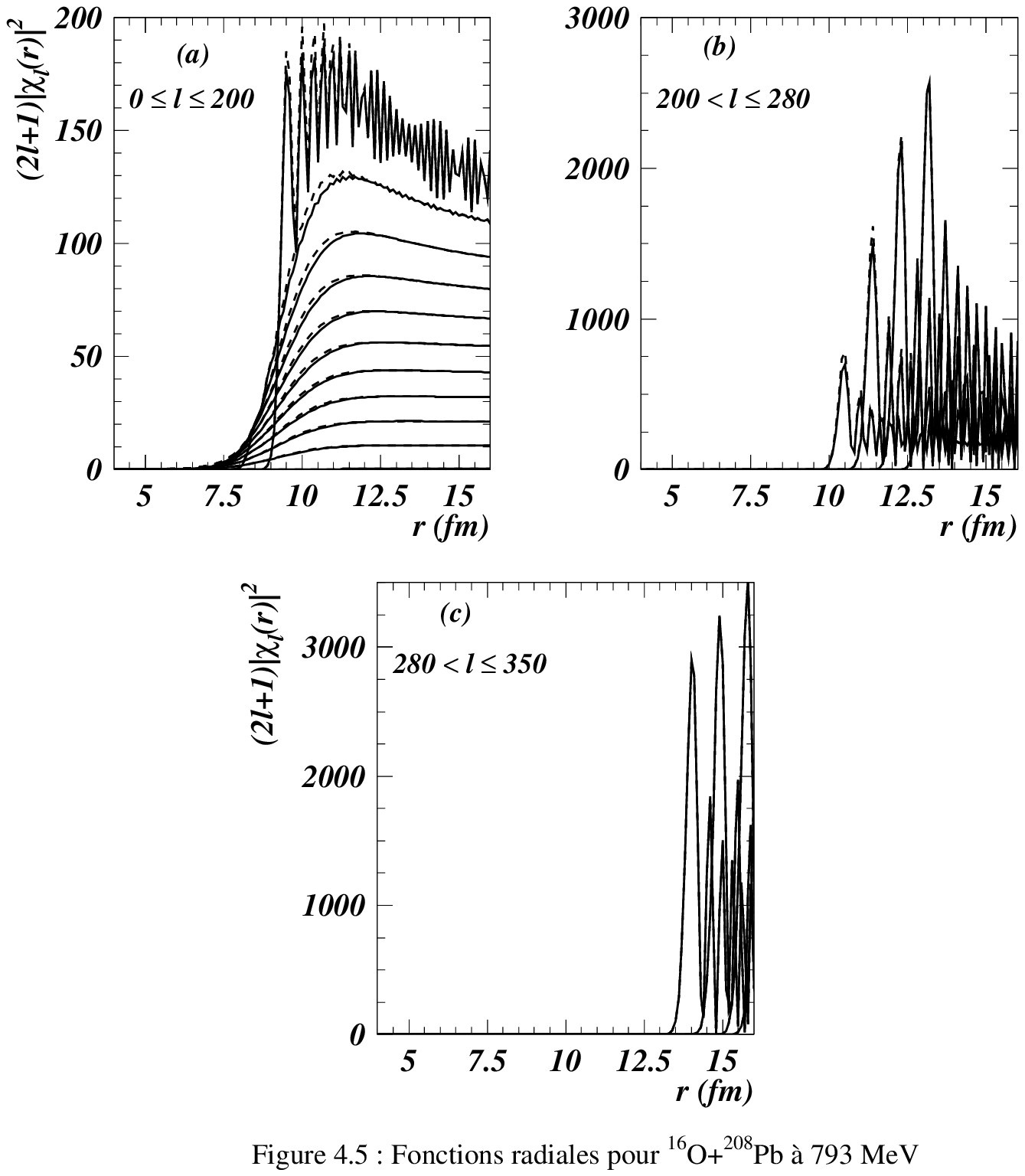}}\end{center}
\vspace{-0.8cm}
\end{figure}
%&&&&&&&&&&&&&&&&&&&&&&&&&&&&&&&&&&&&&&&&&&&&&&&&&&&&&&&&&&&&&&&&&&&&&&&&&&&&&
En effet, aux distances inf\'erieures 
\`a 6 fm  les valeurs n\'egligeables de la fonction $(2l+1)|\chi_l(r)|^2$ 
annihilent le potentiel imaginaire qui, \`a ces distances, est intense; par 
contre, aux distances sup\'erieures \`a 15 fm, c'est la valeur n\'egligeable du 
potentiel imaginaire qui annihile les grandes valeurs de la fonction 
$(2l+1)|\chi_l(r)|^2$. Ceci a pour effet que l'absorption se produit dans un 
domaine de distances o\`u les deux potentiels imaginaires co\"{\i}ncident.
Les calculs r\'ealis\'es \`a 1503 MeV conduisent \`a une situation compl\`etement 
diff\'erente de celle obtenue \`a 793 MeV. En effet, \`a partir du potentiel 
r\'eel d\'eduit de la relation de dispersion 
et du terme microscopique d'absorption 
calcul\'e avec le mod\`ele de l'approximation de fermeture, les calculs de 
diffusion \'elastique r\'ealis\'es \`a 1503 MeV permettent d'obtenir les valeurs de 
$\sigma_R$ et de $\chi^2/n$ que nous indiquons dans le tableau 4.4 et qui sont 
tr\`es diff\'erentes de celles obtenues dans les calculs r\'ealis\'es avec les 
potentiels ph\'enom\'enologiques ou semi-ph\'enom\'enologiques (Cf. tableau 4.1 et 4.3 
repectivement).

%&&&&&&&&&&&&&&&&&&&&&&&&&&&&&&&&&&&&&&&&&&&&&&&&&&&&&&&&&&&&&&&&&&&&&&&&&&&&&
\begin{figure}
\vspace{-25mm}
\begin{center}\mbox{\epsfig{file=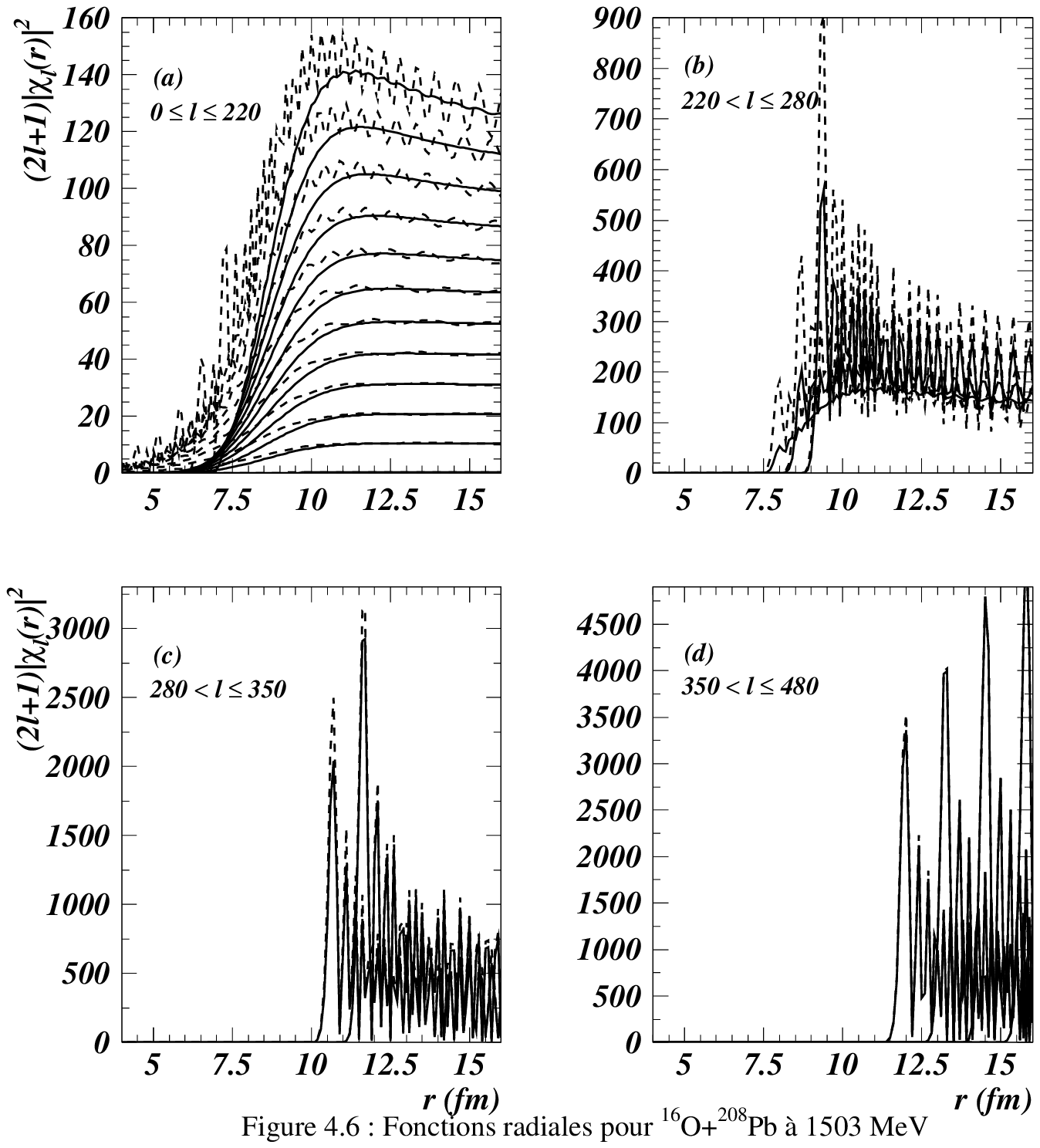}}\end{center}
\vspace{-0.8cm}
\end{figure}
%&&&&&&&&&&&&&&&&&&&&&&&&&&&&&&&&&&&&&&&&&&&&&&&&&&&&&&&&&&&&&&&&&&&&&&&&&&&&&
Dans la figure 4.1.b, nous comparons les mesures exp\'erimentales de diffusion 
\'elastique aux calculs microscopiques (ligne continue). L'accord observ\'e est 
faible; notre potentiel microscopique conduisant \`a une section efficace de 
r\'eaction qui diff\`ere d'environ 10 \% de celle calcul\'ee avec nos 
potentiels  
ph\'enom\'enologiques (tableau 4.1) 
ou semi-ph\'enom\'enologiques (tableau 4.3). 
Dans la partie droite de 
la figure 4.4.a, nous comparons les distributions radiales d'absorption calcul\'ees avec 
l'\'equation 3.10, en utilisant nos potentiels imaginaires ph\'enom\'enologique 
(ligne discontinue) ou microscopique (ligne continue). \'Etant donn\'e que 
notre potentiel imaginaire ph\'enom\'enologique a \'et\'e ajust\'e de 
mani\`ere \`a reproduire les donn\'ees de diffusion \'elastique, il doit 
necessairement inclure la contribution \`a l'absorption due \`a tous les canaux 
ouverts. Comme cel\`a a fait l'objet d'une discussion ant\'erieure; \`a cette 
\'energie, la diffusion nucl\'eon-nucl\'eon devient tr\`es importante conduisant \`a 
une forte contribution au terme \'elastique de volume de courte port\'ee, tandis 
que la contribution des \'etats d'\'energie d'excitation plus basse conduit \`a un 
faible potentiel imaginaire de surface. Lorsque de nouveaux canaux s'ouvrent, 
l'absorption se distribue entre tous les canaux ouverts et la forme du 
potentiel imaginaire doit, par cons\'equent, changer. Ainsi donc, notre 
potentiel imaginaire ph\'enom\'enologique doit avoir une port\'ee 
moyenne inf\'erieure \`a celle du potentiel microscopique et, ce dernier 
surestime les contributions des \'etats de moindre \'energie d'excitation, et ne 
prend pas en consid\'eration la contribution \`a l'absorption des processus de 
diffusion nucl\'eon-nucl\'eon. En effet, dans la partie droite de la 
figure 4.4.b, nous comparons les 
potentiels imaginaires ph\'enom\'enologique (ligne discontinue) et microscopique 
(ligne continu) et nous observons que les deux potentiels sont tr\`es 
diff\'erents dans le domaine de distances o\`u se produit l'absorption. Cette 
grande diff\'erence a pour cons\'equence que les fonctions d'onde radiales du 
mouvement relatif, $\chi_l(r)$, correspondant \`a chaque valeur $l$ du moment 
angulaire relatif, et calcul\'ees avec chacun de ces potentiels sont tr\`es 
diff\'erentes dans la r\'egion o\`u se produit l'absorption. Dans les 
figures 4.6.a-b-c-d sont repr\'esent\'ees les fonctions 
$(2l+1)|\chi_l(r)|^2$ calcul\'ees \`a 1503 MeV avec notre potentiel imaginaire 
ph\'enom\'enologique (lignes discontinues) et microscopique (lignes continues).
Au vu de toutes ces diff\'erences, il n'est pas surprenant qu'il y ait un 
d\'esaccord entre les calculs de diffusion \'elastique et les donn\'ees 
exp\'erimentales. Il n'est pas \'etonnant, non plus, que les sections efficaces de 
r\'eaction diff\`erent de 10\% et que les distributions radiales de l'absorption 
calcul\'ees avec nos potentiels imaginaires, ph\'enom\'enologique et microscopique 
soient diff\'erentes.

\newpage
$ $
\newpage
\chapter{{\Large{\bf{CONSID\'ERATIONS FINALES.}}}}

\section{\large{{\bf{R\'ESULTATS ET CONCLUSIONS}}}}
Le travail de recherche dont cette th\`ese a fait l'objet avait pour but 
l'\'etude des interactions ion-ion aussi bien aux \'energies proches de la 
barri\`ere de Coulomb o\`u les processus fortement collectifs contr\^olent 
l'absorption, qu'aux \'energies plus \'elev\'ees o\`u les ph\'enom\`enes non collectifs 
tels que la diffusion nucl\'eon-nucl\'eon entrent en comp\'etition avec les 
ph\'enom\`enes collectifs et arrivent m\^eme \`a dominer l'absorption. 
Tout ce travail, qui a fait l'objet de quatre publications internationales 
(Nucl. Phys. A586 (1995) 537; Conf\'erence publi\'ee dans la revue 
Nucleus n$^0$ 26 (1999) ISSN 0864-084X; Phys. Rev. C59 (1999) 1518; 
Phys. Rev. C60 (1999) 034612-1), a donn\'e lieu aux r\'esultats et conclusions 
suivants :

1)- Nous avons analys\'e les distributions angulaires de la diffusion \'elastique 
pour diff\'erents syst\`emes \`a des \'energies proches de la barri\`ere de 
Coulomb. Afin d'obtenir des conclusions globalement valides, nous avons selectionn\'e des 
syst\`emes tr\`es diff\'erents, tant du point de vue de leurs masses que de celui 
des caract\'eristiques des structures des deux noyaux en collision. Nous avons, 
ainsi, \'etudi\'e un large \'eventail de syst\`emes allant de ceux form\'es par deux 
noyaux sph\'eriques \`a couches ferm\'ees ($^{16}O$+$^{208}Pb$) \`a ceux 
fortement d\'eform\'es et \`a couches ouvertes ($^{32}S$+$^{24}Mg$), dont les \'etats 
li\'es de plus basses \'energies sont fortement excit\'es y compris aux 
\'energies tr\`es proches de la barri\`ere de Coulomb.

2)- Comme pr\'eliminaire de l'analyse des donn\'ees exp\'erimentales, nous 
avons ajust\'e les distributions angulaires de la diffusion \'elastique au moyen 
d'un potentiel optique dont la partie r\'eelle, convenablement renormalis\'ee, a 
\'et\'e calcul\'ee par double convolution de l'interaction effective M3Y avec les 
densit\'es des noyaux en interaction dans leurs \'etats fondamentaux. Quant au 
terme d'absorption, il a \'et\'e d\'ecrit par un potentiel du type Woods-Saxon dont 
les param\`etres ont \'et\'e soigneusement selectionn\'es de mani\`ere \`a optimiser 
l'accord entre les distributions angulaires mesur\'ees et celles calcul\'ees. 
Les r\'esultats obtenus lors de ces ajustements ont \'et\'e appel\'es 
semiph\'enom\'enologiques et ont \'et\'e adopt\'es comme r\'ef\'erences pour les calculs 
microscopiques post\'erieurs. En particulier, \'etant donn\'e que ces potentiels 
ont \'et\'e ajust\'es afin de reproduire, pour chaque cas, les donn\'ees 
exp\'erimentales, ils incluent tous les effets de structure, de d\'ependance 
de l'\'energie, de la densit\'e ou de l'inter-\'echange de nucl\'eons bien qu'il ne 
soit pas possible de savoir dans quelle proportion influe chaque effet. 
Cependant, les distributions radiales de l'absorption calcul\'ees au moyen de ces 
potentiels montrent sans ambig\"uit\'e aucune qu'aux \'energies proches de la 
barri\`ere de Coulomb l'absorption se produit dans un \'etroit domaine de 
distances de la surface du potentiel noyau-noyau et qu'elle est totalement 
insensible \`a la valeur du terme d'absorption \`a petites distances, 
bien qu'elle s'\'elargisse \`a 
mesure que l'\'energie de la collision augmente. Les sections efficaces de 
r\'eaction d\'eduites des ajustements aux donn\'ees exp\'erimentales se calculent \`a 
partir des distributions radiales de l'absorption et incluent tous les 
processus qui contribuent \`a l'absorption. Ils incluent, en particulier, la 
fusion.

3)- Utilisant les potentiels de convolution renormalis\'es d\'ecrits pr\'ec\'edemment 
et moyennant une mod\'elisation simple de la th\'eorie de Feshbach, nous avons 
calcul\'e la contribution \`a l'absorption due \`a un nombre r\'eduit d'\'etats 
collectifs de surface facilement excitables. \'Etant donn\'e que le propagateur 
qui d\'ecrit le mouvement relatif des deux noyaux en interaction est un 
propagateur complet, et vu que nous avons inclus tous les canaux d'inter\^et, 
notre potentiel imaginaire contient tous les processus de surface directs ou de 
multi-\'etapes qui contribuent de mani\`ere significative \`a l'absorption. 
Pour les syst\`emes plus fortement d\'eform\'es, ($^{32}S+^{24}Mg$, 
$^{35}Cl+^{24}Mg$ et $^{37}Cl+^{24}Mg$) les pr\'edictions obtenues avec les 
potentiels microscopiques reproduisent de mani\`ere ad\'equate les donn\'ees 
exp\'erimentales et sont en total accord avec les r\'esultats obtenus pr\'ec\'edemment 
avec les potentiels semiph\'enom\'enologiques. En particulier, nos calculs 
montrent que l'absorption totale, qui inclut la fusion, peut \^etre reproduite 
avec seulement la contribution au terme d'absorption d'un nombre r\'eduit d'\'etats 
collectifs de surface. Notre hypoth\`ese consistant \`a n'inclure que les 
\'etats vibrationnels de plus basse \'energie et \`a approximer les facteurs 
de forme 
par des termes de surface implique que les voies de volume du noyau compos\'e 
ne 
contribuent pas de mani\`ere significative \`a l'absorption vu que nous n'avons 
pris en consid\'eration que les \'etats de surface. 
Notre mod\'elisation conduit \`a la fusion \`a travers des processus de 
multi-\'etapes. Pour les autres syst\`emes \'etudi\'es, la contribution \`a 
l'absorption due aux 
processus de transfert d'un ou de plusieurs nucl\'eons peut parfois \^etre 
comparable \`a celle due aux processus in\'elastiques. L'inclusion d'un terme 
d'absorption ph\'enom\'enologique pour d\'ecrire des processus plus p\'eriph\'eriques que 
les excitations in\'elastiques, nous a permis de reproduire de mani\`ere 
satisfaisante les distributions angulaires exp\'erimentales de la diffusion 
\'elastique. En particulier, l'augmentation de la section efficace obtenue en 
incluant le terme en question, par rapport \`a celle que l'on obtenait en ne 
consid\'erant que l'absorption due aux voies in\'elastiques, est compatible avec la 
valeur de la section efficace de transfert pour les syst\`emes o\`u cette 
derni\`ere a \'et\'e mesur\'ee.

4)- L'utilisation de potentiels ph\'enom\'enologiques d\'ecrits dans la litt\'erature 
nous a permis de reproduire de fa\c{c}on tr\`es pr\'ecise les diffusions 
\'elastiques $^{16}O+^{208}Pb$ aux \'energies de 793 MeV et 1503 MeV, et de 
d\'eterminer la section efficace de r\'eaction \`a ces \'energies. Pour les raisons 
signal\'ees pr\'ecedemment, ces potentiels, ainsi que les calculs de mod\`ele 
optique effectu\'es avec, seront utilis\'es comme r\'ef\'erence des calculs 
semiph\'enom\'enologiques et microscopiques post\'erieurs.

5)- A partir de potentiels imaginaires du type Woods-Saxon obtenus par 
ajustement aux donn\'ees de la diffusion \'elastique $^{16}O+^{208}Pb$ aux \'energies 
comprises entre 78 MeV et 312.6 MeV, et moyennant une mod\'elisation simple de 
la relation de dispersion {\bf(MA86)}, nous avons utilis\'e le pouvoir de 
pr\'ediction de 
cette derni\`ere pour d\'eterminer la d\'ependance \'energ\'etique du potentiel 
r\'eel \`a 793 MeV et \`a 1503 MeV. La validit\'e de la relation de dispersion 
et l'\'equivalence entre ces 
potentiels et les potentiels 
ph\'enom\'enologiques ont \'et\'e clairement mises en \'evidence. En effet, en 
rempla\c{c}ant les potentiels r\'eels ph\'enom\'enologiques par ceux extraits de la 
relation de dispersion, les calculs de mod\`ele optique nous permettent de 
reproduire les donn\'ees avec une qualit\'e similaire et d'obtenir la m\^eme valeur 
de la section efficace de r\'eaction. Bien que les deux types de potentiel 
aient 
des profondeurs tr\`es diff\'erentes ils pr\'esentent la m\^eme d\'ependance 
\'energ\'etique et co\"{\i}ncident \`a partir de 10 $fm$ \`a 793 MeV et de 9.5 $fm$ 
\`a 1503 MeV. Ceci indique qu'aux petites distances, la valeur du potentiel 
r\'eel n'est pas significative.

6)- En utilisant les potentiels r\'eels extraits de la relation de dispersion et le 
mod\`ele propos\'e par N. Vinh Mau, nous avons calcul\'e le terme d'absorption du 
potentiel. \'Etant donn\'e que le mod\`ele \'evalue de fa\c{c}on globale la 
contribution \`a l'absorption due aux voies non-\'elastiques, l'analyse de la 
diffusion \'elastique $^{16}O+^{208}Pb$ \`a 793 MeV et \`a 1503 MeV peut nous 
indiquer si, \`a ces \'energies, l'absorption reste encore contr\^ol\'ee par les 
processus collectifs inclus dans le mod\`ele et, par cons\'equent, si le mod\`ele 
est toujours valide \`a ces \'energies relativement \'elev\'ees. 
Nos calculs \`a 793 MeV conduisent \`a des r\'esultats identiques \`a ceux obtenus 
avec les potentiels de r\'ef\'erence. Nous observons un excellent accord entre les 
potentiels microscopique et ph\'enom\'enologique \`a partir de 8 $fm$ qui est la 
valeur o\`u d\'ebute la r\'egion dans laquelle se produit l'absorption. Par contre, 
\`a 1503 MeV, le potentiel ph\'enom\'enologique a une port\'ee plus courte que le 
microscopique. Ceci conduit \`a des calculs de mod\`ele optique qui ne 
reproduisent pas convenablement les donn\'ees exp\'erimentales. Ce d\'esaccord n'est 
pas surprenant en soi \'etant donn\'e que le mod\`ele n'inclut pas les processus de 
diffusion nucl\'eon-nucl\'eon qui conduisent \`a une absorption de volume dont 
l'importance cro\^{\i}t avec l'\'energie.\\

\section{\large{{\bf{PERSPECTIVES FUTURES.}}}}

Les r\'esultats obtenus dans notre analyse microscopique aux \'energies proches de 
la barri\`ere de Coulomb ouvrent diff\'erentes possibilit\'es pour continuer ce 
travail de recherche.

En premier lieu, \'etant donn\'e qu'aux \'energies tr\`es proches de la barri\`ere de 
Coulomb les effets de polarisation coulombienne repr\'esentent une partie de 
l'absorption comparable \`a celle d'origine nucl\'eaire, il est necessaire 
d'inclure ces effets dans la mod\'elisation de la th\'eorie de Feshbach du 
potentiel optique que nous avons construite. Ceci peut se faire en calculant 
les \'el\'ements de matrice avec des facteurs de forme d\'ecrivant les deux 
contributions dues aux polarisations d'origine coulombienne et nucl\'eaire.

En deuxi\`eme lieu, et pour maintenir la coh\'erence des calculs, il serait 
interessant d'introduire comme termes de Feshbach, les contributions \`a 
l'absorption dues aux processus de transferts les plus significatifs.

En troisi\`eme lieu, vu que notre mod\'elisation permet d'obtenir non seulement la 
contribution totale \`a l'absorption, mais aussi celle due \`a chacune des voies 
incluses dans le calcul, il est possible de d\'eterminer aussi bien la section 
efficace totale de r\'eaction qu'une distribution \`a travers les diff\'erents 
canaux. Par ailleurs si, tel qu'il a \'et\'e mis en \'evidence dans un travail 
ant\'erieur sur la fusion sous-coulombienne {\bf(VI93)}, ce sont les processus 
multi-\'etapes qui 
conduisent \`a la fusion, il est possible de calculer s\'epar\'ement les 
contributions \`a l'absorption dans chaque canal dues aux processus directs et 
de multi-\'etapes et, par cons\'equent, calculer la contribution \`a la 
section efficace de fusion correspondant \`a chaque voie de r\'eaction.

Enfin, il faudrait signaler que les r\'esultats obtenus lors de l'analyse de la 
diffusion $^{16}O+^{208}Pb$ avec le mod\`ele propos\'e par N. Vinh Mau \`a des 
\'energies allant de 78 MeV \`a, au moins, 793 MeV, ainsi que la capacit\'e 
de pr\'ediction du mod\`ele pour les calculs de fusion sous-coulombienne, nous 
permettent de conclure que dans la perspective de calculs futurs, ils 
pourraient \^etre adopt\'es comme calculs-mod\`eles de r\'ef\'erence, 
comme nous l'avons fait auparavant avec les calculs ph\'enom\'enologiques 
ou semiph\'enom\'enologiques.

\newpage

\newpage
$ $
\newpage
{\bf{\LARGE { R\'ef\'erences :}}}
\vskip 10mm

\noindent{\bf (AN90)} $\> \> \>$ M.A. Andr\'es, F. Catara et F.G. Lanza,\\
\hspace*{20mm} Inst. Conf. Ser. 110 (1990) 231;\\
\hspace*{20mm} Phys. Rev. C44 (1991) 2709. 
\vskip 1mm
\noindent{\bf (AZ85)} $\> \> \>$ M. El Azab Farid et G.R. Satchler, \\
\hspace*{20mm} Nucl. Phys. A438 (1985) 525.
\vskip 1mm
\noindent{\bf (BA84)} $\> \> \>$ A. Baeza, B. Bilwes, R. Bilwes, J. Diaz et 
J.L. Ferrero, \\
\hspace*{20mm} Nucl. Phys. A419 (1984) 412.
\vskip 1mm
\noindent{\bf (BA91)} $\> \> \>$ 	J.M. Barrigon, \\
\hspace*{20mm} Th\`ese de doctorat de l'universit\'e de Extremadura, Espagne 
1991. 
\vskip 1mm
\noindent{\bf (BA92)} $\> \> \>$ J.M. Barrigon, A. Baeza, J.L. Ferrero, J.C. 
Pacheco B. Bilwes et R. Bilwes, \\
\hspace*{20mm} Nucl. Phys. A545 (1992) 720. 
\vskip 1mm
\noindent{\bf (BE77)} $\> \> \>$ G. Bertsch, J. Borisowicz, H. Mc Manus et 
W.G. Love, \\
\hspace*{20mm} Nucl. Phys. A284 (1977) 399. 
\vskip 1mm
\noindent{\bf (BI83)} $ \> \> \> \>$ 	B. Bilwes, R. Bilwes, V. D'Amico, 
J.L. Ferrero, G. Giardina et R. Potenza, \\
\hspace*{20mm} Nucl. Phys. A408 (1983) 173. 
\vskip 1mm
\noindent{\bf (BI86)} $\> \> \> \>$ 	R. Bilwes et B. Bilwes, \\
\hspace*{20mm} Rapport C.R.N. (1986) 13. 
\vskip 1mm
\noindent{\bf (BI87)} $\> \> \>$ 	B. Bilwes, R. Bilwes, F. Ballester, 
J. Diaz, J.L. Ferrero, C. Roldan, \\
\hspace*{20mm} L. Stuttg\'e et F. Sanchez, \\
\hspace*{20mm} Nucl. Phys. A473 (1987) 353. 
\vskip 1mm
\noindent{\bf (BI91)} $\> \> \>$ 	B. Bilwes, R. Bilwes, N. Vinh Mau, 
J.L. Ferrero et J.C. Pacheco, \\
\hspace*{20mm} Nucl. Phys. A526 (1991) 292
\vskip 1mm
\noindent{\bf (BO75)} $\> \> \>$ 	A. Bohr et B.R. Mottelson, \\
\hspace*{20mm} Nuclear Structure, W.A. Benjamin, INC. New York (1975). 
\vskip 1mm
\noindent{\bf (BO81)} $\> \> \>$ A. Bouyssy, N. Vinh Mau et D.M. Brink, \\
\hspace*{20mm} Phys. Lett. 102B (1981) 109. 
\vskip 1mm
\noindent{\bf (BOR75)} $\> \>$  F. Borkowski et G.C. Li, \\
\hspace*{20mm} Z. Phys. (1975) 29 
\vskip 1mm
\noindent{\bf (BOU81)} $ \> \>$  A. Bouyssy, H. Ng\^o et N. Vinh Mau, \\
\hspace*{20mm} Nucl. Phys. A371 (1981) 173. 
\vskip 1mm
\noindent{\bf (BR59)} $\> \> \>$ 	G.E. Brown et M. Bosterli, \\
\hspace*{20mm} Phys. Rev. Lett. 3 (1959) 472. 
\vskip 1mm
\noindent{\bf (BR77)} $\> \> \>$ 	D.M. Brink et N. Tagikawa, \\
\hspace*{20mm} Nucl. Phys. A279 (1977) 159. 
\vskip 1mm
\noindent{\bf (BR80)} $\> \> \>$ 	W.J. Briscoe, H. Crannell et J.C. 
Bergstrom, \\
\hspace*{20mm} Nucl. Phys. A344 (1980) 475. 
\vskip 1mm
\noindent{\bf (BR81)} $\> \> \>$ 	R.A. Broglia et A. Winther, \\
\hspace*{20mm} Heavy ion reactions, The Benjaming Cummings Publ. Co. Read-\\
\hspace*{20mm} ing. Mass. (1981). 
\vskip 1mm
\noindent{\bf (BR85)} $\> \> \>$ 	M. Brack, C. Guet et R.B. Hakansson, \\
\hspace*{20mm} Phys. Rep. 123 ( 1985) 274. 
\vskip 1mm
\noindent{\bf (BRO81)} $ \> \>$  R.A. Broglia, G. Pollarolo et A. Winter, \\
\hspace*{20mm} Nucl. Phys. A361 (1981) 307 
\vskip 1mm
\noindent{\bf (CO82)} $\> \> \>$ 	J. Cook, \\
\hspace*{20mm} Comput. Phys. Commun. 25 (1982) 125. 
\vskip 1mm
\noindent{\bf (DI89)} $\> \> \> \>$ J. Diaz, J.L. Ferrero, J.A. Ruiz, 
B. Bilwes et R. Bilwes, \\
\hspace*{20mm} Nucl. Phys. A494 (1989) 311. 
\vskip 1mm
\noindent{\bf (EL85)} $\> \> \>$ El Itaoui, J.P. Ellis et B.A. Mughrabi,\\
\hspace*{20mm} Nucl. Phys. A441 (1985) 511. 
\vskip 1mm
\noindent{\bf (EN78)} $ \> \>$  P.M. Endt et C. Van der Leun,\\
\hspace*{20mm} Nucl. Phys. A310 (1978) 1 
\vskip 1mm
\noindent{\bf (ER72)} $\> \> \>$ 	R. Erwein, \\
\hspace*{20mm} M\'emoire d'ing\'enieur, CNAM, Mulhouse, (1972). 
\vskip 1mm
\noindent{\bf (FA83)} $\> \> \>$  	A. Faessler, L. Rikus et S.B. 
Khadkikar, \\
\hspace*{20mm} Nucl. Phys. A401 (1983) 157. 
\vskip 1mm
\noindent{\bf (FE58)} $\> \> \>$ 	H. Feshbach, \\
\hspace*{20mm} Ann. of Phys. 5 (1958) 357. 
\vskip 1mm
\noindent{\bf (FE62)} $\> \> \>$ 	H. Feshbach, \\
\hspace*{20mm} Ann. of Phys. 19 (1962) 287. 
\vskip 1mm
\noindent{\bf (FE90)} $\> \> \>$ 	J.L Ferrero, J.C. Pacheco, A. Baeza, 
J.M. Barrigon, B. Bilwes, R. Bilwes \\
\hspace*{20mm} et N.Vinh Mau, \\
\hspace*{20mm} Nucl. Phys. A514 (1990) 367. 
\newpage
\noindent{\bf (FE92)} $\> \> \>$ 	H. Feshbach, \\
\hspace*{20mm} Theoretical nuclear physics : Nuclear reactions, John Wiley 
and\\ 
\hspace*{20mm} Sons, Inc.  U.S.A. (1992). 
\vskip 1mm
\noindent{\bf (FR68)} $\> \> \>$ 	R.F. Frosch, R. Hofstadter, J.S. Mc 
Carthy, G.K. Noldeke, K.J. Van Oostrum,\\
\hspace*{20mm} B.C. Clark, R. Herman et D.G. 
Ravenhall,\\
\hspace*{20mm} Phys. Rev. 174 (1968) 1380. 
\vskip 1mm
\noindent{\bf (GO76)} $\> \> \>$ 	M. Golin, F. Petrovich et D. Robson, \\
\hspace*{20mm} Phys. Lett. 64B (1976) 253. 
\vskip 1mm
\noindent{\bf (GR68)} $\> \> \>$ G.W. Greenless, G.J. Pyle et Y.C. Tang, \\
\hspace*{20mm} Phys. Rev. 171 (1968) 1115. 
\vskip 1mm
\noindent{\bf (HE56)} $\> \> \>$ 	R.H. Helm, \\
\hspace*{20mm} Phys. Rev. 104 (1956) 1466. 
\vskip 1mm
\noindent{\bf (HO63)} $\> \> \>$ 	P.E. Hodgson, \\
\hspace*{20mm} The optical model of elastic scattering, Oxford University \\
\hspace*{20mm} Press, (1963). 
\vskip 1mm
\noindent{\bf (HO80)} $\> \> \>$ 	H. Horinchi, \\
\hspace*{20mm} Prog. Theor. Phys. 64 (1980) 184. 
\vskip 1mm
\noindent{\bf (HU81)} $\> \> \>$ 	B. Humbert, \\
\hspace*{20mm} Th\`ese de doctorat de l'universit\'e L.Pasteur, Strasbourg,\\
\hspace*{20mm} France (1981). 
\vskip 1mm
\noindent{\bf (JA70)} $\> \> \>$  	D.F. Jackson, \\ 
\hspace*{20mm} Nuclear reactions, Methuen and Co LTD, London, 1970. 
\vskip 1mm
\noindent{\bf (JA74)} $\> \> \>$  	C.W. De Jaeger et H. De Vries, \\
\hspace*{20mm} Atomic Data Tables 14 (1974) 479. 
\vskip 1mm
\noindent{\bf (KA73)} $\> \> \>$ 	D. Kalinsky, \\
\hspace*{20mm} Preprint 16/928-6258, The Weizmann Institute of Science (1973). 
\vskip 1mm
\noindent{\bf (KA75)} $\> \> \>$ 	D. Kalinsky, D. Melnic, U. Smilansky, 
N. Trautner, B.A. Watson,\\
\hspace*{20mm} Y. Horowitz, S. Mordechai, G. Baur et D. Pelt, \\
\hspace*{20mm} Nucl. Phys. A250 (1975) 364. 
\vskip 1mm
\noindent{\bf (KA77)} $\> \> \>$ 	D. Kalinsky, D. Melnic, U. Smilansky, 
N. Trautner, Y. Horowitz et S.Mordechai, \\ 
\hspace*{20mm} Nucl. Phys. A289 (1977) 205. 
\newpage
\noindent{\bf (KH81)} $\> \> \>$ 	S.B. Khadkikar, L. Rikus, A. 
Faessler et R. Sartor, \\
\hspace*{20mm} Nucl. Phys. A369 (1981) 495. 
\vskip 1mm
\noindent{\bf (KO84)} $\> \> \>$ A.M. Kobos, B.A. Brown, R. Lindsay et 
G.R. Satchler, \\
\hspace*{20mm} Nucl. Phys. A425 (1984)	233. 
\vskip 1mm
\noindent{\bf (KU81)} $\> \> \>$ 	K.I. Kubo et P.E. Hodgson, \\
\hspace*{20mm} Nucl. Phys. A366 (1981)	320. 
\vskip 1mm
\noindent{\bf (KU91)} $\> \> \>$ 	K.I. Kubo, \\
\hspace*{20mm} Nucl. Phys. A534 (1991)	381. 
\vskip 1mm
\noindent{\bf (LI)  } $\> \> \> \> \> \> \> \>$ I. Linck, \\
\hspace*{20mm} Communication priv\'ee.
\vskip 1mm
\noindent{\bf (LIL)  } $\> \> \> \> \> \> \> \>$ J. Lilley, \\
\hspace*{20mm} Communication priv\'ee.
\vskip 1mm
\noindent{\bf (LI74)} $\> \> \> \>$ 	G.C. Li et M.R. Yerian, \\
\hspace*{20mm} Phys. Rev. C9 (1974) 1861. 
\vskip 1mm
\noindent{\bf (LI85)} $ \> \> \> \>$ J.S. Lilley, B.R. Fulton, M.A. Nagarajan, 
J.J. Thompson et D.W. Banes, \\
\hspace*{20mm} Phys. Lett. B151 (1985) 181. 
\vskip 1mm
\noindent{\bf (LO75)} $\> \> \>$ 	W.G. Love, \\
\hspace*{20mm} Nucl. Phys. A239 (1975) 74. 
\vskip 1mm\noindent{\bf (LO77)} $\> \> \>$ W.G. Love, T. Teresawa et
 G.R. Satchler\\
\hspace*{20mm} Nucl. Phys. A291 (1977) 183. 
\vskip 1mm
\noindent{\bf (MA86)} $\> \> \>$ C. Mahaux, H. Ng\^o et G.R. Satchler, \\
\hspace*{20mm} Nucl. Phys. A449 (1986) 354;\\
\hspace*{20mm} Nucl. Phys. A456 (1986) 134. 
\vskip 1mm
\noindent{\bf (ME87)} $\> \> \>$ M.C. Mermaz et al., \\
\hspace*{20mm} Z. Phys. A326 (1987) 353.
\vskip 1mm
\noindent{\bf (NAG)} $\> \> \>$ M.A. Nagarajan \\
\hspace*{20mm} Private communication. 
\vskip 1mm
\noindent{\bf (NA85)} $\> \> \>$ M.A. Nagarajan, C. Mahaux et G.R. Satchler, \\
\hspace*{20mm} Phys. Lett. 54 (1985) 1136. 
\vskip 1mm
\noindent{\bf (PA91)} $\> \> \>$J.C. Pacheco, J.L. Ferrero, N. Vinh Mau et 
B. Bilwes, \\
\hspace*{20mm} Phys. Lett. B267 (1991) 455. 
\newpage
\noindent{\bf (PA95)} $\> \> \>$J.C. Pacheco, B. Bilwes, F. Sanchez, 
J.A. Ruiz, J. Diaz, J.L. Ferrero\\ 
\hspace*{20mm} et M.D. Kadi-Hanifi, \\
\hspace*{20mm} Nucl. Phys. A588 (1995) 537. 
\vskip 1mm
\noindent{\bf (PA97)} $\> \> \>$ 	J.C. Pacheco, \\
\hspace*{20mm} (1997) non publi\'e. 
\vskip 1mm
\noindent{\bf (PAC95)} $ \> \>$  J.C. Pacheco, \\
\hspace*{20mm} (1995) non publi\'e. 
\vskip 1mm
\noindent{\bf (PE62)} $\> \> \>$ 	F.G. Perey et B. Buck, \\
\hspace*{20mm} Nucl. Phys. 32 (1962) 363. 
\vskip 1mm
\noindent{\bf (PE64)} $\> \> \>$ 	F.G. Perey et D.S Saxon, \\
\hspace*{20mm} Phys. Lett. 10 (1964) 107. 
\vskip 1mm
\noindent{\bf (PE80)} $\> \> \>$ 	R. Peierls et N. Vinh Mau, \\
\hspace*{20mm} Nucl. Phys. A343 (1980) 1. 
\vskip 1mm
\noindent{\bf (PO76)} $\> \> \>$ J.E. Poling, E. Norbeck et R.R. Carlson, \\
\hspace*{20mm} Phys. Rev. C13 (1976) 648. 
\vskip 1mm
\noindent{\bf (PO83)} $ \> \>$  G. Pollarolo, R.A. Broglia, et A. Winther, \\
\hspace*{20mm} Nucl. Phys. A406 (1983) 369 
\vskip 1mm
\noindent{\bf (RA81)} $\> \> \>$ 	J. Raynal, \\
\hspace*{20mm} Phys. Rev. C23 (1981) 2571
\vskip 1mm
\noindent{\bf (RA94)} $\> \> \>$ 	J. Raynal, \\
\hspace*{20mm} Nuclear Data Bank of the OCDE (1994). 
\vskip 1mm
\noindent{\bf (RO)  } $\> \> \> \> \>$ 	P. Roussel, \\
\hspace*{20mm} Communication priv\'ee.
\vskip 1mm
\noindent{\bf (RO86)} $\> \> \>$ 	P. Roussel, \\
\hspace*{20mm} Th\`ese d'\'etat. Universit\'e Paris-Sud. Centre d'Orsay (1986).
\vskip 1mm
\noindent{\bf (RO88)} $\> \> \>$ 	P. Roussel-Chomaz, N. Alamanos, F. 
Auger, J. Barrette, B. Berthier,\\
\hspace*{20mm} B. Fernandez, L. Papineau, H. Doubre et W. Mittig,\\
\hspace*{20mm} Nucl. Phys. A477 (1988) 345
\vskip 1mm
\noindent{\bf (SA79)} $\> \> \>$ 	G.R. Satchler et G.W. Lowe, \\
\hspace*{20mm} Phys Rep. 55 (1979) 183. 
\vskip 1mm
\noindent{\bf (SA82)} $\> \> \>$ 	A.B. Santra et B. Shina, \\
\hspace*{20mm} Phys. Lett. 110B (1982) 359. 
\vskip 1mm
\noindent{\bf (SA83)} $\> \> \>$ 	G.R. Satchler, \\
\hspace*{20mm} Direct Nuclear Reactions (Oxford University Press 1983)	
Sect.\\
\hspace*{20mm} 4.3, p. 126. 
\vskip 1mm
\noindent{\bf (SA89)} $\> \> \>$ 	F. Sanchez, \\
\hspace*{20mm} Th\`ese de doctorat de l'universit\'e de Valencia, Espagne 
(1989). 
\vskip 1mm
\noindent{\bf (SC69)} $\> \> \>$ 	L.I. Schiff, \\
\hspace*{20mm} Quantum Mechanics, Mc Graw-Hill, New York, (1969). 
\vskip 1mm
\noindent{\bf (SO92)} $\> \> \>$ 	J.H. S\"orensen et A. Winther, \\
\hspace*{20mm} Nucl. Phys. A550 (1992) 329. 
\vskip 1mm
\noindent{\bf (ST85)} $\> \> \>$ 	L. Stuttg\'e, \\
\hspace*{20mm} Th\`ese de doctorat de l'universit\'e Louis Pasteur, 
Strasbourg,\\
\hspace*{20mm} France (1985). 
\vskip 1mm
\noindent{\bf (TH85)} $\> \> \>$ I.J. Thompson, M.A. Nagarajan, J.S. Lilley et 
B.R. Fulton, \\
\hspace*{20mm} Phys. Lett. 157B (1985) 250. 
\vskip 1mm
\noindent{\bf (VA75)} $\> \> \> \>$ 	R. Vandenbosch et M.P. Webb,\\
\hspace*{20mm} Univ. of Washington Nuc. Phys. Lab. Annual Rep. (1975) 136,\\ 
\hspace*{20mm} unpublished.
\vskip 1mm
\noindent{\bf (VI77)} $\> \> \> \>$ 	F. Videbaek, R.B. Golgstein, L. 
Grodzins, S.G. Steadman, T.A. Belote\\
\hspace*{20mm} et J.D. Garrett,\\
\hspace*{20mm} Phys. Rev. C15 (1977) 954. 
\vskip 1mm				
\noindent{\bf (VI86)} $\> \> \> \>$ 	N. Vinh Mau, \\
\hspace*{20mm} Nucl. Phys. A457 (1986) 413. 
\vskip 1mm
\noindent{\bf (VI87)} $\> \> \> \>$ 	N. Vinh Mau, \\
\hspace*{20mm} Nucl. Phys. A470 (1987) 406. 
\vskip 1mm
\noindent{\bf (VI90)} $\> \> \> \>$ 	N. Vinh Mau, \\
\hspace*{20mm} Inst. Phys. Conf. Ser. N 110 (1990) 1 
\vskip 1mm
\noindent{\bf (VI91)} $\> \> \> \>$ N. Vinh Mau, J.L. Ferrero, J.C. Pacheco et 
B. Bilwes, \\
\hspace*{20mm} Nucl. Phys. A531 (1991) 435. et Phys. Rev. C47 (1993) 899. 
\vskip 1mm				
\noindent{\bf (VI93)} $\> \> \> \>$ N. Vinh Mau, J.C. Pacheco, J.L. Ferrero et 
B. Bilwes, \\
\hspace*{20mm} Nucl. Phys. A560 (1993) 879. 
\newpage				
\noindent{\bf (VIN93)} $\> \> \> \>$ 	N.Vinh Mau, \\
\hspace*{20mm} Th\'eorie des r\'eactions nucl\'eaires : Mod\`eles 
d'interaction directe,\\
\hspace*{20mm} Institut de Physique Corpusculaire, Valencia, Espagne (1993). 
\vskip 1mm
\noindent{\bf (VR87)} $\> \> \>$ 	H. de Vries, C.W. de Jager et C. de 
Vries,\\
\hspace*{20mm} Atomic. Data and Nuclear Data Tables, 36 (1987) 495. 
\vskip 1mm
\noindent{\bf (WE72)} $\> \> \>$ 	L. Wendling, \\
\hspace*{20mm} M\'emoire d'ing\'enieur, CNAM, Mulhouse, (1972). 
\vskip 1mm
\noindent{\bf (WO54)} $\> \> \>$ 	R.D. Woods et D.S. Saxon, \\
\hspace*{21mm} Phys. Rev. 95 (1954) 577. 
\vskip 1mm
\noindent{\bf (ZE77)} $\> \> \> \>$ 	G. Zenhacker, \\
\hspace*{20mm} Rapport C.R.N., S.A.T.D., Strasbourg, France (1977). 
\vskip 1mm

\newpage
$ $
\newpage
\end{document}